%% file: CircuitQED_LectureNotes_Langford.tex
\documentclass[12pt,a4paper]{article}

\usepackage{lecturesstyleNKL}

\include{lecturesdefinitions}

\begin{document}

\title{Circuit QED --- Lecture Notes}

\author{ Nathan~K.~Langford \\%
\multicolumn{1}{p{.65\textwidth}}{\centering\small{\emph{Department of Physics, Royal Holloway University of London, Egham Hill, Egham, Surrey  TW20 0EX, United Kingdom}}} }

\date{}

\maketitle

\abstract{The new and rapidly growing field of \emph{circuit QED} offers extremely exciting prospects for learning about and exercising intimate control over quantum systems, providing flexible, engineerable design and strong nonlinearities and interactions in systems consisting of microwave radiation fields and fixed artificial ``atoms''.  These notes aim to provide a non-expert introduction to the field of circuit QED, to give a basic appreciation of the promise and challenges of the field, along with a number of key concepts that will hopefully be useful for the reader who is new to the field and beginning to explore the research literature.  They were written as a pedagogical text designed to complement a course delivered to third-year undergraduate students.

After a introductory section which discusses why studying circuit QED might be worthwhile and interesting, I introduce the basic theory tools from quantum optics and quantum information which are needed to understand the key elements of circuit QED.  I also provide a brief overview of superconductivity, focussing on the concepts which are most relevant to operation in the regimes of interest in circuit QED.  I then describe the three main types of superconducting qubits, and finally give a basic introduction to decoherence and mixture and how they relate to quantum behaviour in electronic circuits.}

\tableofcontents

\section*{Preface to the notes}
\addcontentsline{toc}{section}{Preface to the notes}

These notes were prepared as a module in a third-year undergraduate course for students who had not yet necessarily taken an advanced quantum mechanics course.  The notes are therefore designed to be a pedagogical text at undergraduate level, not a research review.  With very few exceptions, I have therefore not tried to provide references to research literature throughout the notes.

As much as possible, I have not assumed prior knowledge of superconductivity, and the notes are written within the context of the pure-state formalism of quantum theory, without using techniques or formalism associated with density matrices and mixed states.  Consequently, a more advanced reader (e.g., at postgraduate level) should be able to skim over or skip substantial portions of the course notes.

It is also perhaps worth noting that the exercises, which were originally provided as Problem Sheets for the students, do \emph{not} generally revise material that has already been explicitly covered in the notes.  They are instead designed to guide the reader through other important distinct concepts, which are complementary to the material in the notes.

For readers who wish to explore the topic in greater detail and rigour than was possible within the constraints of this course, the following references may be useful.  Michel Devoret's seminal lecture notes from \emph{Les Houches}~\cite{DevoretMH1995qfe} provide a detailed description of how to analyse quantum electronic circuits, generalising the earlier work of Yurke and Denker~\cite{YurkeB1984qnt}.  A more specific introduction into using these techniques in the context of superconducting qubits is given in Ref.~\cite{DevoretMH2004sqs}.  More general research reviews of superconducting circuits and circuit QED can be found in~\cite{ClarkeJ2008sqb,YouJQ2011apq,YouJQ2005scq}.

Due to the constraints of the course, I have not included much information about practical circuit fabrication and analysis techniques, but have instead tried to focus on how to build (conceptually) and describe simple quantum systems (the key elements) from electronic circuits.  I have obviously therefore glossed over an important step in how to connect the schematic level to the physical implementations.  For more detail on this aspect, the research reviews cited above~\cite{ClarkeJ2008sqb,YouJQ2011apq,YouJQ2005scq} provide a good place to start.

\section{Motivation: Why should I care about Circuit QED?}

\subsection{Controlling individual quantum systems: Nobel Prize 2012}

The 2012 Nobel prize in Physics went to Serge Haroche and Dave Wineland ``\emph{for ground-breaking experimental methods that enable measuring and manipulation of individual quantum systems}''.

\begin{description}[nosep]
\item[$\quad$] \url{http://www.nobelprize.org/nobel_prizes/physics/laureates/2012/}
\end{description}

In particular, Serge Haroche carried out his Nobel-winning work in the area of cavity quantum electrodynamics (cavity QED), which in turn inspired the field of circuit QED.  Here is the text from the press release:

{\small ``\emph{Serge Haroche and David J. Wineland have independently invented and developed methods for measuring and manipulating individual particles while preserving their quantum-mechanical nature, in ways that were previously thought unattainable.}

``The Nobel Laureates have opened the door to a new era of experimentation with quantum physics by demonstrating the direct observation of individual quantum particles without destroying them. For single particles of light or matter the laws of classical physics cease to apply and quantum physics takes over. But single particles are not easily isolated from their surrounding environment and they lose their mysterious quantum properties as soon as they interact with the outside world. Thus many seemingly bizarre phenomena predicted by quantum physics could not be directly observed, and researchers could only carry out thought experiments that might in principle manifest these bizarre phenomena.

``Through their ingenious laboratory methods Haroche and Wineland together with their research groups have managed to measure and control very fragile quantum states, which were previously thought inaccessible for direct observation. The new methods allow them to examine, control and count the particles.

``Their methods have many things in common. David Wineland traps electrically charged atoms, or ions, controlling and measuring them with light, or photons.

``Serge Haroche takes the opposite approach: he controls and measures trapped photons, or particles of light, by sending atoms through a trap.''}

Okay, so that's all well and good, but why is it so interesting to be able to control individual quantum systems?  Well, the first reason is just ``because it's there''.  We believe that quantum mechanics is the way our world works and yet almost everything we see on a day-to-day basis conforms quite happily to what we think of as classical physics.  So being able to pick up and manipulate individual quantum systems is just inherently cool.  But there are other reasons too---this ability is important for both \emph{fundamental} and \emph{application-based} research.  (This is science jargon for research that helps us \emph{understand} stuff and research that helps us \emph{do} stuff.)

\textbf{Fundamental research}:  Although quantum mechanics is the theory that we believes governs almost all aspects of physics (barring the complication of fitting it in with general relativity and gravity), it is still difficult to study it directly.  Because most of what we see and do on an everyday basis is governed by classical physics, we generally only see the effects of quantum mechanics indirectly, e.g., if we're looking at the spectrum of black-body radiation.  Studying individual quantum systems provides us with the means of directly testing the foundations of the quantum theory.

Most of the remarkable predictions of quantum mechanics arise from two underlying principles:  the principles of \emph{superposition} and \emph{entanglement}.  These two features give rise to some very bizarre consequences that seem to contradict our very basic physical intuitions about the nature of reality, such as the existence of nonlocal quantum interactions which seem to violate the no-signalling principle of special relativity (that no information can be transferred faster than the speed of light).  This gave rise to the famous Einstein-Podolsky-Rosen paradox.  As a result, many people have tried to formulate alternative theories to quantum mechanics which predict the same answers without breaking our comfortable beliefs about the reality and locality, etc.  The difficulty is how do we test these theories?  In the last 50 years, many people have therefore tried to come up with experiments that can distinguish between the predictions of quantum mechanics and these alternative theories.  However, because these differences are very subtle, these tests normally require the ability to exert very precise control over quantum systems in order to be able to prove they are behaving quantum mechanically and not otherwise.

\begin{itemize}[topsep=0mm,partopsep=0mm]

\item \emph{Local realism}:  Probably the most famous fundamental test of quantum mechanics is related to the debate between Schr\"odinger and Einstein, Podolsky and Rosen over the nonlocality predicted by entanglement, and which led to the EPR paradox.  The question is: are measurements on entangled states really random, or are they really predetermined by some underlying theory which we just don't happen to know about?  In 1964, John Bell devised an experimental test that could distinguish between these two questions, which required being able to prepare special entangled states which became known as Bell states (which we will learn about later).  These experimental tests are known as \emph{Bell tests} and people talk about ``violating Bell's inequalities''.  Bell experiments have since been performed many times, but no one has yet managed to close all practical ``loopholes'' to conclusively demonstrate a Bell violation.  

\item \emph{Other tests}: Other tests include \emph{EPR steering} (local realism), \emph{Kochen-Specker tests} (noncontextuality), and \emph{Leggett-Garg tests} (macroscopic realism).

\end{itemize}

\textbf{Applications}:  If we accept that quantum mechanics is correct, the next question to ask is what can we do with it?  Well, it turns out that there we can in fact use quantum systems to do certain things that are completely impossible with classical physics.  These so-called \emph{quantum technologies} have largely grown out of a field which is now called \emph{quantum information science}, which incorporates not only theory, but also a whole range of areas in experimental physics, including photonic, trapped-ion, cold-atom and quantum electronics (including circuit QED) systems.  These technologies are something quite different and new.

Over the last century, we have seen the birth of many technologies which somehow rely on quantum mechanics, e.g., lasers and computer transistors.  Even these two examples are used absolutely everywhere.  Lasers are used in CD, DVD and Blu-Ray players, in supermarket price scanners, and in global telecommunications.  And transistors are, well, everywhere we have a computer chip.  At the microscopic level, the details of a laser's or transistor's behaviour rely on quantum physics, in that we need quantum physics to accurately predict their relevant operating parameters, such as frequency for a laser or optimal voltage bias for a transistor.  However, these are still ultimately classical machines, because if we ignore their microscopic details and just look at them as black-box machines, they still obey classical physics.

The quantum technologies I'm talking about, however, are fundamentally quantum machines, because even at the black-box level, they operate according to quantum physics.  This requires extreme levels of control over complex quantum systems (perhaps made of many individual quantum systems) and this is very difficult to achieve.  Most experiments so far have only provided ``proof-of-principle'' demonstrations of how such fundamentally quantum machines might operate.

Some examples of quantum technologies are:

\begin{itemize}[topsep=0mm,partopsep=0mm]

\item Quantum computers

\item Quantum communication: quantum key distribution

\item Quantum simulations

\item Quantum metrology

\end{itemize}

\subsection{Why is it so hard?}

If everything behaves according to quantum mechanics at the most fundamental level, why is it so hard to see these effects and control them?  Well, the problem is that quantum effects are very, very fragile and the real world is a very, very noisy place.  We are talking about capturing a single photon, or a single electron, or a single atom, or a single ion and making it do exactly what we want.  But all the while, that single quantum system is being bumped and jostled and bombarding by photons, fields and other atoms from the surrounding environment.  And the environment is a really big place, with a lot of heat (vibrations), light and stray fields all over the place.  Imagine trying to do a very delicate, difficult acrobatics trick in the middle of a huge crowd of people at a rock concert or a football match!

So what we really need to do is build a closed quantum system that is completely isolated from its environment.  Not so difficult, surely?  Just put it in a vacuum and shield it from external radiation and fields by dunking it in a refrigerator and shielding it inside a completely closed metal box...  But this is the key to the problem.  Actually, we \emph{don't want} a completely isolated quantum system.  If it were completely closed off from the environment, how would we be able to prepare its initial conditions, manipulate it and measure it?  All these things require us to be able to interact with the system.

This is the great challenge:  we want to build a quantum system which is completely isolated from the environment, \emph{except} via a few, very specific knobs that we can control, \emph{but} these control knobs have to be designed is a very special way that they don't just immediately allow the bad effects of the environment to leak straight back in.

The unwanted effects of the environment are usually called \emph{decoherence} and we will talk about these more later.

\subsection{Natural quantum systems vs engineered quantum systems}

There are two basic approaches to building individual quantum systems: \emph{nature} vs \emph{engineering}.

\textbf{The first approach} is to look for systems that already exist in nature and carefully isolate individual parts of those systems which are small enough to naturally exhibit quantum effects and use them.

\begin{itemize}[topsep=0mm,partopsep=0mm]

\item \emph{Natural quantum systems}: Examples are photons, single atoms and single ions.

\item \emph{Key advantages}: 1) isolating a single, fundamental particle, like a photon or an ion, can give a system which is very strongly quantum mechanical---they are fundamental particles which are so small that only quantum physics can describe their behaviour; and 2) it may be possible to achieve very good isolation from the environment.

\item \emph{Key disadvantages}: 1) we are stuck with what nature provides us---e.g., we don't get to choose the energy levels of our atoms; 2) we often only have access to fairly weak methods for interacting with the system---e.g., it is hard to shine a laser on an atom or ion in such a way as to achieve a very strong coupling between the two; and 3) because the systems are so small, isolating, controlling and measuring an individual system can require incredibly precise tools, which are hard to make and hard to operate at the required level.

\end{itemize}

\textbf{The second approach} is to build man-made systems that, by some trick, behave in a collective manner which is much simpler than the underlying atomic-scale complexity of the systems.  

\begin{itemize}[topsep=0mm,partopsep=0mm]

\item \emph{Engineered quantum systems}: Examples are superconducting electronic circuits and ``quantum'' mechanical systems (i.e., mechanical systems which behave ``quantumly''---physical oscillators, like very small springs or styluses).

\item \emph{Key advantages}: 1) we get to build them ourselves and this gives us a huge amount of flexibility in how we design them---subject only to the engineering limitations of our fabrication techniques, we can really tune and tweak all the operating parameters to exactly what we want them to be; 2) the collective degrees of freedom which govern their quantum behaviour can couple very strongly to our control tools, so that we can manipulate them very precisely; and 3) we can study quantum mechanical effects in ``macroscopic'' (large, visible) systems, which opens up new territory for quantum mechanics.

\item \emph{Key disadvantages}: 1) because the systems are large and complex, they often interact very strongly with the environment, which can create lots of noise and decoherence, and getting them into a regime where the quantum effects dominate the environmental noise can be very challenging, such as requiring cooling to very low temperatures (like the millikelvin temperatures in a dilution refrigerator); 2) building systems which exhibit strong quantum effects requires incredibly precise fabrication techniques, so that the engineered systems are still very ``pure'' in some sense (this can help limit environmental noise); and 3) the size of noise in the systems often still makes it difficult to precisely measure the systems without destroying the fragile quantum effects, meaning that the measurement, preparation and manipulation tools required still need to be very precise.

\end{itemize}

One thing we need to be very careful to avoid for both natural and engineered quantum systems is energy dissipation.  Quantum effects are very fragile and one of the easiest ways for them to be destroyed is for the energy stored in the quantum system to vanish (leak away) into the environment.  

Things look pretty bright for engineered quantum systems, however.  For example, superconductivity provides a classical example of how a complex physical system can behave in an extremely simple collective manner which belies the atomic complexity of the material.  We'll talk about this in more detail much later, but essentially, when a metal ``goes superconducting'', all of the free electrons couple together in special pairs (called Cooper pairs), which are indistinguishable from each other (they are bosons, like photons), and this allows them to form a large-scale \emph{condensate}, which is a single, macroscopic quantum state, so all of the electrons behave as if they are somehow synchronised together.

Also, because superconducting electrons don't experience any resistance, there is no energy dissipation created by normal electronic resistance to destroy the quantum effects.  (There are other forms of energy dissipation, of course, but this is a great head start.)

Finally, because of this collective behaviour, it turns out that it's possible to build simple electronic circuits which behave very simply, even at a quantum mechanical level, despite the fact that they are in fact made up of billions of atoms.  For example, you probably all at least vaguely remember the LC electronic oscillator, which is formed when you connect a inductor and capacitor in parallel with each other.  Well, it turns out that, if you build such an oscillator and then cool it down to very low temperatures, the simple, collective oscillator degree of freedom doesn't interact with any other degrees of freedom in the circuit (like atomic vibrations, etc), and the circuit starts behaving like an isolated quantum mechanical oscillator.

This is just a taste of how this all works.

\subsection{Cavity QED and Circuit QED}

\begin{figure}[ht]
\includegraphics[width=\textwidth]{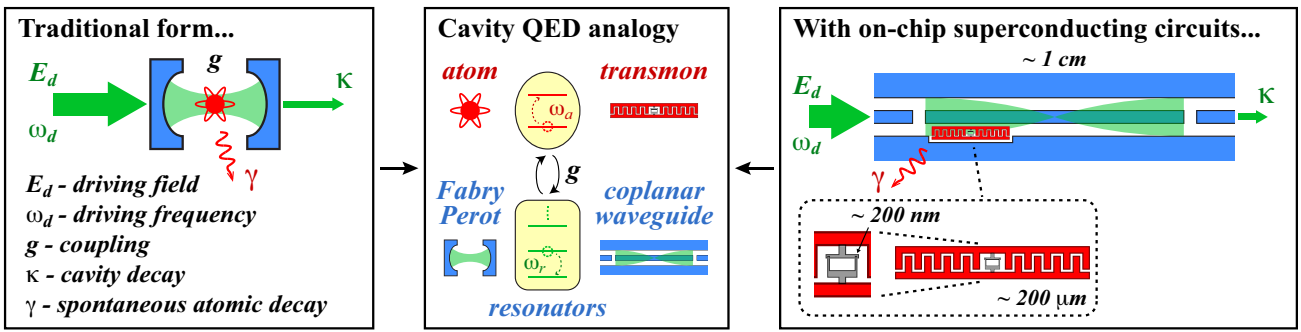}
\captionsetup{labelfont=bf,font={small},margin=6pt}
\caption{\emph{\textbf{ Analogy between traditional cavity QED and its superconducting circuit-based form.}} The cavity ``mirrors'' in the circuit form are capacitors in the waveguide (gaps in the central conductor). The artificial atom shown, a transmon, is an anharmonic electronic oscillator (a parallel capacitor and nonlinear SQUID inductor); its two lowest energy levels form a qubit, coupling to the waveguide via its electric dipole (rough ``on-chip'' sizes indicated). The large capacitor makes it largely immune to fluctuating charges in the surrounding environment. The figure shows a nonlinear resonator, created by inserting a SQUID in the waveguide's central conductor, illustrating the platform's flexibility.}
\end{figure}

Comparing the traditional and circuit-based forms of cavity QED provides a classic example of the parallel between the \emph{nature} vs \emph{engineering} approaches to building quantum systems.

Quantum electrodynamics (QED) is the study of the fully quantum interaction between light (the electromagnetic field) and atoms.  The original goal of QED was to study atoms and to better understand their energy levels and internal electronic structure.  This then became a key underlying part of the broader field of quantum optics, when people started becoming interested the underlying quantum nature of the light itself, because QED describes in some sense the interaction which powers the effects in quantum optics.  Interest in quantum optics arose out of the invention of the laser and the possibilities which it opened up for studying new states of light.

The main goal of cavity QED is to study at the most detailed level the interaction between light and matter and to study the quantum states of each.  In the Nobel prize experiments by Serge Haroche, he uses individual atoms as a way to probe the behaviour of the light field, which is in his case in the microwave range.  There are also other cavity QED systems and experiments which use optical wavelengths. 

The outline is:
\begin{itemize}[topsep=0mm,partopsep=0mm]

\item Interaction between a single atom and a beam of light is very weak.  This makes it very hard to see the effects which arise at the single-particle level (atom or photon).

\item Solution:  confine the light in a cavity, so that it bounces back and forth.  This means that a single photon can interact with an atom many, many times before they decay away, amplifying the effects of the interaction.

\item This requires very high ``quality'' cavities.  The more reflective the mirrors, the more times the light will bounce back and forth.

\item The key elements in cavity QED are an ``atomic'' element and a field mode.

\item The challenges for traditional cavity QED are to get an atom to stay in the one place to interact with the electromagnetic field for a long time, and to match the size of the field mode closely enough to the size of the atom to make sure that the atom is interacting more strongly with the cavity mode than any other stray electromagnetic radiation that may be flying around.

\item In circuit QED, the ``atomic'' element is an artificial atom made from a small-scale electronic circuit which can be positioned very precisely wherever we want it and it doesn't move around, like normal atoms.  In their simplest form, these artificial atoms are designed to behave like ideal two-level quantum systems: quantum bits, or \emph{qubits}.  There are many different types of artificial atoms or qubits, e.g., the Cooper-pair box qubit, the current-biased phase qubit and flux qubits.

\item In circuit QED, the microwave radiation is very strongly confined (spatially) in a waveguide, which enables a very strong coupling between the microwave field and the artificial atom.

\end{itemize}

\subsection{Simple on-chip electronic components for circuit QED}

There are two key types of on-chip electronic components that we can use:  \emph{lumped elements} and \emph{distributed elements}.  The two types are distinguished by how large they are by comparison with the wavelength of microwave radiation at the relevant frequency.  For lumped elements, the size of the component, $l$, is much smaller than the wavelength, $l \ll \lambda$, whereas for distributed elements, the size of the component is roughly the same size as the wavelength, or bigger, $l \gtrsim \lambda$.

You probably recall that electronic signals are completely tied up with electromagnetic radiation.  For example, antennas are components which are designed to convert electronic signals to non-confined, travelling electromagnetic radiation (transmitters) or, vice versa, to convert radiation into electronic signals (receivers).  The same is true of microwave radiation and electronic signals in the electronic circuits we are going to talk about.

Physically, it is the radiation that characterises the difference between the two regimes.  If a component is much smaller than the relevant wavelength (\emph{a lumped element}), this means that it isn't necessary to think about the field which mediates the interaction between the different electron movements that define the current flow in the circuit.  The field has such a short distance to propagate that you don't need to consider how it travels from one place to another.  The behaviour of the component is completely defined by the currents and voltages and you don't need to think about radiation.  In some sense, it is the motion of the electrons that carries the signals (the information).  In this regime, the circuit operates much as you are familiar with from standard electronics.  You can think about capacitance, inductance, resistance, etc.


If, however, a component is similar in size to the relevant microwave wavelength, or larger (\emph{a distributed element}), the propagation of the fields from one part of the circuit to another becomes a critical effect and you need to start thinking about waves and phases.  Now you need to start thinking about the radiation and its behaviour.  It is now the radiation that carries the signals (the information).


\textbf{Lumped elements}
\begin{description}[topsep=0mm,parsep=0mm,partopsep=0mm,leftmargin=1cm,style=sameline,font=\it]

\item[Capacitors:]  As usual, these are elements which can store energy in the electric field.

\item[Inductors:]  As usual, these are elements which can store energy in the magnetic field.

\item[LC resonators:]  These are simple harmonic oscillators.

\item[Resistors:]  These elements dissipate energy.  We generally try to avoid these entirely in circuit QED, because energy dissipation is so detrimental to quantum effects.

\item[Josephson junctions:]  These are non-dissipative (no energy loss), nonlinear elements, and it is these which allow us to create artificial atoms.  We will talk about these in much more detail later.

\end{description}

\textbf{Distributed elements}
\begin{description}[topsep=0mm,parsep=0mm,partopsep=0mm,leftmargin=1cm,style=sameline,font=\it]

\item[Coplanar waveguides:]  Like coaxial cable.  Used to guide microwave radiation around on the electronic chip.  Energy is carried in both electric and magnetic fields.  To do detailed calculations, we need to think about capacitance per unit length and inductance per unit length.

\item[Waveguide cavities:]  Waveguide resonators behave like musical pipes or optical Fabry-Perot cavities.  They have multiple resonances at different harmonics of the cavity.

\item[Cavity ``mirrors'':]  The cavity ``mirrors'' can be made from either capacitors inserted into the central conductor (this creates a node in the cavity magnetic field).  In this case, microwave radiation is partially transmitted through the break in the central conductor.  The amount of transmission is determined by the amount of overlap between evanescent fields of the radiation from either side of the gap.  Alternatively, the cavity can be completely capped by connecting the central conductor to ground at one end.  This creates a node in the electric field of the cavity at that point.  Radiation is then perfectly reflected.

\end{description}

\section{Basic theory tools}

Cavity QED is the study of the interaction between atoms and the field modes inside a cavity.  Both atoms and cavities are systems that have resonances, so we start by looking at the simplest resonant system, the simple harmonic oscillator.

\subsection{The quantum simple harmonic oscillator}

Consider a mass $m$ on a spring with spring constant $k$.  The position of the mass is $x$ and the velocity is $v = dx/dt$.  We can solve the classical equations of motion as follows:
\begin{align}
F = ma &= -kx \\
\label{eq:sho-eqnmotion}
m \ddot{x} & = -kx \\
\ddot{x} & = -\frac{k}{m} x \\
\gives  x(t) &= A \cos \omega_0 t  + B \sin \omega_0 t \\
& \text{where } \omega_0 = \sqrt{k/m}
\end{align}
You can check this is a solution by substituting $x(t)$ back into Eq.~\ref{eq:sho-eqnmotion}.  But basically, we already see the key feature of a simple harmonic oscillator, namely that it is a resonant system (i.e., it exhibits periodic, sinusoidal oscillations) with a characteristic frequency given by $\omega_0 = \sqrt{k/m}$.

To look at the quantum case, we write down the total energy, i.e. the Hamiltonian:
\begin{align}
H &= KE + PE \\
&= \frac{1}{2} mv^2 + \frac{1}{2} kx^2 \\
&= \frac{p^2}{2m} + \frac{1}{2} m \omega_0^2 x^2
\end{align}
At this point, we can already make some guess as to what is going to happen.  This system looks like a particle moving freely in a quadratic potential.  Whenever a quantum particle is moving in a potential well, we get quantised energy levels.

Why do boundary conditions lead to quantised energy levels?  It's just wave mechanics, exactly like organ pipes (sound waves) or bathtubs (water waves), where you get resonances only at certain discrete frequencies.  It's just that, in quantum mechanics, the particles behave like waves too.

The question is what do those energy levels look like?

To make the harmonic oscillator quantum, we simply replace $x$ and $p$ by the usual position and momentum operators, $\hat{x}$ and $\hat{p}$.  Since they do not commute, we can also write down the usual commutator, $\sqbr{x,p}=i\hbar$.  In other words, the quantum simple harmonic oscillator is described by the quantum Hamiltonian:
\begin{align}
\hat{H} &= \frac{\hat{p}^2}{2m} + \frac{1}{2} m \omega_0^2 \hat{x}^2
\end{align}
Throughout these notes, I will often drop the ``hats'' on the operators, when it is clear from context that it is an operator and not a complex number.

Now, while all the magic of quantum mechanics is already contained inside these equations (the Hamiltonian and the commutator), we still don't really know what it means in terms of the specific energy levels.  A simple way to solve this is to make the following substitutions:
\begin{align}
\hat{x} &= \sqrt{\frac{\hbar}{2m\omega_0}} (a^\dag + a) \\
\hat{p} &= i\sqrt{\frac{\hbar m\omega_0}{2}} (a^\dag - a)
\end{align}
Here, $a$ and $a^\dag$ are called the \emph{annihilation} and \emph{creation} operators, for reasons which will become clear.

We now start by calculating the commutation relations between these new operators from the commutator between $x$ and $p$.
\begin{align}
i\hbar &= \sqbr{x,p} \\
&= i \sqrt{\frac{\hbar}{2m\omega_0}} \sqrt{\frac{\hbar m\omega_0}{2}} \sqbr{a^\dag + a, a^\dag - a} \\
&= \frac{i\hbar}{2} \br{ \sqbr{a^\dag,a^\dag} - \sqbr{a^\dag,a} + \sqbr{a,a^\dag} + \sqbr{a,a} }
\end{align}
At this point, we recall the commutator identities that $\sqbr{A,A} = 0$ and $\sqbr{A,B} = -\sqbr{B,A}$.  Thus, we have:
\begin{align}
\sqbr{a,a^\dag} = 1
\end{align}
So, like $x$ and $p$, $a$ and $a^\dag$ are also non-commuting observables, which means that they contain complementary information that cannot be simultaneously accessed by measurement.

We now calculate the new Hamiltonian in terms of annihilation and creation operators:
\begin{align}
H &= -\frac{1}{2m} \frac{\hbar m\omega_0}{2} (a^\dag - a)^2 + \frac{m \omega_0^2}{2} \frac{\hbar}{2m\omega_0} (a^\dag + a)^2 \\
&= \frac{\hbar \omega_0}{2} (a a^\dag + a^\dag a) \\
\label{eq:sho-hamiltonian}
&= \hbar \omega_0 (a^\dag a + \half)
\end{align}

Finally, we also define the energy eigenstates and eigenvalues via the equation:
\begin{align}
\hat{H} \ket{n} = E_n \ket{n}
\end{align}
For a simple harmonic oscillator, these states are called Fock states and we will be able to show that $n$ is the number of excitations.  There are many different types of excitations: e.g., photons (for electromagnetic radiation), phonons (for vibrations), etc.

But we still need to work out what the energy eigenstates are.  We can solve this by asking what $a$ and $a^\dag$ actually do.  Using the commutation relations, we can show that:
\begin{align}
\label{eq:sho-raising}
\hat{H} \hat{a}^\dag \ket{n} &= (E_n+\hbar \omega_0) \hat{a}^\dag \ket{n} \\
\label{eq:sho-lowering}
\hat{H} \hat{a} \ket{n} &= (E_n-\hbar \omega_0) \hat{a} \ket{n}
\end{align}
In other words, for any given energy eigenstate $\ket{n}$ with eigenvalue $E_n$, we can show that $\hat{a}^\dag\ket{n}$ is another energy eigenstate with energy $E_n+\hbar\omega_0$.  This shows us that the energy eigenstates form an infinite ladder of energy levels, where each energy level is $\hbar\omega_0$ away from the adjacent levels---the energy spacing between adjacent levels is a fixed, constant value related to the characteristic frequency of the oscillator.

Likewise, we know that $\hat{a}\ket{n}$ is another energy eigenstate with energy $E_n-\hbar\omega_0$.  However, this can't go down forever.  We know that negative (total) energies are impossible and there must therefore be some lowest-energy ground state, $\ket{0}$.  But if $\ket{0}$ is the lowest possible energy level, that means that there is no eigenstate $\hat{a}\ket{0}$ and we must have $\hat{a}\ket{0}=0$, for consistency.  Combining this fact with Eq.~\ref{eq:sho-raising}, we can now calculate the eigenenergies:
\begin{align}
E_n = (n+\half) \hbar\omega_0
\end{align}
and identify $n$ with the number of energy excitations (packets) in the oscillator.  Comparing this equation with Eq.~\ref{eq:sho-hamiltonian}, we can also define the number operator $\hat{n}=\hat{a}^\dag \hat{a}$ via the equation:
\begin{align}
\label{eq:sho-numberop}
\hat{n} \ket{n} = n \ket{n}
\end{align}

Using the concept that $n$ represents the number of energy excitations, each containing energy $\hbar\omega_0$, Eqs~\ref{eq:sho-raising} and~\ref{eq:sho-lowering} show us that $a\ket{n} \propto \ket{n-1}$ and $a^\dag\ket{n} \propto \ket{n+1}$. In fact, more specifically, we can show:
\begin{align}
\hat{a}^\dag \ket{n} &= \sqrt{n+1} \ket{n+1} \\
\hat{a} \ket{n} &= \sqrt{n} \ket{n-1}
\end{align}
where the coefficients on the right-hand side can be calculated from the fact that $\bra{n} \hat{a}^\dag \hat{a} \ket{n} = \bra{n} \hat{n} \ket{n} = n$.  This explains why they are called annihilation and creation operators, because they \emph{annihilate} and \emph{create} a single energy excitation in the oscillator.  They are often also called \emph{raising} and \emph{lowering} operators, for obvious reasons.

\subsection{Evolution of the quantum simple harmonic oscillator}

Now we know what the energy levels are, the next thing to look at is how the energy eigenstates evolve in time.

The following is a brief digression to remind you about quantum evolution.  Quantum states evolve according to the Schr\"odinger equation:
\begin{align}
i\hbar \frac{d}{dt} \ket{\psi(t)} &= H \ket{\psi(t)}.
\end{align}
This is \emph{Hamiltonian evolution}, but it is often also called \emph{unitary evolution}, because, given an input state $\ket{\psi(0)}$, we can always define the output state $\ket{\psi(t)}$ in terms of a unitary operator, $U(t)$, via:
\begin{align}
\ket{\psi(t)} = U(t) \ket{\psi(0)}
\end{align}
Substituting this into the Schr\"odinger equation, we get:
\begin{align}
i\hbar \frac{d}{dt} U(t) \ket{\psi(0)} &= H U(t) \ket{\psi(0)}
\end{align}
Now, since this must be true for all initial states, we therefore get a differential equation in the operators only, which we can then solve:
\begin{align}
i\hbar \frac{d}{dt} U(t) &= H U(t) \\
\frac{d}{dt} U(t) &= \frac{-iH}{\hbar} U(t) \\
\label{eq:unitaryevolution}
\gives U(t) &= \exp\sqbr{ \frac{-iHt}{\hbar} }
\end{align}

Let us assume now that we have a simple harmonic oscillator initially prepared in the energy eigenstate, or Fock state, $\ket{n}$.  We can therefore calculate the state as a function of time using Eq.~\ref{eq:unitaryevolution} and the Hamiltonian we derived above:
\begin{align}
\ket{\psi(t)} = \exp\sqbr{ \frac{-i\hat{H}t}{\hbar} } \ket{\psi(0)} &= \exp\sqbr{ \frac{-it}{\hbar} \hbar\omega_0 \hat{n} } \ket{n} \\
&= \exp\sqbr{ -i n \omega_0 t } \ket{n}.
\end{align}
where we have used the fact that $\exp(\hat{n}) \ket{n} = \exp(n) \ket{n}$.  We have also ignored here the vacuum point energy ($\frac{1}{2} \hbar \omega_0$), since this just produces a global phase for every state.  It is a worthwhile exercise to go through and prove this identity (you will need to use Eq.~\ref{eq:sho-numberop}).

The key point to note here is that the quantum phase of the eigenstate, $\ket{n}$, evolves at a rate which is proportional to $n$.  So if we instead initially prepared a superposition state, $\ket{\psi(0)} = \sum_n c_n \ket{n}$, then each amplitude in the superposition, $c_n$, would evolve at different rates and the superposition would change in time.

\subsection{The quantum LC oscillator}

Consider a simple electronic circuit, with a capacitor, $C$, in parallel with an inductor, $L$.  From standard electromagnetic theory and electronics, we can use the capacitance and inductance to relate the classical currents and voltages via simple first-order differential equations:
\begin{align}
I_C &= C \frac{dV_C}{dt} \\
V_L &= L \frac{dI_L}{dt}
\end{align}
Using Kirchoff's rules, we also know that $I_C = I_L$ (current junction rule) and $V_C+V_L = 0$ (voltage loop rule).  Defining $I_C \equiv I$ and $V_C \equiv V$, we therefore derive the circuit ``equation of motion'':
\begin{align}
I &= C \frac{d}{dt} \br{-L \frac{dI}{dt}} \\
&= -LC \frac{d^2I}{dt^2}  \\
\gives \frac{d^2I}{dt^2} &= -\frac{1}{LC} I \\
\gives I(t) &= A \cos \omega_0 t  + B \sin \omega_0 t \\
& \text{where } \omega_0 = \frac{1}{\sqrt{LC}}
\end{align}
In other words, this simple LC circuit behaves exactly like a simple harmonic oscillator.  This becomes more obvious when we look at the energy of the circuit.  The capacitor stores electrical energy of $\half CV^2$, while the inductor stores magnetic energy of $\half LI^2$.  Rewriting these in terms of the charge on the capacitor, $Q$ (using $Q=CV$ and $\dot{Q} = I$), we get:
\begin{align}
\label{eq:quantumLC-Hcharge}
H = \frac{L}{2}\dot{Q}^2 + \frac{1}{2C} Q^2.
\end{align}
Comparing this with the mechanical oscillator described above, we can define the following relationships:
{\renewcommand{\arraystretch}{1.3}
\begin{table}[h!]
\begin{center}
\begin{tabular}{cc}
mechanical oscillator & LC oscillator \\
\hline
$m$ & $L$ \\
$k$ & $1/C$ \\
$\omega_0 \equiv \sqrt{k/m}$ & $\omega_0 \equiv \sqrt{1/LC}$ \\
$p \equiv mv$ & $p \equiv L\dot{Q} = LI = \Phi$
\end{tabular}\end{center}
\end{table}}
\\ Interestingly, the last line of this table shows that the electrical equivalent of the momentum, in this situation, is just the magnetic flux through the inductor, so the conjugate variables in this simple circuit are the charge on the capacitor and the flux through the inductor.

The key point to realise from this calculation is that the quantum LC circuit is just a quantum simple harmonic oscillator, with quadratic potential energy, quantised, equally spaced energy levels, annihilation and creation operators and Fock states.  All of the standard tools described above follow through in exactly the same way.

\subsection{The waveguide / Fabry-Perot cavity}

The waveguide cavity (or Fabry-Perot cavity) has many resonances with a fundamental mode and higher harmonics, just like a hollow pipe or vibrating string used in musical instruments.  For example, consider a coplanar waveguide cavity formed between two capacitors in the central conducting strip of the waveguide.  As discussed above, this means that there will be an antinode in the electric field at each end of the cavity.  The shortest cavity that can produce a standing-wave resonance is therefore a half-wave cavity, i.e., $L = \lambda/2$.  If we call the resonant frequency of this fundamental mode $\omega_0 = \pi c / L$, then the higher harmonics will have frequencies $\omega_m = (m+1)\omega_0$.  By connecting the central to the ground planes at one end of the cavity, it is possible to create a cavity with an antinode in electric field at one end and a node at the other (the grounded end).  This forms a quarter-wave cavity, i.e., $L = \lambda/4$, with a fundamental resonant frequency $\omega_0 = \pi c / L$, and the higher-order modes have frequencies $\omega_m = (2m+1)\omega_0$.

While we won't go through the rigorous derivation, it is possible to show that each of these modes or harmonics behaves like a simple quantum harmonic oscillator, which can again be described using all of the standard tools.  This time, they are modes of the electromagnetic field (in this case, a standing microwave field), and the resonant frequency is the frequency of the radiation.

If a microwave field is incident upon the resonator which is ``in resonance'' with the cavity (i.e., it has the same frequency), then it will be completely transmitted through the cavity.  If, however, the field is ``out of resonance'' (i.e., its frequency is substantially detuned from any of the cavity's harmonic frequencies), then it will be completely reflected from the cavity.

For convenience, however, because the underlying resonance is still just a simple harmonic oscillator, it is still sometimes useful to model the standing-wave resonance by an equivalent lumped-element LC oscillator, but a different resonator is required for each mode.

\subsection{Atoms vs harmonic oscillators}

Atoms have distinct energy level spacings, which allows us to address individual transitions in isolation from the other transitions.  As with the cavity case, if light impinges on an atom ``out of resonance'' (detuned) from any of the atom's relevant energy level transitions, then it will not induce any change in the atom's state.  The atom's spectrum provides a unique fingerprint for the species of the atom (a fingerprint which describes what frequencies of light the atom will interact with).  In the field of spectroscopy, this fingerprint is used to identify the types of atoms that may be present inside a material, e.g., to identify the different types of atoms inside the sun or other stars.

Being able to address individual transitions is critical for many systems, like lasers.  It allows us to isolate individual parts of the atom.  This is also critical for exercising precise quantum control over the atom.

Although they still show resonances, atoms, which have energy levels with different spacings, obviously represent quite a different situation from harmonic oscillators, which have equally spaced energy levels.  The key element of the harmonic oscillator is that it has a linear restoring force, e.g., $F = -kx$ in the case of a mass on a spring.  In fact, any system with a linear restoring force will result in a harmonic oscillator and therefore have equally spaced levels.  In order to get unequal spacings, we need to have a nonlinear system---e.g., an oscillator with a nonlinear restoring force.  Equivalently, we need to distort the oscillator's potential energy, so that it isn't purely quadratic, because a quadratic potential automatically provides a linear restoring force.

So where do we get a nonlinearity from?  In the Bohr model of the atom, the nonlinearity arises from the Coulomb force, which obeys an inverse square law (which is nonlinear).  In circuit QED, we generally create nonlinearities using a special electronic component, called the Josephson junction, which we'll look at more later.  The Josephson junction is one of the most important elements of circuit QED systems, exactly because it provides this nonlinearity.

\subsection{Qubit basics}

The simplest ``atom-like'' system we can imagine is a two-level system.  This is exactly what is used in Haroche's Nobel prize cavity QED experiments.  We call such a system a quantum bit, or \emph{qubit}.

There are two ways of creating a qubit.  One is to use a finite-dimensional system which intrinsically has two discrete energy levels.  Simple examples of this would be to use the polarisation of a photon or the spin of an elementary spin-$\half$ particle like an electron.  Such systems don't tend to arise naturally in engineered systems, however, which have many internal degrees of freedom and don't lead naturally to systems with highly restricted dimensionality.

An alternative approach to getting a qubit is to use a multilevel system (like an oscillator or atom) which is sufficiently nonlinear to allow us to isolate two levels from its larger overall Hilbert space.  This is the technique used predominantly in experimental systems like trapped ions, cold atoms and superconducting qubits (in circuit QED).

The qubit and the cavity are the two fundamental units of cavity QED, so we will now develop the basic tools we will need to describe a qubit system.

At the simplest level, a qubit is any quantum system which has just two basis states which describe its state space (Hilbert space).  We call these states the \emph{logical basis states} or \emph{computational basis states}, and write them as $\ket{\zero}$ and $\ket{\one}$.  In quantum information, these logical states are not necessarily connected to an obvious eigenstate of the system (although they can be)---they are simply the basis which we define to represent or \emph{encode} (or carry) the quantum information.  This abstraction is part of the power of quantum information, because it 1) allows completely different experimental systems to attack the same underlying problems and compare their outcomes meaningfully, and 2) enables us to think more clearly about what is happening to the information without the baggage of what the specific experimental structure is.  However, in experimental quantum information, it is always necessary to be aware of your own experimental context and to understand how to relate the abstract information to the physical realisation.

Throughout these course notes, I will write the logical basis states in bold to distinguish them from resonator states with 0 or 1 excitations.  This is not a standard convention, but it will be a helpful clarification for the reader for the purposes of these notes.

So how do we deal with qubits?  In fact, you're probably quite familiar with many aspects of these systems already, even if you don't already know it.

Using the computational basis states, an arbitrary qubit state can be written as:
\begin{align}
\ket{\psi} = \alpha_\zero \ket{\zero} + \alpha_\one \ket{\one}
\end{align}
where $\alpha_\zero$ and $\alpha_\one$ are complex numbers that satisfy the normalisation criterion $|\alpha_\zero|^2 + |\alpha_\one|^2 = 1$.  In fact, there are three standard bases that probe different aspects of the qubit state.  The first is the computational basis, $\{ \ket{\zero}, \ket{\one} \}$, and the other two are formed from different types of equal superpositions of the computational basis states: $\{ \ket{\pm} \equiv (\ket{\zero} \pm \ket{\one})/\sqrt{2} \}$ and $\{ \ket{{\pm}i} \equiv (\ket{\zero} \pm i \ket{\one})/\sqrt{2} \}$.

To get more intuition about these relationships, it is useful to write the general state in terms of real amplitude and phase parameters.  Ignoring the irrelevant global phase, an arbitrary qubit state is given by:
\begin{align}
\label{eq:qubit-realparams}
\ket{\psi} = \alpha_\zero \ket{\zero} + \alpha_\one e^{i \Delta} \ket{\one}
\end{align}
where $\alpha_{\rm\bf j}$ are now real parameters and $\Delta$ is the relative phase between the two basis states.  The $\{ \ket{\pm} \}$ therefore gives information about the real part of the superposition ($\Delta=0$) and the $\{ \ket{{\pm}i} \}$ basis gives information about the imaginary part of the superposition ($\Delta=\pi/2$).

\subsection{Qubits and the Bloch sphere}

\begin{figure}
\begin{center}\begin{tabular}{cc}
(a) \includegraphics[width=40mm]{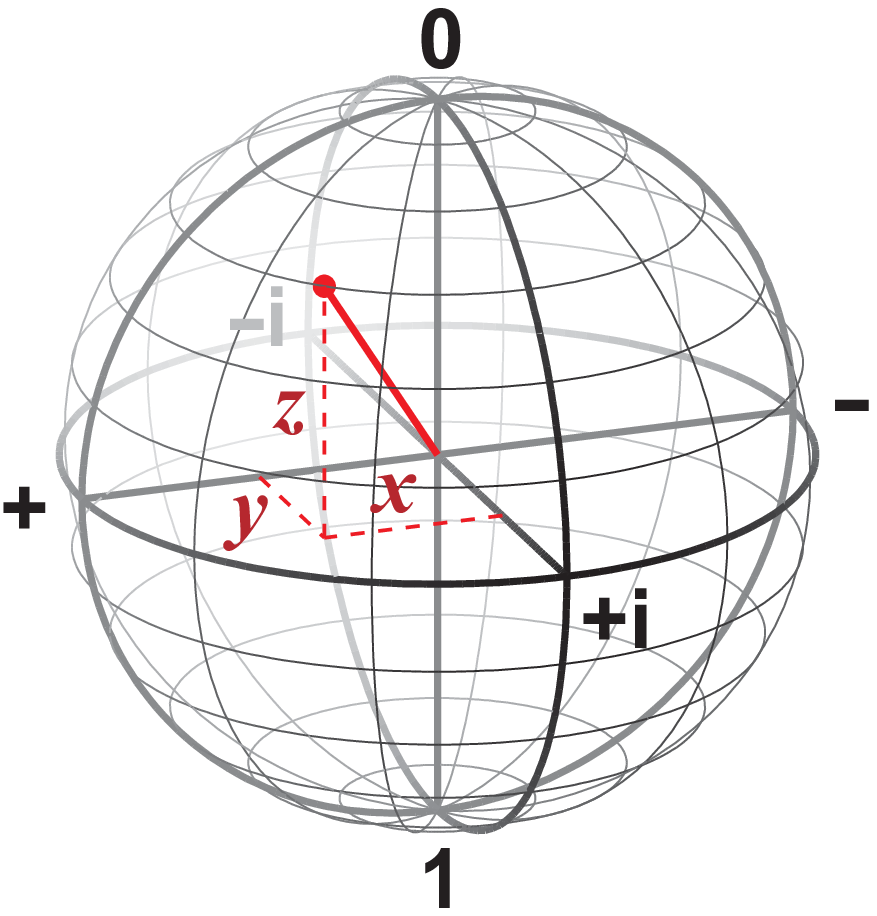} &
(b) \includegraphics[width=40mm]{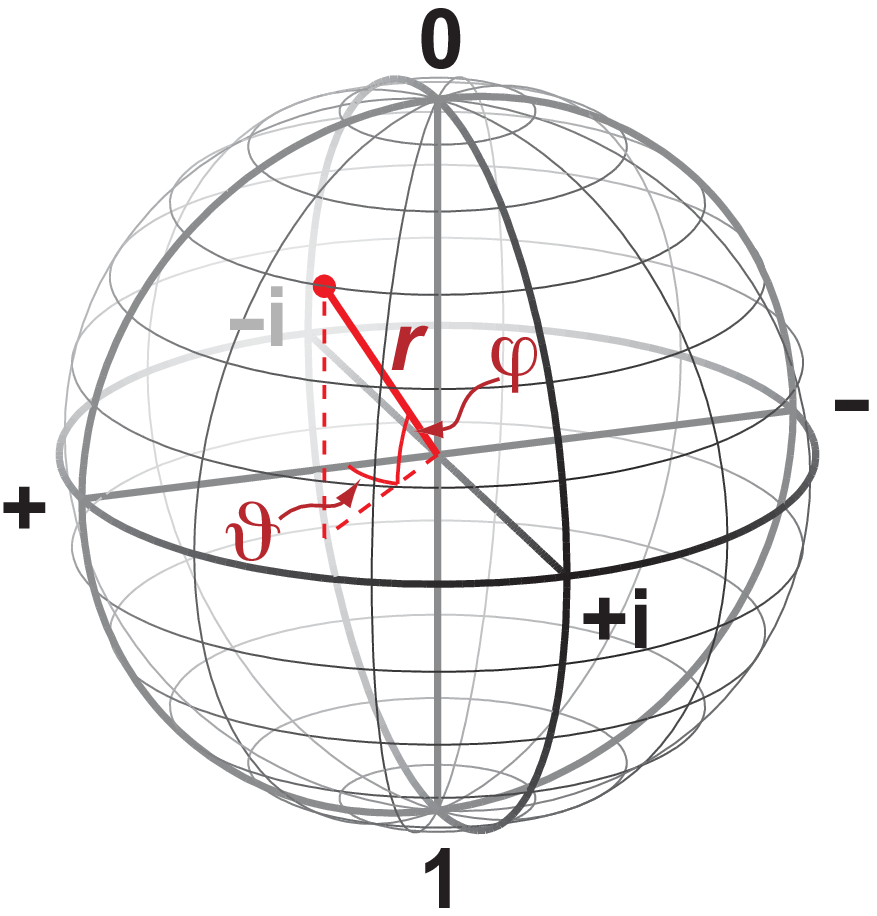}
\end{tabular}\end{center}
\captionsetup{labelfont=bf,font={small},margin=6pt}
\caption{\emph{\textbf{The Bloch sphere}} in: (a) Cartesian coordinates; (b) polar coordinates.}
\label{fig:blochsphere}
\end{figure}

Fortunately we're not just left at the mercy of abstract algebra here.  There is also a very useful method for visualising the state of a qubit and Eq.~\ref{eq:qubit-realparams} provides a clue as to how this is done.  First consider the case where $\Delta = 0$.  Because the states are normalised (here, $\alpha_\zero^2 + \alpha_\one^2 = 1$), the set of possible states can be represented geometrically by the points on a unit circle.  Let's now ignore $\alpha_{\rm\bf j}$ for a moment and remember that complex numbers can also be represented on a two-dimensional plane, where the $x$ and $y$ variables are the real and imaginary parts.  The set of all possible phases defined by the complex number $e^{i \Delta}$ can therefore also be represented by the points on a unit circle, this time in the complex plane.  But a full ``circle'' of points is possible for any given choice of $\alpha_{\rm\bf j}$, so these two geometrical representations provide complementary information.  In fact, considering now the completely general case, we can represent the set of all possible qubit states by the surface of a unit sphere (i.e., with unit radius) in 3D space.  This is generally called the \emph{Bloch sphere}%
\footnote{For reference, this is also called the Poincar\'e sphere in the context of polarisation optics, but that is not relevant for this course.}
(Fig.~\ref{fig:blochsphere}).  The diagrammes show that the complementary information provided by the two circles described above end up as orthogonal circles on the Bloch sphere.  The $\Delta=0$ case describes the great circle passing through the two poles ($\ket{\zero}$ and $\ket{\one}$) as well as the two real superposition states, $\ket{\pm}$ (you can think of this as the primary line of longitude of the Bloch sphere ``planet'').  For each point on this great circle, the complex phase is then a circle running laterally around the sphere (a line of latitude).  Note that, in the Bloch sphere representation, orthogonal states are points on opposite sides of the sphere.

It now becomes clear why we chose those three measurement bases and why they provide complementary information.  It is not too difficult to show that the $x$, $y$ and $z$ coordinates of a point (i.e., a quantum state) on the surface of the Bloch sphere are given by the following relations:
\begin{align}
x &= |\braket{+}{\psi}|^2-|\braket{-}{\psi}|^2 \\
y &= |\braket{{+}i}{\psi}|^2-|\braket{{-}i}{\psi}|^2 \\
z &= |\braket{\zero}{\psi}|^2-|\braket{\one}{\psi}|^2
\end{align}

But the usefulness of this representation goes even further, because it also tells us how to understand measurements and state manipulation (the effect of operators).

In the Bloch sphere representation, an arbitrary basis $\{ \ket{\phi}, \ket{\phi_\perp} \}$ ($\phi$ and $\phi_\perp$ are orthogonal) is represented by an axis of the sphere running through $\phi$ and $\phi_\perp$.  Now, given a quantum state $\ket{\psi}$, which is represented by a point $\uvec{\psi}(x,y,z)$ on the Bloch sphere, the expected measurement outcome probabilities for a measurement basis $\{ \ket{\phi}, \ket{\phi_\perp} \}$ are simply related to the component of $\uvec{\psi}(x,y,z)$ in the direction of the $\phi$--$\phi_\perp$ axis (a value of 1 says that the measurement will always give the result $\phi$ and a value of $-1$ says that the measurement will always give the result $\phi_\perp$).

In terms of state manipulation, single-qubit unitary operations are simply given by rotations of the Bloch sphere.  Consider the standard Pauli spin operators:
\begin{equation}
\begin{array}{ccc}
\sigma_x \equiv X = \sqbr{\begin{matrix} 0 & 1 \\ 1 & 0 \end{matrix}}, &
\sigma_y \equiv Y = \sqbr{\begin{matrix} 0 & -i \\ i & 0 \end{matrix}}, &
\sigma_z \equiv Z = \sqbr{\begin{matrix} 1 & 0 \\ 0 & -1 \end{matrix}}.
\end{array}
\end{equation}
As an example, it is fairly straightforward to show that:
\begin{align}
\begin{array}{cccc}
\sigma_x \ket{\zero} = \ket{\one}, &
\sigma_x \ket{\one} = \ket{\zero}, &
\sigma_x \ket{\pm} = \ket{\pm} &
\text{\& }
\sigma_x \ket{{\pm}i} = \ket{{\mp}i}.
\end{array}
\end{align}
In other words, the $\sigma_x$ spin operator implements a $180^\circ$ rotation of the Bloch sphere around the axis defined by its eigenstates, $\ket{\pm}$.  The same is true also for the other Pauli spin operators.

More generally, it is possible to show that any single-qubit unitary rotates the Bloch sphere by a certain angle around its ``eigenaxis''.  Consider, for example, a qubit system with an energy gap of $\Delta$ between its two computational basis states.  It turns out that the Hamiltonian describing its free evolution is $H_0 = -\frac{1}{2} \hbar \Delta \, \sigma_z$.  Using Eq.~\ref{eq:unitaryevolution} which we derived above for the unitary evolution operator, it is possible to show that:
\begin{align}
U_0(t) = \left[ \begin{matrix} e^{i\Delta t/2} & 0 \\ 0 & e^{-i\Delta t/2} \end{matrix} \right]
\end{align}
If we allow the state to evolve for a time $t=\pi/2\Delta$, then we can show that (ignoring global phases):
\begin{align}
\begin{array}{cccccc}
U_0 \ket{\zero} = \ket{\zero}, &
U_0 \ket{\one} = \ket{\one}, &
U_0 \ket{+} = \ket{{-}i} &
U_0 \ket{{-}i} = \ket{-} &
U_0 \ket{-} = \ket{{+}i} &
\text{\& }
U_0 \ket{{+}i} = \ket{+}.
\end{array}
\end{align}
$U_0$ is therefore a $90^\circ$ rotation of the Bloch sphere around its eigenaxis, the $\zero$--$\one$ axis%
\footnote{Using this convention for the free evolution Hamiltonian, the rotation is a right-handed rotation around the axis point from $\zero$ to $\one$ on the Bloch sphere, i.e., opposite to the $z$ direction defined in Fig.~\ref{fig:blochsphere} which corresponds to the above forms of the Pauli matrices.  In other words, a (positive) $\sigma_z$ evolution gives a right-handed rotation around the positive $z$ direction, but the qubit Hamiltonian here is proportional to $-\sigma_z$, because $\ket{\zero}$ is assumed to have a lower energy than $\ket{\one}$.}.

Using the ideas underlying these simple examples, it is even possible to show that a completely arbitrary single-qubit unitary can be written in the very elegant form:
\begin{align}
U_{\uvec{n}}(\theta) =\exp{(-i\frac{\theta}{2}\uvec{n} \cdot \uvec{\sigma})},
\end{align}
where $\uvec{n}$ is a 3D unit vector ($|\uvec{n}|=1$; nothing to do with the number operator in this instance), $\uvec{\sigma}=(\sigma_x,\sigma_y,\sigma_z)$, and $\uvec{n} \cdot \uvec{\sigma} = \sum_{j{=}x,y,z} n_j \sigma_j$.  Here, $U_{\uvec{n}}(\theta)$ describes a rotation of the Bloch sphere%
\footnote{In this more general form, the operator describes a right-handed rotation around the positive direction of $\uvec{n}$.}
by an angle $\theta$ around the axis defined by $\uvec{n}$.

\subsection{Exercises---microwave resonators}

We have already that the microwave-frequency resonators in circuit QED (whether lumped-element or distributed waveguide resonators) can be described in terms of simple harmonic oscillators with a resonant frequency, $\omega_0$, and a free-evolution Hamiltonian given by (ignoring the zero point energy):
\begin{align}
\hat{\mathcal{H}}_0 = \hbar \omega_0 \hat{a}^\dag \hat{a},
\end{align}
where $\hat{a}$ and $\hat{a}^\dag$ are the annihilation and creation operators, which obey the commutation relation $\sqbr{\hat{a},\hat{a}^\dag} = 1$, and can be defined by:
\begin{align}
\hat{a} \ket{n} &= \sqrt{n} \ket{n-1} \\
\hat{a}^\dag \ket{n} &= \sqrt{n+1} \ket{n+1}
\end{align}
along with $\hat{a}\ket{0} = 0$, where the number (Fock) state, $\ket{n}$, is the eigenstate of the number operator $\hat{n} = \hat{a}^\dag \hat{a}$ containing $n$ microwave excitations, and is defined by $\hat{n} \ket{n} = n \ket{n}$.

\subsubsection{Coherent states}

A coherent state, $\ket{\alpha}$, is a special quantum state of a resonator, which is very useful in quantum optics and circuit QED, because it behaves almost classically.  It can be defined to be the eigenstate of the annihilation operator:
\begin{align}
\hat{a} \ket{\alpha} = \alpha \ket{\alpha}
\end{align}
or the conjugate identity:
\begin{align}
\bra{\alpha} \hat{a}^\dag = \bra{\alpha} \alpha^*
\end{align}

\begin{enumerate}[label=(\alph*)]

\item Starting with this definition, show that the coherent state can be written in the number-state basis as:
\begin{align}
\nn
\ket{\alpha} = \sum_{n{=}0}^\infty \exp(-|\alpha|^2/2) \frac{\alpha^n}{\sqrt{n!}} \ket{n}.
\end{align}
\emph{Use the following steps:}
\begin{enumerate}[label=\roman*.]
\item Writing the coherent state as an arbitrary superposition of number states, $\ket{\alpha} = \sum_{n{=}0}^\infty c_n \ket{n}$, substitute this expression into the eigenstate equation above to derive the recursion relation for the coefficients, $c_n$:
\begin{align}
\nn
c_{n{+}1} = \frac{\alpha}{\sqrt{n+1}} c_n
\end{align}
and hence show that:
\begin{align}
\nn
\ket{\alpha} = \sum_{n{=}0}^\infty \frac{\alpha^n}{\sqrt{n!}} c_0 \ket{n}.
\end{align}
\item Use the normalisation of the coherent state, $\braket{\alpha}{\alpha} = 1$, to show that:
\begin{align}
\nn
c_0 = \exp(-|\alpha|^2/2).
\end{align}
\end{enumerate}

\item Using the definition of the coherent state, show that the average excitation number $\expect{\hat{n}} = |\alpha|^2$.

\item Starting again from the definition of the coherent state, and using the result of the previous question, show that the variance (width) of the excitation-number distribution is also $\Delta \hat{n} = |\alpha|^2$. 

[Hint: Use the commutator of $\hat{a}$ and $\hat{a}^\dag$ to show that $\hat{a}^\dag \hat{a} \hat{a}^\dag \hat{a} = \hat{a}^\dag (1+ \hat{a}^\dag \hat{a}) \hat{a}$ and recall that the variance of an observable, $\hat{A}$, is defined by $\Delta \hat{A} = \expect{\hat{A}^2} - \expect{\hat{A}}^2$.]

\item The goal of this question is to study how a coherent state evolves under the oscillator's free evolution Hamiltonian.

Firstly, given an initial Fock state, $\ket{\psi(0)} = \ket{n}$, show that:
\begin{align}
\nn
\ket{\psi(t)} = \exp(i n \omega_0 t) \ket{n}.
\end{align}
[Hint: Use the operator form of the Taylor expansion for an exponential, $\exp(\hat{A}) = \sum_{k{=}0}^\infty \hat{A}^k / k!$.]

Secondly, given a coherent state as initial state, $\ket{\psi(0)} = \ket{\alpha}$, use this previous result in conjunction with the number-state decomposition from (a) to show that:
\begin{align}
\nn
\ket{\psi(t)} = \ket{\alpha \exp(i \omega_0 t)} \equiv \ket{\alpha(t)}.
\end{align}

\emph{Note: This result underpins the ``classical'' behaviour of the coherent state, because its form and evolution are both completely described a single classical complex number.}

\end{enumerate}

\subsubsection{The phase space of an oscillator}

A complex number, $\alpha$, can be described in terms of an amplitude and phase, via
\begin{align}
\alpha = |\alpha| e^{i\theta}.
\end{align}
It can also be described in terms of its real and imaginary parts:
\begin{align}
\real{\alpha} &= \frac{1}{2} \br{\alpha + \alpha^*} = |\alpha| \cos\theta \\
\imag{\alpha} &= -\frac{i}{2} \br{\alpha - \alpha^*}  = |\alpha| \sin\theta
\end{align}
which oscillate sinusoidally as a function of the phase, $\theta$.  A complex number can then be represented pictorially on a 2D plane, with the real and imaginary parts representing the $x$ and $y$ coordinate variables.

It turns out that the annihilation and creation operators are quantum analogues of a complex number, representing the quantum amplitude of a field in the resonator mode.  In this analogy, the number operator, $\hat{n} = \hat{a}^\dag \hat{a}$, can be seen to be the equivalent of the ``modulus squared'' of a complex number.  We can therefore define two new operators for the real and imaginary parts of this amplitude:
\begin{align}
\nn
\hat{X}_1 &= \frac{1}{2} \br{\hat{a} + \hat{a}^\dag} \\
\nn
\hat{X}_2 &= -\frac{i}{2} \br{\hat{a} - \hat{a}^\dag}
\end{align}
These new operators now represent the phase-space coordinates for the quantum field, often called the ``quadrature'' operators (for historical reasons).  They also define the coordinates of a 2D pictorial phase-space representation of the quantum oscillator mode.

\begin{enumerate}[label=(\alph*)]

\item Show that the commutator $\sqbr{\hat{X}_1, \hat{X}_2} = i/2$.

\item Consider a system in the vacuum or ground state, $\ket{\psi} = \ket{0}$.  Show that $\expect{\hat{X}_1} = \expect{\hat{X}_2} = 0$ and that $\Delta \hat{X}_1^2 = \Delta \hat{X}_2^2 = 1/4$. [Hint: Use again the definition for variance, $\Delta \hat{A} = \expect{\hat{A}^2} - \expect{\hat{A}}^2$.]

\item Consider a system in a Fock state, $\ket{\psi} = \ket{n}$.  Show that $\expect{\hat{X}_1} = \expect{\hat{X}_2} = 0$ and that $\Delta \hat{X}_1^2 = \Delta \hat{X}_2^2 = n/2 + 1/4$.

\item Consider a system in a coherent state, as defined in the previous question, $\ket{\psi} = \ket{\alpha}$.
\begin{enumerate}[label=\roman*.]
\item Show that
\begin{align}
\nn
\expect{\hat{X}_1} &= |\alpha| \cos\theta \\
\nn
\expect{\hat{X}_2} &= |\alpha| \sin\theta
\end{align}
\item Show also that $\Delta \hat{X}_1^2 = \Delta \hat{X}_2^2 = 1/4$.
\end{enumerate}
\end{enumerate}

\emph{Note: These results highlight the different, complementary nature of the number states and coherent states.}

\emph{The number states, being eigenstates of the number operator, by definition have an average number value of $n$ and variance of 0.  By contrast, the number state's average quadrature values are 0, displaying no dependence on phase, and the quadrature variances are large ($\propto n$).  In fact, the number states have a completely random phase (uniform distribution in phase), reflecting the underlying fact that \emph{number} and \emph{phase} are conjugate physical variables.  (In phase space, a number state is pictured by a circle of radius $\sqrt{n}$.)}

\emph{The coherent state, on the other hand, has average quadrature values that oscillate sinusoidally with the phase of the coherent state amplitude, $\alpha$.  This highlights the fact that the coherent state has a well-defined phase.  The quadrature variances, however, are equal to each other and also equal to the vacuum-state variance, indicated that the coherent states are just circular ``blobs'' in phase space, displaced from the origin by a distance $|\alpha|$ at an angle $\theta$.}

\subsection{Exercises---qubits}

For the following questions you will need to use the standard qubit basis states:
\begin{align}
\ket{\pm} &\equiv (\ket{\zero} \pm \ket{\one})/\sqrt{2} \\
\ket{{\pm}i} &\equiv (\ket{\zero} \pm i \ket{\one})/\sqrt{2}
\end{align}
and the form of the standard Pauli spin operators:
\begin{equation}
\begin{array}{ccc}
\sigma_x = \sqbr{\begin{matrix} 0 & 1 \\ 1 & 0 \end{matrix}}, &
\sigma_y = \sqbr{\begin{matrix} 0 & -i \\ i & 0 \end{matrix}}, &
\sigma_z = \sqbr{\begin{matrix} 1 & 0 \\ 0 & -1 \end{matrix}}.
\end{array}
\end{equation}

\subsubsection{Rabi oscillations}

A Rabi experiment describes the evolution of a qubit when its two energy levels are coupled together by an interaction Hamiltonian.  Consider a qubit which is prepared initially in the ground state, $\ket{\psi(0)} = \ket{0}$, and then a field is applied which couples the two energy levels together, such that the qubit state evolves for a variable time $t$ according to the coupling Hamiltonian:
\begin{align}
\mathcal{H} = \half \hbar \Omega \sigma_x
\end{align}
where $\Omega$ is the ``Rabi frequency'', which describes the strength of the coupling.  After a time $t$, the population of the qubit in the ground state is then measured again.

\begin{enumerate}[label=(\alph*)]

\item Show that the unitary operator describing the qubit evolution under the coupling Hamiltonian is given by:
\begin{align}
\nn
U(t) = \exp (-\half i \Omega t \, \sigma_x)
\end{align}
[Hint:  Use the unitary form of the Schr\"odinger equation derived in the notes above.]

\item Using the operator form of the Taylor expansion for an exponential, $\exp(\hat{A}) = \sum_{k{=}0}^\infty \hat{A}^k / k!$, show that the unitary evolution operator, $U(t)$, is given by the matrix:
\begin{align}
\nn
U(t) = \sqbr{ \begin{matrix} \cos \half\Omega t & -i \sin \half\Omega t \\ -i \sin \half\Omega t & \cos \half\Omega t \end{matrix}}
\end{align}
[Hint: Verify the identity that $\sigma_x^2 = I$ and then divide the Taylor expansion into a sum of terms involving $I$ and a sum of terms involving $\sigma_x$.]

\item Given the initial state $\ket{\psi(0)} = \ket{0}$, hence show that the output state after time $t$ is:
\begin{align}
\nn
\ket{\psi(t)} = \sqbr{ \begin{matrix} \cos \half\Omega t \\ -i \sin \half\Omega t \end{matrix}}
\end{align}
and that the ground and excited state populations are:
\begin{align}
\nn
p_\zero(t) &= \frac{1}{2}(1 + \cos \Omega t) \\
\nn
p_\one(t) &= \frac{1}{2}(1 - \cos \Omega t)
\end{align}

\item Sketch the ground state population as a function of delay time, $t$.

\end{enumerate}

\subsubsection{Ramsey fringes}

A Ramsey experiment describes the evolution of a superposition state of a qubit which is evolving freely according to its ``rest'' Hamiltonian:
\begin{align}
\mathcal{H}_0 = -\half \hbar \Delta \, \sigma_z
\end{align}
where $\hbar \Delta$ is the energy splitting of the two qubit levels.  Given a qubit that is prepared initially in the ground state, $\ket{\psi(0)} = \ket{0}$, an equal superposition $\ket{+}$ state can be prepared using a so-called ``Hadamard'' gate (a $\pi/2$ rotation of the Bloch sphere):
\begin{align}
H = \frac{1}{\sqrt{2}} \sqbr{ \begin{matrix} 1 & 1 \\ 1 & -1 \end{matrix}}
\end{align}
For reference, a Hadamard gate can be implemented by applying the $\sigma_y$ coupling Hamiltonian, $\mathcal{H}_y = \half \hbar \Omega_y \, \sigma_y$, for a time $t = \pi/2\Omega_y$ (sometimes called a \emph{$\pi$/2 pulse}).

In brief, a Ramsey oscillation experiment is implemented as follows:
\begin{itemize}
\item Step 1: Starting with a qubit in the ground state, prepare a superposition $\ket{+}$ state by applying a Hadamard gate.
\item Step 2: Allow the state to evolve freely for a variable time $t$.
\item Step 3: Rotate the final state back by a second Hadamard gate and measure the ground state population, defined by the probability $p_\zero$.
\end{itemize}

\begin{enumerate}[label=(\alph*)]

\item Verify that, like the Pauli spin operators, $H^2 = I$.

\item Using the unitary form of the Schr\"odinger equation, show that the unitary evolution operator for free evolution is given by:
\begin{align}
\nn
U_0(t) = \sqbr{ \begin{matrix} \exp \br{i \Delta t/2} & 0 \\ 0 & \exp \br{-i \Delta t/2} \end{matrix}}
\end{align}
[Hint: Use the fact that, for any diagonal operator, $D=\sum_i d_i |i\rangle\langle i|$, the exponential of the operator is defined by $\exp(D) =\sum_i \exp(d_i) |i\rangle\langle i|$.]

\item Using this result, calculate the final output state after Step 3 of the procedure outlined above and show that the measured probability is $p_\zero(t) = \frac{1}{2}(1 + \cos \Delta t)$.

\end{enumerate}

\subsubsection{The density matrix---what is the density matrix?}

\begin{enumerate}[label=(\alph*)]

\item For a pure state, $\ket{\psi}$, the density operator is defined by $\rho = \ket{\psi}\bra{\psi}$.  For an arbitrary qubit state, $\ket{\psi} = \alpha \ket{\zero} + \beta \ket{\one}$, show that the density operator is given by:
\begin{align}
\nn
\rho = |\alpha|^2 \ket{\zero}\bra{\zero} + \alpha \beta^* \ket{\zero}\bra{\one} + \alpha^* \beta \ket{\one}\bra{\zero} + |\beta|^2 \ket{\one}\bra{\one}.
\end{align}

\item We can define the matrix form of the density operator in the usual way, by writing $\rho_{ij} = \matelem{i}{\rho}{j}$.  Using this, show that the density matrix is given by:
\begin{align}
\nn
\rho = \sqbr{ \begin{matrix} |\alpha|^2 & \alpha \beta^* \\ \alpha^* \beta & |\beta|^2 \end{matrix}}
\end{align}

\item Consider an operator, $A$, which takes an initial state $\ket{\psi_1}$ to a final state $\ket{\psi_2}$ (i.e., $\ket{\psi_2} = A \ket{\psi_1}$).  Using the definition for a density operator given above, show that, in density matrix form, this evolution is described by:
\begin{align}
\nn
\rho_2 = A \rho_1 A^\dag
\end{align}

\item Consider the operator $\uvec{n} \cdot \uvec{\sigma} = \sum_{j=x,y,z} n_j \sigma_j$ (introduced in the notes above), where $\uvec{n}$ is a 3D unit vector ($|\uvec{n}|=1$).  Write down the matrix representation of $\uvec{n} \cdot \uvec{\sigma}$. Show that $\uvec{n} \cdot \uvec{\sigma}$ is a unitary operator.

\item Using the formula for the evolution of a density matrix from (c), show that:
\begin{align}
\rho_1 = \frac{1}{2} \sqbr{ \begin{matrix} 1+n_z & n_x-i n_y \\ n_x + i n_y & 1-n_z \end{matrix}}
\end{align}
is an eigenstate of $\uvec{n} \cdot \uvec{\sigma}$.  [Hint: Show that $(\uvec{n} \cdot \uvec{\sigma}) \, \rho_1 \, (\uvec{n} \cdot \uvec{\sigma})^\dag = \rho_1$.]

\end{enumerate}

\section{Superconductivity}

Superconductivity is an extremely varied and complex phenomenon.  The following is a brief summary of some of a superconductor's properties:
\begin{description}[topsep=0mm,parsep=0mm,partopsep=0mm,leftmargin=1cm,style=sameline,font=\it]

\item[Zero resistance:]  Far below a critical temperature, critical magnetic field and a certain cut-off frequency, superconductors act like a perfect conductor, i.e., they conduct electricity with zero resistance.

\item[Two-fluid behaviour:]  In its superconducting state, a superconductor's behaviour can generally be modelled as a mixture of two independent electron fluids, a normal (i.e., resistive) electron sea and an ideal superconducting fluid.  The fraction of electrons participating in the normal fluid state increases with temperature until the critical temperature ($T_{\rm c}$), where all electrons behave normally.  Thus, the critical temperature is the point where we start to see superconducting effects, but they become more pronounced as the material is cooled.

\item[Optical properties:]  The appearance of a superconductor doesn't change as it is cooled below its critical temperature.  This means that its resistivity doesn't change at optical frequencies.  This ``cut-off'' frequency for most superconductors is around $10^{11}$ Hz.  It is usually called its \emph{plasma frequency}.

\item[Different metals:]  Not all metals appear to become superconducting at low temperatures (at least at temperatures reached to date).  Metals that are superconductors are generally poor conductors in their normal state.  Most good conductors (e.g., copper, gold, silver, etc.) are not superconductors.

\item[Crystal lattice structure:]  The crystal structure of superconductors does not change as they enter the superconducting state.

\item[Different isotopes:]  Although the crystal structure of superconductors doesn't change, the critical temperature of a isotopically pure metal varies with the isotopic mass, so the mass of the lattice particles does seem to play a role.

\item[Energy gap:]  At low temperatures (well below the critical temperature), the contribution to specific heat from the conduction electrons in a superconductor decays exponentially with increasing temperature, i.e., $C \sim \exp (-b/kT)$.  This is consistent with the heating, dissipative process being separated by an energy gap from a \emph{superconducting ground state}.

\item[Meissner effect:]  A superconducting material does not just resist changes in magnetic field the way a normal (or even perfect) conductor does.  It actually expels all magnetic field completely as it becomes superconducting.  All electric and magnetic fields decay exponentially inside a very small surface depth known as the penetration depth.  A related effect is that currents can only flow on the surface of a superconductor (again within the penetration depth of the material's surface).

\item[Long-range coherence:]  The phase of the conduction electrons in a superconductor is coherent over a large distance, much larger than their typical interaction length and potentially macroscopic, even on the size scale of an electronic circuit.

\end{description}

\subsection{Superconductivity in circuit QED}

Superconductivity was first discovered experimentally in 1911, but it was almost 50 years later, in 1957, before a detailed quantum theory was proposed which could explain the phenomenon properly at a microscopic level.  Discovered by Bardeen, Cooper and Schrieffer, this is generally referred to as the BCS theory of superconductivity~\cite{BardeenJ1957ts}.  More recently, however, in the 1980s, it was discovered that such a full microscopic theory is not necessary in order to explain the quantum behaviour of an electronic circuit.  Instead, it turns out that it is only necessary to describe the behaviour of the collective macroscopic degrees of freedom which are related to the standard variables of classical electronic circuit theory~\cite{LeggettAJ1980mqs, YurkeB1984qnt}.  As a result, for the purpose of circuit QED, many of the subtleties of superconductivity are not really relevant%
\footnote{If you are interested in learning more details about superconductivity than the brief summary I will include here to help guide your intuition in later sections, I would strongly recommend ``Introduction to Superconductivity'' by Rose-Innes and Rhoderick~\cite{Rose-InnesRhoderick}.  It is extremely well written and provides a good detailed physical description, while avoiding unnecessary mathematical complexity.}.

The main simplifying factor for circuit QED is that experiments are run at very low temperatures, much less than the critical temperature, $T \ll T_{\rm c}$.  This means that any ``normal conductor'' effects are negligible and the circuits behave like ideal superconductors.  
\begin{description}[topsep=0mm,parsep=0mm,partopsep=0mm,leftmargin=1cm,style=sameline,font=\it]

\item[Zero resistance:]  The most important reason why superconductors are used in circuit QED is that they provide a way to build resistance-free, and therefore energy dissipation-free electronic circuits.  Because of the very low operating temperatures, the ideal superconductivity allow the circuits to behave as good quantum systems with long coherence times and the usual classical electronic variables can be treated quantum mechanically.

\item[Energy gap:]  The existence of an energy gap between the \emph{superconducting ground state} and the dissipative ``normal'' conduction processes, again combined with operating temperatures far smaller than this gap, makes these circuits substantially resistant to thermal noise, because thermal processes are unable to excite dissipation.

\item[Optical properties:]  Because the superconducting behaviour only survives below a certain cut-off frequency, this limits circuit QED experiments to operating far below this frequency.  Such experiments are usually carried out in the range 1--30 GHz (and more commonly below 15 GHz).  This brings some disadvantages.  In fact, working in this microwave regime sets a much more stringent requirement on low-temperature operation than the superconducting critical temperature.  For example, at around 170 mK, the peak of the black-body radiation spectrum is at around 10 GHz, and at 20 mK, the peak of the black-body spectrum is around 1.2 GHz.  This is one motivation for working in cavities, since this limits the bandwidth of thermal radiation that is seen by the superconducting qubits.

\item[Long-range coherence:]  Long-range phase coherence of the superconducting electrons is another critical property for circuit QED experiments, which require phase coherence over the entire macroscopic scale of the circuit.  Without this property, circuit QED would be restricted to only very simple, small circuit structures.

\end{description}

\subsection{The superconducting condensate and Cooper pairs}

\begin{figure}
\begin{center}
\includegraphics[width=140mm]{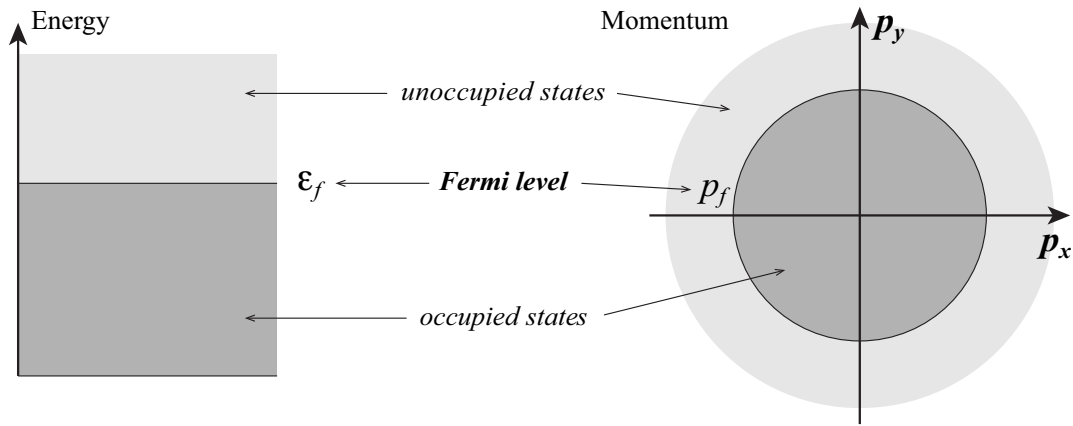}
\end{center}
\captionsetup{labelfont=bf,font={small},margin=6pt}
\caption{\emph{\textbf{Electronic band structure of a normal metal.}}}
\label{fig:bandstructure-nc}
\end{figure}

In a normal metal, the conduction electrons behave approximately like plane waves travelling through the medium%
\footnote{More accurately, the electron wave-functions are Bloch waves.  See Kittel~\cite{Kittel}, if you are interested in more details.}.
Because electrons are fermions (spin 1/2), they always remain distinguishable particles and the Pauli exclusion principle prevents them from occupying exactly the same quantum state.  Consequently, the electrons fill up the available states one at a time up to an energy known as the Fermi level, the point where all the electrons are used up.  In a bulk material, the available density of states is essentially continuous and this is kind of like filling a glass with a fixed amount of water, and this is called a ``sea'' of electrons, or more specifically the Fermi sea (see Fig.~\ref{fig:bandstructure-nc}).  A conductor is characterised by the fact that there is no gap in the available states near the Fermi level, so it is very easy for applied fields to induce a nett current flow in the electrons near the surface of the Fermi sea.

In a perfect lattice, such electron waves should be able to flow through the metal without losing any energy.  This is obviously not true in the real world, however.  In real, imperfect lattices, electrons experience a resistance (and thus energy dissipation) because of continuous inelastic scattering events off irregularities in the lattice.  These irregularities can be caused by either lattice vibrations (phonon scattering) or impurities.  Since the background lattice phonons are thermally excited, they disappear as the metal is cooled towards absolute zero temperature.  Thus, a pure metal would become a perfect conductor at zero kelvin.  In an impure metal, the resistance would instead approach a constant background value caused by the lattice impurities.

In the right conditions, it turns out that fermionic electrons can instead start behaving like bosons.  Specifically, pairs of electrons (fermions) with opposite spins can form a bound, bosonic ``molecule'', known as a \emph{Cooper pair}~\cite{CooperLN1956bep}, which has spin 0 (spin ${+}\frac{1}{2}$ + spin ${-}\frac{1}{2}$).  The attractive force which binds this molecule is induced by an exchange of phonons (vibrational lattice excitations) between two phonon scattering events.  Interestingly, these scattering events, which are responsible for causing resistance in normal metal conduction, are now a key ingredient in superconducting behaviour.  This is why superconductors are poor conductors when in the normal state.

As I have already mentioned, understanding the details of the microscopic theory of superconductivity is generally almost entirely unnecessary for understanding circuit QED.  Ignoring the details of this pairing process, therefore, the key take-away point is the fact that these paired electrons are able to behave completely differently from unpaired electrons, which are subject to all the usual constraint imposed by the Pauli exclusion principle.  Instead of remaining distinguishable and occupying distinct quantum states, the superconductor's electrons instead form a giant condensate, in which each Cooper pair is formed of electrons with opposite spin and equal and opposite momenta.  This condensate represents a macroscopic collective ground state for the superconducting system.

Although it isn't necessary to understand the full BCS theory, it will perhaps be helpful to write down the form of the Cooper pair wave-function in the superconducting ground state to highlight a couple of key features.  In the superconducting condensate, each Cooper pair occupies a wave-function of the following form:
\begin{align}
\label{eq:sc-cooperpair}
\Phi(\bvec{r}_1, \bvec{r}_2) = \sum
a_j \, \phi(\bvec{r}_1,\bvec{P}_j,\uparrow) \, \phi(\bvec{r}_2,-\bvec{P}_j,\downarrow)
\end{align}
where, as usual, $|\Phi(\bvec{r}_1, \bvec{r}_2)|^2$ is the probability density of finding the electrons at $\bvec{r}_1$ and $\bvec{r}_2$, and where $\phi(\bvec{r},\bvec{P},\uparrow)$ is the single-electron wave-function with momentum $\bvec{P}$ and spin in the ``up'' direction.  Note that this equation explicitly incorporates the fact that the Cooper-pair electrons have opposite spin and equal and opposite momenta (zero total momentum).

This state is a superposition of states with many different momenta and this is where some of the unique properties of the superconductor come from.  As described above, the resistance in normal metals results from the fact that electrons continuously scatter off lattice irregularities.  In superconductors, the electrons still undergo these same scattering events, but within each pair, the phonon that is created at one electron is immediately absorbed by the other.  Because total momentum is conserved during these interactions, this just converts between different terms on the right of the above wave-function.  In some sense, because all of these different states are already present in the large superposition anyway, these continuous scattering events don't change the overall quantum state of the pair and no energy is lost from the system.

Essentially, the strong symmetry of the Cooper-pair wave-function enforces symmetric scattering events.  Creating an excitation in the superconducting ground state requires breaking this symmetry in the form of breaking a Cooper pair and creating two individual unpaired electrons, which therefore again obey fermionic statistics.  The energy barrier that must be overcome to do this therefore creates an energy gap between ground states and excited states.

\subsection{The superconducting phase and current flow}

The wave-function in Eq.~\ref{eq:sc-cooperpair} describes a perfectly isotropic distribution of electron momenta, where for every electron travelling with a momentum $\bvec{P}$, there is another travelling at $-\bvec{P}$.  This describes a superconductor ``at rest'', where there is no current flow.  But what happens if we apply an external field to the superconductor?  In this situation, it turns out that we can write down an alternative wave-function:
\begin{align}
\label{eq:sc-cooperpair-current}
\Phi_\bvec{P} (\bvec{r}_1, \bvec{r}_2) = \sum a_j \, \phi\sqbr{ \bvec{r}_1,(\bvec{P}_j{+}\bvec{P}/2), \uparrow} \, \phi\sqbr{ \bvec{r}_2,(-\bvec{P}_j{+}\bvec{P}/2), \downarrow}
\end{align}
This wave-function retains all of the collective symmetry properties which underlie the superconducting behaviour, but each Cooper pair now carries a resultant momentum $\bvec{P}$, \emph{which is the same for all pairs and all states in the wave-function}.  Because of the plane-wave-like form of the underlying single-electron wave-functions $\phi \propto \exp (i \bvec{P}\cdot \bvec{r}/\hbar)$, it is therefore possible to combine the contributions from the overall momentum in a global term outside the summation:
\begin{align}
\Phi_\bvec{P} (\bvec{r}_1, \bvec{r}_2) &= \sum a_j \, \phi(\bvec{r}_1,\bvec{P}_j,\uparrow) \, \phi(\bvec{r}_2,-\bvec{P}_j,\downarrow) \; e^{i \bvec{P} \cdot \bvec{r}_1 / 2\hbar} e^{i \bvec{P} \cdot \bvec{r}_2 / 2\hbar} \\
\label{eq:sc-cooperpair-currentphase}
&= \Phi(\bvec{r}_1, \bvec{r}_2) \; \exp \br{i \bvec{P} \cdot \bvec{r} / \hbar}
\end{align}
where $\bvec{r} \equiv (\bvec{r}_1 + \bvec{r}_2)/2$ is the position of the Cooper pair's centre of mass.  Thus, each pair can be treated as a single ``particle'' with momentum $\bvec{P}$ at its centre-of-mass position $\bvec{r}$, with mass $2m$ and charge $2e$.

The very interesting thing here is that the global condensate wave-function sees an evolving phase as a result of a flow in the superconducting Cooper pairs.  This actually makes complete sense if we think of the Cooper pair as a quantum particle with momentum $\bvec{P}$ and mass $2m$.  We know that quantum particles behave like waves and that their wavelengths are determined by the de Broglie wavelength formula, $\lambda = h/p$.  Therefore, if $\bvec{P}=0$, this means that the condensate wavelength is infinite and that the condensate phase must be constant throughout the condensate.  But if $\bvec{P} \neq 0$, then the wavelength is finite and this causes \emph{phase gradients} in the condensate.

All of this shows that the \emph{phase} of a condensate is a physically important and defining feature of the condensate wave-function.  And since the phase must be constant throughout a condensate which is ``at rest'', we can therefore write:
\begin{align}
\Phi(\bvec{r}_1, \bvec{r}_2) &= |\Phi(\bvec{r}_1, \bvec{r}_2)| e^{i\varphi}
\end{align}
where $\varphi$ does not depend on the positions $(\bvec{r}_1, \bvec{r}_2)$ and any variation of the phase in space is described by the additional $\exp \br{i \bvec{P} \cdot \bvec{r} / \hbar}$ term in Eq.~\ref{eq:sc-cooperpair-currentphase}.

Of course, nonzero momentum is equivalent to current flow, so it is also possible to relate phase gradients in a superconducting condensate to the current that is flowing.  The momentum is related to the velocity of the particles, via $\bvec{P} = (2m) {\bf v}$, where $(2m)$ is the effective mass of a single Cooper pair.  Using conservation of charge, we can write the total current flowing passed a point (defined by a small surface element $A$):
\begin{align}
I = n_p (2e) A v
\end{align}
where $n_p$ is the Cooper pair density and $(2e)$ is the effective charge of a single Cooper pair.  We can therefore now write the vectorial current \emph{density}:
\begin{align}
\bvec{J} = n_p (2e) \frac{\bvec{P}}{(2m)} = \frac{n_p e}{m} \bvec{P}
\end{align}

\subsection{The many-electron condensate wave-function}

We saw above that $\Phi_\bvec{P}$ describes a single Cooper-pair plane wave with wavelength $\lambda = h/P$ and probability density $|\Phi_\bvec{P}|^2$.  But we also know that all Cooper pairs have exactly the same wavelength (momentum), $\lambda$ ($\bvec{P}$), because all pairs occupy the same condensate wave-function.  We can therefore write a \emph{total} condensate wave-function in a similar form---after all, if all Cooper-pairs have the same wavelength, then the total wave-function must also have the same wavelength and must therefore also have the same waveform: i.e.,
\begin{align}
\label{eq:sc-condensate-currentphase}
\Psi_\bvec{P} = \Psi \exp \br{i \bvec{P} \cdot \bvec{r} / \hbar}
\end{align}
where, provided there are many electrons, we can now interpret $|\Psi_\bvec{P}|^2 = |\Psi|^2$ as the \emph{number density} of Cooper pairs.  Note that $\bvec{P}$ is still the momentum \emph{per Cooper pair} in this formula.

Technically, this is a semiclassical approximation to the true many-electron condensate.  An heuristic argument justifying such an interpretation goes something like this.  Suppose we start from the single Cooper-pair wave-function, $\Phi_\bvec{P}(\bvec{r}_1, \bvec{r}_2)$, where $|\Phi_\bvec{P}(\bvec{r}_1, \bvec{r}_2)|^2$ is the probability density of finding a single Cooper pair at the positions $(\bvec{r}_1, \bvec{r}_2)$.  If we have $N$ Cooper pairs in the condensate, then the \emph{average} number density of Cooper pairs is therefore $N|\Phi_\bvec{P}(\bvec{r}_1, \bvec{r}_2)|^2$.  But if $N$ is very large, then the uncertainty in this number will be very small, much smaller than $N$ itself, and we can therefore think of this number as representing the \emph{actual} number density of Cooper pairs.  Replacing a probability density by a physical density in this way ignores the concept of quantum mechanical number fluctuations.

In the true quantum case, both the number density (or number) and the phase of the total many-electron condensate wave-function become quantum mechanical observables instead of well-defined classical numbers.  In fact, it even turns out that the number and phase are conjugate quantum variables, i.e., they do not commute and are subject to the Heisenberg uncertainty principle.  This means that a condensate with a precisely specified phase must contain a completely uncertain number of Cooper pairs, and likewise, a condensate with a precise number of pairs must have a completely uncertain phase.

\subsection{Superconducting circuits}

Up until this point, we have been discussing the phenomenon of superconductivity in bulk, isotropic superconducting materials.  Next, we want to consider a slightly more complex scenario and look at superconductivity in electronic circuits.

Suppose that we have some electronic circuit which is made from superconducting material and a supercurrent is flowing around the circuit along a path which connects two arbitrary points, $X$ and $Y$.  We are now considering a more general scenario where $\bvec{P} = \bvec{P}(\bvec{r})$.  In other words, the supercurrent doesn't have to flow in a straight line, but we will at least assume that it follows the geometry of the circuit (e.g., it flows along superconducting wires) and that our path connecting the two points just follows along with the current flow, i.e., that $\bvec{P}(\bvec{r})$ is always \emph{parallel} to $\bvec{r}$.

Integrating the resulting phase gradient along the path allows us to write:
\begin{align}
\label{eq:sc-pathphase-momentum}
\Delta \varphi_{XY} = \frac{1}{\hbar} \int_X^Y \bvec{P}(\bvec{r}) \cdot d\bvec{r} = \frac{1}{\hbar} \int_X^Y P(\bvec{r}) dr
\end{align}
We can simplify this still further if we also assume that the current flow is uniform around the circuit, i.e., $P(\bvec{r}) = P$:
\begin{align}
\gives \Delta \varphi_{XY} = \frac{Pl}{\hbar}
\end{align}
where $l$ is the total path length between $X$ and $Y$.  In terms of current instead of momentum, we have:
\begin{align}
\label{eq:sc-pathphase-current}
\Delta \varphi_{XY} = \frac{m}{en_p\hbar} \int_X^Y \bvec{J}(\bvec{r}) \cdot d\bvec{r} = \frac{mJl}{en_p\hbar}
\end{align}

In many cases, it is also important to consider the effect how a circuit will behave in the presence of an externally applied magnetic field.  From classical electromagnetism and Maxwell's equations, we know that when a magnetic field is changing, it induces an electric field via:
\begin{align}
\vcurl \bvec{E} = -\frac{d\bvec{B}}{dt}.
\end{align}
This field can in turn exert a force on a charge according to $\bvec{F} = q\bvec{E}$, giving the charge a momentum kick.  In electromagnetism, it is therefore not the standard ``mv'' momentum ($\bvec{P} = m\bvec{v}$) that is conserved, but rather a composite momentum, which has one component from the particle motion and an extra contribution from the field.  We can account for this by replacing the ``mv'' momentum in our usual equations with a composite version via%
\footnote{See Feynman's chapter on Superconductivity~\cite[Ch.~21]{FeynmanLectures3} or Rose-Innes~\cite{Rose-InnesRhoderick} for a more complete justification of this process.}:
\begin{align}
\bvec{P} = m\bvec{v} \quad\gives \bvec{P} = m\bvec{v} + q\bvec{A}
\end{align}
where $\bvec{A}$ is the vector potential defined through $\vcurl \bvec{A} = \bvec{B}$.  This new composite momentum \emph{is} a conserved quantity.  For a single Cooper pair, we have:
\begin{align}
\bvec{P} = 2m\bvec{v} + 2e\bvec{A}
\end{align}
Substituting this into Eq.~\ref{eq:sc-pathphase-momentum} gives:
\begin{align}
\Delta \varphi_{XY} &= \frac{1}{\hbar} \int_X^Y \bvec{P}(\bvec{r}) \cdot d\bvec{r} \\
\label{eq:sc-pathphase-Bfield}
&= \frac{m}{en_p\hbar} \int_X^Y \bvec{J}(\bvec{r}) \cdot d\bvec{r} + \frac{2e}{\hbar} \int_X^Y \bvec{A}(\bvec{r}) \cdot d\bvec{r}
\end{align}
where the first term is the current-induced phase we derived above and the second term is an extra field-induced phase.

\subsection{Flux quantisation}

One of the most important consequences of superconductivity for quantum electronic circuits relates to the fact that electronic circuits are generally defined around closed loops.  Consider a supercurrent which is flowing around a simple ring of superconducting material.  Because the material is superconducting, the condensate phase must always be ``single-valued''.  That is, the condensate must have a single well-defined phase at every point.  So if we follow the current flow in a closed path, $C$, around the superconducting loop, then when we return back to the starting position, the total phase difference around the path must be an integer multiple of $2\pi$:
\begin{align}
\Delta \varphi_C = n2\pi
\end{align}
In other words, the phase difference accrued around a closed superconducting loop must be quantised.

Using the results from the previous section, we can relate this to the phase change around the closed loop calculated based on the path taken by the supercurrent.  As we know from classical electromagnetism and electronic circuit theory, it is particularly important to include the effect of magnetic fields when considering closed loops in space.  From Eq.~\ref{eq:sc-pathphase-Bfield}, we have:
\begin{align}
\Delta \varphi_C &= \frac{m}{en_p\hbar} \oint_C \bvec{J}(\bvec{r}) \cdot d\bvec{r} + \frac{2e}{\hbar} \oint_C \bvec{A}(\bvec{r}) \cdot d\bvec{r} = n2\pi
\end{align}
where $\bvec{A}$ is again the magnetic vector potential defined by $\vcurl \bvec{A} = \bvec{B}$.  But by Stokes' theorem, we can relate path integrals to surface integrals:
\begin{align}
\oint_C \bvec{A} \cdot d\bvec{r} = \dint_S \vcurl \bvec{A} \cdot d\bvec{S} = \dint_S \bvec{B} \cdot d\bvec{S} = \Phi_S
\end{align}
where $\Phi_S$ is the total magnetic flux passing through $S$ (inside the curve $C$).  We can therefore define the new quantised parameter, known as the \emph{fluxoid}, $\Phi\dash$:
\begin{align}
\label{eq:sc-fluxoid-quantisation}
\Phi\dash \equiv  \frac{m}{2n_pe^2} \oint_C \bvec{J} \cdot d\bvec{r} + \Phi_S = n \Phi_0
\end{align}
where $\Phi_0 = h/2e$ is known as the \emph{superconducting flux quantum}.  Since the second term in this equation is a flux (that enclosed by the curve $C$), we can identify this new parameter as another flux-based quantity, hence the name fluxoid.  \emph{Essentially, the fluxoid quantisation relation implies that the total flux enclosed by a superconducting ring must be quantised in units of the superconducting flux quantum%
\footnote{The term involving the current flowing around $C$ accounts for the fact that some flux is able to penetrate a small distance into the body of the superconductor, up to the penetration depth; see the introductory overview at the very beginning of this section.}.}

At this point, it is worthwhile noting that the fluxoid is related to the condensate phase difference via:
\begin{align}
\Phi\dash = \varphi_0 \Delta \varphi_C
\end{align}
where $\varphi_0 = \Phi_0 / 2\pi$ is sometimes known as the \emph{reduced flux quantum}.  In fact, this relationship turns out to be useful for analysing circuits, because as we will see in later sections, a more general concept of flux can be related to circuit voltages.  We can define a generalised connection for open circuit segments (i.e., not only closed loops) relating phase difference to a so-called \emph{branch flux} variable:
\begin{align}
\Phi = \varphi_0 \Delta \varphi
\end{align}
While it is not going to be necessary for us to understand the significance of this connection at any deep level in this course, the main point to take away is that we can often think of flux and condensate phase difference as interchangeable concepts (barring a multiplicative constant) when analysing quantum electronic circuits.

\subsection{Electronic band structure of a superconductor}

\begin{figure}
\begin{center}
\includegraphics[width=100mm]{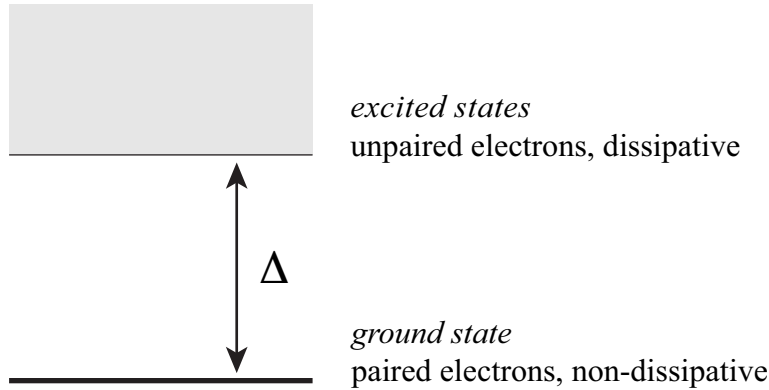}
\end{center}
\captionsetup{labelfont=bf,font={small},margin=6pt}
\caption{\emph{\textbf{Electronic band structure of a superconductor.}}}
\label{fig:bandstructure-sc}
\end{figure}

There are several different ways to understand the band structure of a superconductor, but here we will follow the picture described by Rose-Innes and Rhoderick~\cite{Rose-InnesRhoderick}.  The complexity arises because a superconductor exhibits characteristics of both fermionic unpaired electrons and effectively bosonic Cooper pairs.  The superconducting condensate ground state is a single shared quantum state which can be occupied by many Cooper pairs, although not unpaired electrons, but at momenta greater than those of the states involved in the condensate (near the Fermi level), there are still bands of unoccupied states which can accommodate normal conduction processes.  However, as a result of the extraordinary symmetry of the condensate wave-function, there is a substantial energy barrier which must be overcome to split up a Cooper pair and create two unpaired electrons in the available region of single-electron states, which opens up a gap in the energy band structure between the superconducting ground state and the unoccupied valence band.  Figure~\ref{fig:bandstructure-sc} illustrates this understanding.

\subsection{Electron and Cooper-pair tunnelling}

Tunnelling is one of the most interesting features of quantum physics, and plays a critical role in the behaviour of quantum electronic circuits.  This is largely because it underlies the behaviour of the Josephson junction, which provides the strong, non-dissipative nonlinearities which have made circuit QED so successful.

Electron tunnelling occurs when two conducting materials are separated by a thin gap of insulating material, which creates a potential barrier between the electron conduction states on either side of the gap.  However, because the electrons are quantum particles, their wave-functions penetrate some distance through the potential barrier into the insulating region which would otherwise be off-limits (forbidden) according to classical mechanics.  Under the right conditions, the electron wave-functions from either side of the gap may penetrate through the barrier far enough to create a significant overlap between them, which in turn induces a coupling between the different states (i.e., electron on one side or the other) that allows electrons to tunnel completely through the forbidden region from one side to the other.

The two basic principles underlying quantum tunnelling of electrons are that energy must be conserved during the process (only the electrons are doing anything, so there is no opportunity for extra energy to come in or leave the system) and that there must be unoccupied states available on the other side for the electrons to move into.

\begin{figure}
\begin{center}
\includegraphics[width=100mm]{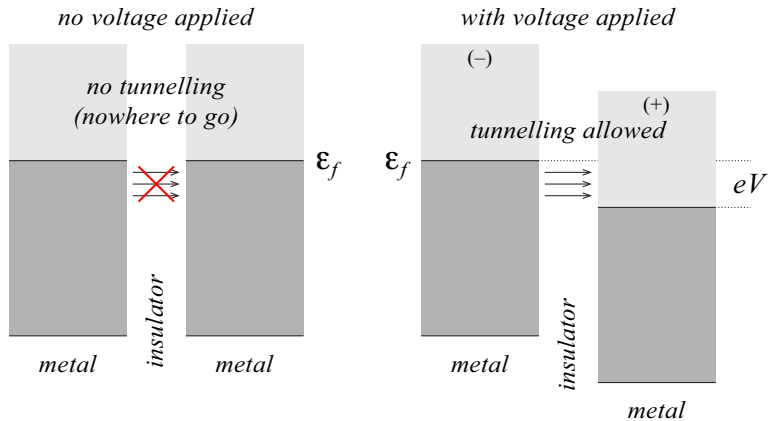}
\end{center}
\captionsetup{labelfont=bf,font={small},margin=6pt}
\caption{\emph{\textbf{Electron tunnelling in normal metals.}}}
\label{fig:tunnelling-nc}
\end{figure}

Consider a tunnel barrier between two identical normal metals (Fig.~\ref{fig:tunnelling-nc}).  As we have done previously, we will assume that the temperature is at absolute zero, so the surface of the Fermi sea is sharply defined (there are no thermal excitations).  Because tunnelling conserves total energy, the electrons can only move exactly sideways.  However, when the two conductors are exactly balanced in voltage, because the metals are identical, they have exactly the same Fermi level.  Consequently, at energies where there are electrons present on one side of the barrier, there are no available states on the other side of the barrier (all states are completely filled up to the Fermi level.  But when a voltage is applied across the tunnel barrier, this lowers the energy levels of one electrode with respect to the other side and there are now electrons on one side of the barrier that can tunnelling into empty levels on the other side without changing energy.  Therefore, tunnelling is allowed and the tunnel current will increase with applied voltage, because increasing the voltage opens up more electrons that can tunnel.

\begin{figure}
\begin{center}
\includegraphics[width=\textwidth]{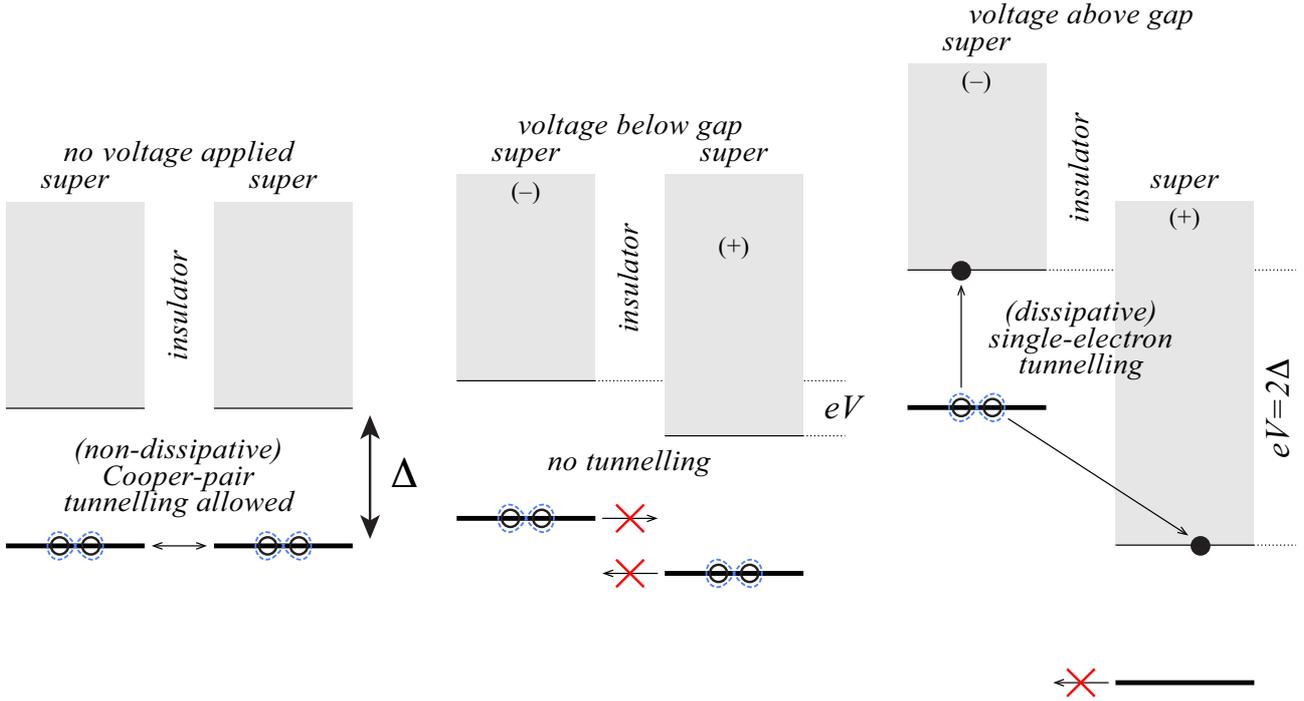}
\end{center}
\captionsetup{labelfont=bf,font={small},margin=6pt}
\caption{\emph{\textbf{Electron tunnelling in superconductors.}}}
\label{fig:tunnelling-sc}
\end{figure}

Now consider a tunnel barrier between two identical superconductors (Fig.~\ref{fig:tunnelling-sc}).  Because the condensate state is effectively bosonic, even though it is a single state, extra Cooper pairs can be easily transferred in and out of it.  Therefore, when the electrodes are balanced (equal voltage), it is possible for Cooper pairs to tunnelling back and forth through the junction.  We will look more at how this occurs later.

However, if a small voltage is applied ($V < 2\Delta / e$), the ground states on either side of the barrier are no longer degenerate and Cooper pairs can no longer tunnelling while maintaining energy conservation.  Interestingly, Cooper-pair tunnelling can still be stimulated in this scenario by irradiating the junction with microwave radiation which is resonant with the voltage-induced gap between the two ground states (i.e., the energy of the microwave photons matches the energy a Cooper pair would gain or lose by tunnelling across the barrier: $\hbar \omega_0 = (2e)V$).

Finally, if a voltage greater than twice the gap is applied ($V\ge 2\Delta/e$), a new single-electron tunnelling process becomes possible.  At zero temperature, there are no unpaired electrons in the condensate, so this process requires breaking up a Cooper pair.  However, at this voltage, the ground-state energy on one side of the barrier is more than $\Delta$ larger than the lowest unoccupied single-electron on the other side, so the enough energy can be lost by one electron tunnelling through the barrier to compensate for the extra energy required for its partner remaining behind to be kicked up into an unoccupied single-electron state.  Overall, the total energy can therefore still be conserved and this tunnelling process is allowed.  As the voltage is further increased, this opens up new channels for single-electron tunnelling (there are more combinations of available states which allow conservation of energy) and the tunnelling current therefore increases.  However, unlike Cooper-pair tunnelling, this is a dissipative process, because the single unpaired electrons no longer benefit from the collective properties of the condensate and are subject to normal resistive scattering processes.

In the next sections, we will now explore the Cooper-pair tunnelling process in more detail.

\subsection{Quantum description of an isolated Josephson junction}

To understand the quantum behaviour of Josephson junctions in more detail, we first consider the behaviour of an isolated Josephson junction, which is where we have two regions (islands) of superconducting material connected by a Josephson tunnel junction.  For the purposes of this course, we approach this from an explicitly quantum mechanical perspective from the beginning (this treatment is based on the one given in Ref.~\cite{DevoretMH1995qfe}).  We start by setting out a number of definitions and deriving several identities that we will need to analyse the junction.

At zero temperature and zero voltage, charge can only flow through the junction in units of Cooper pairs (i.e., the total tunnelling charge $Q(t) = -2e N(t)$).  We therefore look to describe the behaviour of the junction in terms of a number of Cooper pairs.  Specifically, we define an operator $\hat{N}$ which represents the number of Cooper pairs on one side of the tunnel junction, along with its corresponding eigenstates, $\ket{N}$, according to the usual relation:
\begin{align}
\hat{N}\ket{N} = N\ket{N}.
\end{align}

Taking a cue from our earlier discussion of how the condensate wave-function can be described in terms of conjugate number and phase variables (e.g., in Eq.~\ref{eq:sc-condensate-currentphase}), we also define an alternative state basis via:
\begin{align}
\label{eq:sc-number2phase}
\ket{\varphi} = \sum_{N{=}-\infty}^{\infty} e^{iN\varphi} \ket{N}
\end{align}
and the inverse relation:
\begin{align}
\label{eq:sc-phase2number}
\ket{N} = \frac{1}{2\pi} \int_0^{2\pi} d\varphi \, e^{-iN\varphi} \ket{\varphi}
\end{align}
These transformations look and behave much like Fourier transformations.  These variables are also similar to the photon number and phase of an optical mode.  The main difference to note is the convention we use here that $N$ runs from $-\infty$ to $+\infty$.  The intuition behind this choice is that there is some ``neutral charge'' condition where the number of electrons on the island exactly equals the number of protons in the atomic nuclei in the metal lattice, which we label $N=0$.  We then assume that the number of electrons in the condensate is so large that it is reasonable to approximate the lower bound on the summation by $-\infty$.

The first thing we need to check is that Eq.~\ref{eq:sc-phase2number} is indeed the inverse of Eq.~\ref{eq:sc-number2phase}.  This we do by direct calculation.
\begin{align}
\ket{N} &= \frac{1}{2\pi} \int_0^{2\pi} d\varphi \, e^{-iN\varphi} \sum_{M{=}-\infty}^{\infty} e^{iM\varphi} \ket{N} \\
&= \sum_{M{=}-\infty}^{\infty} \ket{N} \br{ \frac{1}{2\pi} \int_0^{2\pi} d\varphi \, e^{-i(N{-}M)\varphi} } \\
&= \sum_{M{=}-\infty}^{\infty} \ket{N} \delta_{NM} \\
&= \ket{N}
\end{align}
where we have used one of the definitions for the Kronecker delta function to prove the required result, namely:
\begin{align}
\delta_{NM} = \frac{1}{2\pi} \int_0^{2\pi} d\varphi \, e^{-i(N{-}M)\varphi}
\end{align}

Next, since $N$ is a standard number variable, we can assume that $\braket{N}{M} = \delta_{NM}$, but we also need to work out what $\braket{\varphi}{\varphi\dash}$ is:
\begin{align}
\braket{\varphi}{\varphi\dash} &= \sum_{MN} e^{-iN\varphi} e^{iM\varphi\dash} \braket{N}{M} \\
&= \sum_{MN} e^{-iN\varphi} e^{iM\varphi\dash} \delta_{NM} \\
&= \sum_N e^{-iN(\varphi-\varphi\dash)} \\
&= 2\pi \, \delta(\varphi-\varphi\dash)
\end{align}
where this time we have used one of the definitions of the Dirac delta function:
\begin{align}
\delta(\varphi-\varphi\dash) = \frac{1}{2\pi} \sum_N e^{-iN(\varphi-\varphi\dash)}.
\end{align}

Working from our guess that $\varphi$ is a phase variable, we next define the following operator:
\begin{align}
e^{i\hat{\varphi}} = \frac{1}{2\pi} \int_0^{2\pi} d\varphi\dash \, e^{i\varphi\dash} \ket{\varphi\dash}\bra{\varphi\dash}
\end{align}
So what is this operator?  How does it operate on the basis state $\ket{\varphi}$?  If these are going to be useful definitions, then we would like to hope that it operates in a reasonable intuitive fashion, e.g., that:
\begin{align}
e^{i\hat{\varphi}} \ket{\varphi} = e^{i\varphi} \ket{\varphi}.
\end{align}
Our next step is therefore to check that this is indeed the case, again by direct calculation:
\begin{align}
e^{i\hat{\varphi}} \ket{\varphi} &= \frac{1}{2\pi} \int_0^{2\pi} d\varphi\dash \, e^{i\varphi\dash} \ket{\varphi\dash}\braket{\varphi\dash}{\varphi} \\
&= \frac{1}{2\pi} \int_0^{2\pi} d\varphi\dash \, e^{i\varphi\dash} \ket{\varphi\dash} 2\pi \, \delta(\varphi-\varphi\dash) \\
&=  e^{i\varphi} \ket{\varphi}
\end{align}
Let us now calculate how this operator acts on a number eigenstate:
\begin{align}
e^{i\hat{\varphi}} \ket{N} &= \frac{1}{2\pi} \int_0^{2\pi} d\varphi \, e^{i\varphi} \ket{\varphi}\braket{\varphi}{N} \\
&= \frac{1}{2\pi} \int_0^{2\pi} d\varphi \, e^{i\varphi} \ket{\varphi} \sum_M e^{-iM\varphi} \braket{M}{N} \\
&= \frac{1}{2\pi} \int_0^{2\pi} d\varphi \, e^{i\varphi} \ket{\varphi} \sum_M e^{-iM\varphi} \delta_{MN} \\
&= \frac{1}{2\pi} \int_0^{2\pi} d\varphi \, e^{i\varphi} e^{-iN\varphi} \ket{\varphi} \\
&= \frac{1}{2\pi} \int_0^{2\pi} d\varphi \, e^{-i(N-1)\varphi} \ket{\varphi} \\
&= \ket{N-1}
\end{align}
This gives us an alternative number-state representation for the operator.  Because this result is true for all number states, $\ket{N}$, we can therefore write:
\begin{align}
e^{i\hat{\varphi}} &= \sum_N \ket{N-1}\bra{N} \\
\&\quad e^{-i\hat{\varphi}} &= \sum_N \ket{N}\bra{N-1}
\end{align}

We are now in a position to derive the key result for calculating the quantum dynamics of a Josephson junction.  We have already seen in earlier sections that the key step in understanding the dynamics of a quantum system is deriving its Hamiltonian.  This can then be used to solve the dynamics of the system through the Schr\"odinger equation.  Our goal here is to study the form of the coupling Hamiltonian for a Josephson junction.

To begin, we know that the non-dissipative Josephson coupling between the two superconducting islands is mediated by the tunnelling of individual Cooper pairs from one electrode to the other.  The simplest coupling Josephson Hamiltonian should therefore describe a single Cooper-pair tunnelling event.  If a single Cooper pair tunnels from one electrode to the other, then the number of Cooper pairs on one electrode will decrease by one, while the number of pairs on the other will correspondingly increase by one.  One consequence of this, which is at least true for this simple example of an isolated Josephson junction, is that the number of pairs on one electrode is completely determined by the number of pairs on the other.  As a result, the state of the entire system can be specified by the number of pairs on just one of the electrodes.  In the number basis, such a single tunnelling event is represented by terms of the form $\ket{N}\bra{N+1}$.  However, it isn't actually necessary to assume that there is a precise number of Cooper pairs on the superconducting island for a single tunnelling event to take place, which means that a single tunnelling event should actually involve an equally weighted sum of such terms over all $N$, i.e.,
\begin{align}
\mathcal{H}_J \sim \sum_N \ket{N}\bra{N+1}
\end{align}
Finally, the Hamiltonian must be a physical observable, so it also needs to be Hermitian, and we need to include a parameter to allow a varying coupling strength.  Combining all of these factors, we get the total coupling Hamiltonian:
\begin{align}
\label{eq:sc-JJ-numberbasis}
\mathcal{H}_J = -\frac{E_J}{2} \sum_N \ket{N}\bra{N+1} + \ket{N+1}\bra{N}
\end{align}
To confirm that this Hamiltonian does operate as expected, we can calculate:
\begin{align}
\mathcal{H}_J \ket{N} = -E_J/2 \, (\ket{N+1}+\ket{N-1})
\end{align}
which gives a sensible result for a single tunnelling event in an unknown direction (i.e., forwards or backwards across the junction).

Using the results we derived above for the phase operator, we can now rewrite the coupling Hamiltonian in the phase basis:
\begin{align}
\label{eq:sc-JJ-phasebasis}
\mathcal{H}_J = -\frac{E_J}{2} \sqbr{e^{i\hat{\varphi}} + e^{-i\hat{\varphi}}} = -E_J \cos \hat{\varphi} = -\frac{E_J}{2\pi} \int_0^{2\pi} d\varphi \, \cos (\varphi) \, \ket{\varphi}\bra{\varphi}
\end{align}
So while the Hamiltonian is a pure coupling Hamiltonian in the number basis, it is diagonal in the phase basis, so the phase states defined above are eigenstates of the Josephson coupling Hamiltonian.  We will see later that the Josephson coupling can be interpreted as a sinusoidal potential energy term and it is this that provides the nonlinearity for qubits in circuit QED.

\subsection{Real-world Josephson junctions connected in electronic circuits}

The most common way to fabricate Josephson junctions in circuit QED is in planar (two-dimensional structures) on a chip.  The standard technique is to lay down one electrode from aluminium using nanolithography, oxidise the surface of the electrode to create an insulating layer of aluminium oxide, before laying down a second electrode which partially overlaps the first.  The tunnel junction is therefore formed from the overlapping region, with the Cooper pairs being able to tunnel through the oxide layer from one electrode to the other.

At its most basic level, a real-world Josephson junction is just two conducting electrodes, separated by an insulator.  This means that a Josephson junction always in part behaves like a capacitor in addition to the quantum tunnel coupling we derived above.  In fact, for the purposes of circuit QED, a real junction can always be represented by a pure capacitor in parallel with a pure tunnelling element.  As a result, we need to add a capacitive term to the tunnelling term we derived in the previous section in order to correctly model the behaviour of a real junction in a circuit.  Written in terms of the charge across the junction capacitance and the flux across the tunnel junction, the more complete Hamiltonian is therefore
\begin{align}
\label{eq:sc-JJ-realworld}
\mathcal{H} = \frac{\hat{Q}^2}{2C} - E_J \cos \br{ 2\pi \hat{\Phi} / \Phi_0}
\end{align}
Thus, a real-world Josephson junction is defined by two parameters, its tunnel coupling energy, $E_J$, and its junction capacitance, $C_J$.  The junction capacitance can also be rewritten as a capacitive energy scale, $E_C = (2e)^2 / 2C_J$, which is energy required to store a single Cooper pair's worth of charge across the junction capacitance.

This equation also gives the first clue as to what sort of role the junction plays in a circuit theory context.  As we explore in more detail in the exercises, expanding the cosine term of the tunnelling element to lowest nontrivial order as a Taylor series gives the Hamiltonian:
\begin{align}
\mathcal{H} = \frac{\hat{Q}^2}{2C} - E_J + \frac{E_J (2\pi)^2}{2 \Phi_0^2} \hat{\Phi}^2
\end{align}
Comparing this with the Hamiltonian we derived in previous sections for the quantum LC oscillator, we see that, to lowest order, the pure tunnelling element behaves like a linear inductor of inductance $L_{J0} = \Phi_0^2 / 4\pi^2 E_J$.  (Of course, when higher-order terms in the expansion are also considered, we see that the tunnelling element is in fact a nonlinear inductor.)  Consequently, a single real-world Josephson junction can be approximately represented by a capacitor in parallel with a (nonlinear) inductor---it is a (nonlinear) quantum LC circuit in a single element.  In other words, a Josephson junction is a resonant circuit, whose resonant frequency is just $\omega_J = 1/\sqrt{C_J L_J}$.

The other important modification we need to consider is what happens when we connect the junction up to an external source.  The most common external source to use in circuit QED experiments is a constant current source.  Specifically, for a real junction biased (or shunted) by a constant current, $I$, the junction's contribution to the Hamiltonian is given by:
\begin{align}
\label{eq:sc-JJ-currentbiased}
\mathcal{H} = \frac{\hat{Q}^2}{2C} - I\hat{\Phi} - E_J \cos \br{ 2\pi \hat{\Phi} / \Phi_0}
\end{align}
As we will see in the next section of the course, we can interpret the last two terms of this Hamiltonian as a nonlinear potential energy:
\begin{align}
\label{eq:sc-JJ-tiltedwashboard}
\mathcal{U}(\hat{\Phi}) = - I\hat{\Phi} - E_J \cos \br{ 2\pi \hat{\Phi} / \Phi_0}
\end{align}
where $\Phi$ is playing the role of an effective canonical (Hamiltonian) ``position'' variable.  This is often known as a ``tilted washboard potential'', for reasons which are obvious when it is plotted as a function of $\Phi$.  We won't derive this result in this course, but this form of the Hamiltonian will become particularly relevant in the next sections when we look at phase qubits.

For the moment, we will just briefly explore what this potential can tell us in a so-called ``semiclassical'' context.  When we introduced the superconducting condensate wave-function earlier in this section, we discussed the fact that, in many circumstances, a superconducting condensate can be well described by a well-defined Cooper-pair number density and condensate phase.  Because these parameters are just numbers rather than noncommuting operators, they in fact behave as classical variables, even though they are describing an explicitly quantum mechanical phenomenon in the form of superconductivity.  This juxtaposition of classical and quantum concepts is an intriguing feature of superconductivity and is tied up with the fact that superconducting condensates are macroscopic collective systems.

For the purposes of this discussion, we just need to realise that, in this semiclassical context, the dynamics of a current-biased Josephson junction is described by the pre-quantum version of Eq.~\ref{eq:sc-JJ-currentbiased} (i.e., no hats on operators).  The junction can therefore be thought of as a ``classical'' particle with position $\Phi$ and momentum $Q$, moving in the potential given by Eq.~\ref{eq:sc-JJ-tiltedwashboard} (again, no hats).  As with any particle moving in a potential, it will tend to gravitate towards an equilibrium or ``at rest'' position at the bottom of a potential minimum.  In the exercises, we show that the position of this potential minimum is simply given by:
\begin{align}
\Phi = \frac{\Phi_0}{2\pi} \arcsin \br{ \frac{\Phi_0}{2\pi E_J} I }.
\end{align}
When $I=0$, the potential minimum is simply located at $\Phi=0$, which can also be seen by inspection directly from the cosine form of the potential.  As $I$ approaches a critical value of $I_0 = 2\pi E_J/\Phi_0$, however, we can see that there will no longer be a valid solution for $\Phi$ and this corresponds to the point where the effect of the current bias is greater than the corrugation of the cosine potential and there will no longer be a potential minimum to trap the classical particle.  At the critical current value, the potential reaches a point of horizontal inflection at $\Phi = \Phi_0/4$.  Interestingly, it turns out that rearranging this equation for $\Phi$ just gives the standard Josephson current relation for a tunnel junction:
\begin{align}
\label{eq:sc-JJ-currentrelation}
I = I_0 \sin \br{ 2\pi \Phi / \Phi_0} = I_0 \sin \varphi.
\end{align}
We therefore see that the Josephson current relation describes the semiclassical behaviour of a Josephson junction, in the sense that it defines the position of the potential minimum at which point a classical particle would come to rest.

For further discussion of the standard Josephson phenomena, look at the exercises, or alternatively the chapter on Superconductivity in Volume III of the Feynman Lectures~\cite[Ch.~21]{FeynmanLectures3}, Chapter 10 in Kittel's ``Introduction to Solid-State Physics''~\cite{Kittel}, or Chapter 11 in Rose-Innes and Rhoderick~\cite{Rose-InnesRhoderick}.

\subsection{Exercises---superconductivity}

\subsubsection{The semiclassical Josephson effect}

We saw above that a superconducting condensate can be described (in a semiclassical limit) by a many-electron wave-function, $\Psi$, where $|\Psi|^2$ is the density or number of electron pairs ($\equiv n$), and where $\Psi$ is characterised by a macroscopic condensate phase, via $\Psi = \sqrt{n} \, e^{i\theta}$.

Consider a superconducting tunnel junction, where two small isolated islands of identical superconducting material are separated by a small layer of insulating material.  The superconducting wave-functions inside the two islands are described by $\Psi_1 = \sqrt{n_1} \, e^{i \theta_1}$ and $\Psi_2 = \sqrt{n_2} \, e^{i \theta_2}$.  Here, because the superconducting regions are small, isolated ``lumps'', we can consider $n_j$ to represent the total number of electron pairs in the region.

The evanescent overlap of the two superconducting wave-functions inside the ``forbidden'' insulating region induces tunnelling between the two islands.  In terms of these wave-functions, this coupling can be represented by the following coupled differential equations:
\begin{align}
i\hbar \frac{d\Psi_1}{dt} &= \frac{E}{2} \Psi_2 \\
i\hbar \frac{d\Psi_2}{dt} &= \frac{E}{2} \Psi_1
\end{align}
These coupled equations are derived from the Schr\"odinger equation.

\begin{enumerate}[label=(\alph*)]

\item Substituting the definitions of $\Psi_1$ and $\Psi_2$ into these coupled equations, show that the independent differential equations for the number and phase variables are:
\begin{align}
\nn
\frac{dn_1}{dt} &= \frac{E}{\hbar} \sqrt{n_1 n_2} \sin \delta \\
\nn
\frac{d\theta_1}{dt} &= -\frac{E}{2\hbar} \sqrt{\frac{n_2}{n_1}} \cos \delta \\
\nn
\frac{dn_2}{dt} &= -\frac{E}{\hbar} \sqrt{n_1 n_2} \sin \delta \\
\nn
\frac{d\theta_2}{dt} &= -\frac{E}{2\hbar} \sqrt{\frac{n_1}{n_2}} \cos \delta
\end{align}
where $\delta = \theta_2-\theta_1$ is the condensate phase difference across the junction.

[Hint: Rewrite the coupled differential equations in terms of $\delta$ and equate the real and imaginary parts to isolate the desired derivative terms.]

\item Although the islands are small, say, by comparison with the condensate wavelength, they still contain an enormous number of Cooper pairs.  Since they are made from identical superconductors, we can therefore make the approximation $n_1 \approx n_2 \equiv n_0$.

Use this approximation to show that: (i) $d\delta/dt = 0$, and (ii) $dn_1/dt = -dn_2/dt$.

\item Defining the current flowing through the junction to be $I = 2e \, dn_1/dt$ (where $2e$ is the total charge of a Cooper pair), use the above relations to derive the DC Josephson current relation (defining $I_0 = \frac{2e}{\hbar} n_0 E$):
\begin{align}
I = I_0 \sin \delta.
\end{align}

\end{enumerate}

\subsubsection{Understanding Josephson junctions}

In the notes above, starting with the simple assumption that a Josephson junction was described by a Hamiltonian coupling term which involved the tunnelling of single Cooper pairs across an insulating gap, we showed that this could be rewritten as a periodic potential energy term oscillating as a function of the superconducting phase difference across the junction, $\varphi$:
\begin{align}
\mathcal{H} = -E_J \cos \varphi.
\end{align}

\begin{enumerate}[label=(\alph*)]

\item \emph{Linear Josephson inductance}: Comparing this Josephson coupling with the standard Hamiltonian potential energy term for a linear inductor, $\mathcal{H} = \Phi^2 / 2L$, show that the Josephson junction is approximately equivalent to a linear inductor with inductance:
\begin{align}
\nn
L_{J0} = \frac{\Phi_0^2}{4\pi^2 E_J},
\end{align}
where $\Phi_0 = h/2e$ is the superconducting flux quantum.  [Hints: Recall that the superconducting phase variable, $\varphi$, is related to the flux, $\Phi$, via $\varphi = 2\pi \Phi / \Phi_0$.  Expand the cosine using a Taylor expansion to lowest nontrivial order and ignore any constant energy offset terms.]

\item \emph{Nonlinear Josephson inductance}: Using the standard potential term for a linear inductor, it is straightforward to show that $\partial^2\mathcal{H}/\partial\Phi^2 = 1/L$, which is a constant, independent of $\Phi$.  Based on this, we can define a general circuit inductance:
\begin{align}
L(\Phi) = \br{ \dasq{H}{\Phi} }^{-1}
\end{align}
Using this formula, show that the general nonlinear inductance for a Josephson element is:
\begin{align}
\nn
L_J(\Phi) = \frac{L_{J0}}{\cos (2\pi \Phi/\Phi_0)}
\end{align}

\item \emph{Current-biased Josephson junction}: If a constant current is passed through a Josephson junction, the Hamiltonian potential energy must be modified to give the following new form:
\begin{align}
\mathcal{H} = -\frac{I\Phi_0}{2\pi} \varphi - E_J \cos \varphi \equiv \mathcal{U}(\varphi)
\end{align}
where $\mathcal{U}(\varphi)$ is a so-called ``tilted washboard potential'', because it looks like a corrugated potential angled at an average slope determined by the bias current, $I$.

\begin{enumerate}[label=\roman*.]
\item At zero bias current, the first potential minimum is found at $\varphi = 0$.  However, as the potential is tilted by the bias current, the location of the first potential minimum moves away from zero.  For nonzero bias current, show that the location of the first potential minimum is given by:
\begin{align}
\nn
\varphi = \arcsin \br{ \frac{\Phi_0}{2\pi E_J} I }
\end{align}
[Hint: Remember that the potential is at a minimum when $d\mathcal{U}/d\varphi = 0$.]

\item Hence show that when the current is larger than a critical value given by $I_0 = 2\pi E_J / \Phi_0$, the potential no longer has any local minima.

\item Using this critical current, show that the above formula is simply the standard DC Josephson relation, $I = I_0 \sin \varphi$.

\emph{Note: These results show us that the DC Josephson relation simply describes the position of the first minimum in the Hamiltonian potential energy term for a current-biased Josephson junction.  Thus, it turns out that the DC Josephson relation actually describes the behaviour of a superconducting Josephson junction in a ``semiclassical'' picture, where the superconducting phase can be interpreted as a classical ``particle'' moving in a tilted wash-board potential.  In this picture, the location of the potential minimum always gives the ``at rest'' position of particle.  This interpretation clashes with the genuinely quantum picture, however, where the superconducting phase cannot be represented by a point particle, but must instead by described by a wave-function, whose probability distribution is spread out in the potential well, and which has discrete energy levels.  This genuinely quantum picture is critical for describing qubits in circuit QED.}

\item In the semiclassical picture, when the bias current is larger than the critical current, the ``particle'' representing the superconducting phase is no longer confined to a potential well, but transitions into the so-called ``free-running'' state where it gradually accelerates down the potential slope, and the phase evolves according to a trajectory $\varphi = \varphi(t)$.  In this scenario, the rate of work done in accelerating the evolution of the superconducting phase is simply given by the negative rate of change of the potential energy, $-d\mathcal{U}(\varphi)/dt$.

Firstly, show that this is equal to:
\begin{align}
\text{rate of work done} \equiv -d\mathcal{U}(\varphi)/dt = \frac{\Phi_0}{2\pi} \br{ I - I_0 \sin \varphi } \frac{d\varphi}{dt}
\end{align}

Secondly, when the bias current is much larger than the critical current, $I \gg I_0$, show that:
\begin{align}
V = \frac{\Phi_0}{2\pi} \frac{d\varphi}{dt}
\end{align}
This is the second Josephson relation.  [Hint: Use the fact that the rate of work done is also the power used in accelerating the evolution of superconducting phase and is therefore also given by $\text{work rate}=IV$ (current times voltage).]

\emph{Note:  This exactly reproduces, for Josephson junctions, the equation $V = d(LI)/dt = d\Phi/dt$, which we derived in the notes above for the standard linear inductor.}

\end{enumerate}

\end{enumerate}

\subsubsection{Tunable Josephson junctions}

\begin{figure}
\begin{center}
\includegraphics[width=70mm]{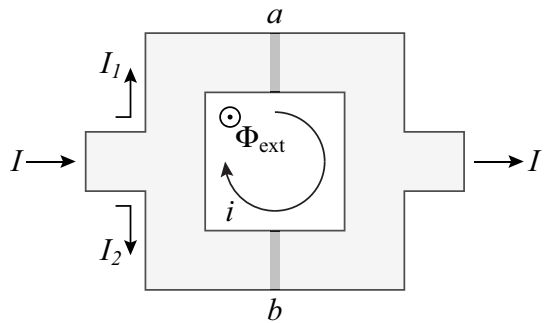}
\end{center}
\caption{A tunable Josephson junction.}
\label{tunablejunction}
\end{figure}

In circuit QED, Josephson junctions are the key element which provides the lossless nonlinearities required for making qubits.  It is often very useful, however, to be able to tune the characteristics of the junction ``in situ'', i.e., while the experiment is in progress.  The standard way to do this is to use a ``split Josephson junction'', a SQUID.  Such a circuit (shown in Fig.~\ref{tunablejunction}) consists of two, usually identical junctions in parallel, which introduces a superconducting loop into the circuit whose properties can be adjusted by applying a magnetic field through the loop.

In this section, we saw that there are three effects which can give rise to a superconducting phase difference around a circuit: 1) as a result of the thin insulating layer, the Josephson junctions themselves have a phase difference across them which relates to the tunnelling supercurrent passing through them via the DC Josephson relation which was derived in Question 1: $I = I_0 \sin \varphi$; 2) currents passing through a superconducting material intrinsically lead to phase gradients along the path of current flow; 3) applied magnetic fields lead to an extra component of phase change relating to the enclosed magnetic flux.

When the loop is superconducting, it must have a definite, single phase value at every point.  This means that the total phase shift around the loop must be quantised in integer multiples of $2\pi$:  $\Delta \varphi_{\rm tot} = n 2\pi$.  It can also be shown that the total phase around the loop is given by:
\begin{align}
\Delta \varphi_{\rm tot} = \varphi_a + \varphi_b + \Delta \varphi(\Phi\dash),
\end{align}
where $\varphi_j$ is the phase difference across junction $j(=a,b)$ in the defined direction of the circulating current, and where $\Delta \varphi(\Phi\dash)$ is the phase shift for the rest of the loop, which is related to the nett flux enclosed by the loop, $\Phi\dash$.  This flux can have contributions from both the externally applied flux and the flux due to circulating currents.

In circuit QED, these split junctions are operated in the regime where the junctions' critical currents and the inductance of the loop are both very small (i.e., small junctions and small loop).  In this regime, where the flux due to circulating currents is always much smaller than the superconducting flux quantum, $\Phi_0 = h/2e$, it turns out that only the applied flux contributes significantly to the nett flux through the loop, so we can write:
\begin{align}
\Delta \varphi = 2\pi \frac{\Phi_{\rm ext}}{\Phi_0}.
\end{align}

Combining these relations, we therefore have the following overall phase quantisation relation for the split-junction:
\begin{align}
\varphi_a + \varphi_b + 2\pi \frac{\Phi_{\rm ext}}{\Phi_0} = n2\pi.
\end{align}

The other design simplification which is usually applied when using split junctions in circuit QED is to make the two junctions identical.  As a result, they will have the same critical current $I_0$, and, since the two alternative paths through the loop (through junction $a$ and junction $b$) are therefore also identical, the overall current flowing through the split junction, $I=I_1+I_2$, will divide equally between the two paths, i.e., $I_1 = I_2 = I/2$.

\begin{enumerate}[label=(\alph*)]

\item Consider first the case where there is no overall current flowing through the split junction (i.e., $I=0$), so that the only current flowing in the circuit is the circulating current which results from the applied magnetic field.

\begin{enumerate}[label=\roman*.]
\item What is the nett current flowing through each junction (in the clock-wise direction)?
\item Using the DC Josephson relation, show that $\varphi_a (I{=}0) = \varphi_b (I{=}0)$.
\item Using the phase quantisation relation, now show that:
\begin{align}
\nn
\varphi_a (I{=}0) = \varphi_b (I{=}0) = \pi \br{ n - \Phi_{\rm ext} / \Phi_0 }
\end{align}
\end{enumerate}

\item Consider now the case where an overall current is flowing through the split junction (i.e., $I \neq 0$).  The nett current flowing through each junction is then: $I_a = i + I/2$ and $I_b = i - I/2$.  Because these tunnelling supercurrents are no longer equal, the phase jumps across the two junctions must also be different.  However, the phase quantisation relation shows us that the sum of these phases must be constant for a given applied field, i.e., $\varphi_a + \varphi_b = \text{constant}$.  Using the results from the previous question, we can therefore define the new junction phases:
\begin{align}
\varphi_a &= \pi \br{ n - \Phi_{\rm ext} / \Phi_0 } + \delta\varphi \\
\varphi_b &= \pi \br{ n - \Phi_{\rm ext} / \Phi_0 } - \delta\varphi
\end{align}
where $\delta\varphi$ is a newly defined offset phase, as well as the corresponding Josephson relations:
\begin{align}
I_a &= i + I/2 = I_0 \sin \br{ \pi \br{ n - \Phi_{\rm ext} / \Phi_0 } + \delta\varphi } \\
I_b &= i - I/2 = I_0 \sin \br{ \pi \br{ n - \Phi_{\rm ext} / \Phi_0 } - \delta\varphi }
\end{align}
Ultimately, however, we don't care about the internal degrees of freedom of this circuit, such as $I_a$, $I_b$ and $i$, but only the behaviour of the external degrees of freedom, $I$.

\begin{enumerate}[label=\roman*.]
\item Using the above relations, show that the effective Josephson relation for the overall split-junction element is:
\begin{align}
\nn
I = 2I_0 \cos  \sqbr{ \pi \br{ n - \Phi_{\rm ext} / \Phi_0 } } \sin \delta\varphi.
\end{align}
[Hint: Use the standard trigonometric relations, $\sin (A \pm B) = \sin(A) \cos(B) \pm \cos(A) \sin(B)$.]
\item Comparing this equation with the basic DC Josephson relation, justify how this split-junction element can be interpreted as a single tunable Josephson junction.  Identify the variable which represents the effective phase difference across the tunable junction and write down the form of the tunable critical current.
\end{enumerate}

\end{enumerate}

\section{Quantum circuits and qubits}

The overarching goal of this course is to learn about cavity QED, particularly in the context of circuit QED.  As we have already discussed, cavity QED is the study of strong quantum interactions between \emph{light} and \emph{matter}.  The following table briefly summarises some of the key concepts we have covered so far in a simple flow chart.

\begin{table}[h!]
\begin{center}\begin{tabular}{c|c}
\uline{Light} & \uline{Matter} \\
& \\
radiation / resonators & atoms / qubits (simplest form) \\
$\downarrow$ & $\downarrow$ \\
simple harmonic oscillators: $V = \half k x^2$ & need to address \\
quadratic potential = linear restoring force & individual transitions \\
(linear system) & $\downarrow$ \\
$\downarrow$ & $\downarrow$ \\
equal energy level spacings & need unequal energy spacings \\
$\downarrow$ & $\downarrow$ \\
$\downarrow$ & need nonlinear system (non-quadratic potential) \\
$\downarrow$ & $\downarrow$ \\
(For Circuit QED) & (For Circuit QED) \\
LC resonators & Josephson junctions \\
waveguide resonators & $V=-E_J \cos \varphi = -E_J \br{1 - \frac{\varphi^2}{2!} + \frac{\varphi^4}{4!} - \ldots}$ \\
& $\Rightarrow$ nonlinearity
\end{tabular}\end{center}
\end{table}

In the last decade, there have been many circuit QED experiments demonstrating many different ``species'' of artificial qubits, a flexibility which has one of the strengths of circuit QED as an architecture for designer quantum systems.  However, there are three key types of qubit that encompass the main concepts, commonly referred to as \emph{charge qubits}, \emph{phase qubits} and \emph{flux qubits}.  In this course, we will focus on charge qubits in order to understand the key concepts and develop our basic tools.  We will then briefly discuss the two other types in a sort of cartoon picture.

\subsection{Classical electronics}

As we have seen in earlier sections, in classical electronics, we solve the dynamics of a circuit using Kirchoff's rules, with the voltage ($V$) and current ($I$) as the relevant variables.  We then use the appropriate current voltage relations for the individual elements:
\begin{align}
\text{Capacitor:}&\quad I = C \frac{dV}{dt} = \frac{d}{dt}(CV) = \frac{dQ}{dt} \\
\text{Inductor:}&\quad V = L \frac{dI}{dt} = \frac{d}{dt}(LI) = \frac{d\Phi}{dt}
\end{align}
where we have used the relations $Q = CV$, which is the definition of a linear capacitance, and $\Phi = LI$, which is the definition of a linear inductance.  These equations show how the dynamical voltage and current circuit variables can also be expressed in terms of charge, $Q$, and flux, $\Phi$, variables.  This provides some underlying intuition as to why $Q$ and $\Phi$ can be used to give an alternative, equivalent description of circuit dynamics.  It is also worth noting that the voltage-flux relation for a linear inductor, $V = \dot{\Phi}$, is exactly the same as the second Josephson relation (see exercises).  This connection also gives an intuitive notion of why this can be used to provide a more generalised definition of flux which is not restricted to closed circuit loops.

We have already seen that the Hamiltonian for a LC resonator circuit is given by:
\begin{align}
\mathcal{H} = \half CV^2 + \half LI^2.
\end{align}
From this starting point, we derived Eq.~\ref{eq:quantumLC-Hcharge} by making the substitutions $V=Q/C$ and $I=\dot{Q}$, giving:
\begin{align}
H = \frac{L}{2}\dot{Q}^2 + \frac{1}{2C} Q^2.
\end{align}
Comparing this with the Hamiltonian for a simple mechanical oscillator leads to a natural identification of $Q$ as an effective ``position'' variable, making the first term a ``kinetic energy'' term and the second a potential energy term.  Furthermore, we can use these identifications to determine the corresponding momentum to be $p = mv = L\dot{Q} = LI = \Phi$, giving:
\begin{align}
H = \frac{\Phi^2}{2L} + \frac{Q^2}{2C}.
\end{align}
But in the previous section, we saw that the nonlinear coupling term for a Josephson junction is $V_J(\Phi) = -E_J \cos \br{2\pi \Phi / \Phi_0}$.  If we interpret $\Phi$ as a momentum, then this is some strange nonlinear kinetic energy term, which is not a familiar concept.  Instead, we'd much rather have a standard kinetic energy term with a nonlinear \emph{potential energy}, since we're very familiar with how to solve the dynamics of quantum particles in complex potentials.  We therefore go back to the original voltage-current form of the Hamiltonian and make the alternative, although perhaps slightly less intuitive substitutions $V = \dot{\Phi}$ and $I = \Phi/L$, giving:
\begin{align}
H = \frac{1}{2L} \Phi^2 + \frac{C}{2}\dot{\Phi}^2.
\end{align}
We can now identify $\Phi$ as the new canonical position variable, and this time, the conjugate canonical momentum is $p = mv = C\dot{\Phi} = CV = Q$.  As it must, this of course leads to exactly the same Hamiltonian as above, but this time we identify the inductive (flux) term as a potential energy and the capacitive (charge) term as a kinetic energy.

\subsection{Charge qubits: a.k.a. Cooper-pair box qubits}

\subsubsection{The isolated Cooper-pair box}

We start by considering an isolated Cooper-pair box (CPB), which is a small island of superconducting material connected to a superconducting reservoir (ground) via a tunnel junction with capacitance $C_J$ and Josephson energy $E_J$.  The key conceptual feature of a CPB is that the island must be small enough (i.e., the islands capacitance to the outside world must be small enough) that adding or removing even one Cooper pair from the island can make a significant contribution to the energy of the system.

Because only one side of the junction is connected to the outside world (via the superconducting reservoir), the full CPB Hamiltonian is exactly the same as the one we derived for an isolated Josephson junction in the previous section, i.e.:
\begin{align}
\mathcal{H} = \frac{\hat{Q}^2}{2C_J} - E_J \cos \br{ 2\pi \hat{\Phi} / \Phi_0}
\end{align}
where $\Phi$ and $Q$ are now conjugate quantum variables obeying the commutation relation: $\sqbr{ \Phi, Q } = i\hbar$.  We can also re-express this Hamiltonian in a completely equivalent form using the alternative number and phase variables, defined by $\hat{Q} = 2e \hat{N}$ and $\Phi = \varphi_0 \hat{\varphi}$:
\begin{align}
\mathcal{H} = E_C \hat{N}^2 - E_J \cos \hat{\varphi}
\end{align}
where $E_C = (2e)^2/2C_J$ is the junction charging energy and $N$ and $\varphi$ obey the commutation relation: $\sqbr{\varphi,N} = i$.

Before moving on, it is worth noting that the capacitive term defines the charging energy associated with adding extra Cooper pairs onto the small CPB, while the Josephson term defines the energy associated with Cooper pairs crossing the tunnel junction.

\subsubsection{The voltage-biased Cooper-pair box}

A problem with the very simple isolated CPB is that we don't have any method for controlling the qubit.  We therefore consider a slightly generalisation where we bias the electrostatic potential of the CPB by applying a ``gate voltage'' to a nearby electrode, which is capacitively coupled to the island.

Typically, such a voltage biasing can be achieved simply by bringing the end of a superconducting wire near to the CPB island.  The island is, by definition, extremely sensitive to changes in the charge environment, since the CPB is designed to be strongly affected even by a single Cooper-pair.  However, in order to develop an intuition for how this will affect the system Hamiltonian, let us assume that the island is one lump of superconducting material with two very small superconducting wires attached to it.  One wire connects the CPB to the reservoir via a tunnel junction, while the other connects the CPB to a gate capacitor, $C_g$, which is in turn connected directly to a gate voltage.  Using the normal circuit rules, it is not too difficult to show that applying a gate voltage, $V_g$, induces a ``gate charge'', $Q_g$ on the electrode of the gate capacitor which is attached to the island.  As a result, this charge has been drawn off the CPB island and the CPB charging energy is now determined only by the $Q-Q_g$ which remains on the island itself.  The modified Hamiltonian is therefore:
\begin{align}
\mathcal{H} &= \frac{\br{\hat{Q}-Q_g}^2}{2C_\Sigma} - E_J \cos \br{ 2\pi \hat{\Phi} / \Phi_0} \\
&= E_C \br{\hat{N}-N_g}^2 - E_J \cos \hat{\varphi}
\end{align}
where $C_\Sigma = C_g + C_J$ is the total capacitance from the island to the rest of the world and $N_g = Q_g/2e$.  We therefore see that the gate voltage allows us to set an arbitrary offset charge for the CPB, defining a new ``zero charge'' reference point.  The basic intuition is that if you apply a positive gate voltage near the CPB, then it becomes more energetically favourable to have extra negatively charged Cooper pairs on the island, while if you apply a negative gate voltage, the reverse is true.

To understand this system in more detail, we will now calculate its energy eigenstates, i.e., the states $\ket{E}$ such that $\mathcal{H} \ket{E} = E\ket{E}$.  To do so, it is convenient to first rewrite the Hamiltonian explicitly in the Cooper-pair number basis.  We already know what the coupling term looks like in this form (Eq.~\ref{eq:sc-JJ-numberbasis}).  For the other, we use the definition of $\hat{N} = \sum_N N\ket{N}\bra{N}$ to show that $\hat{N}^2 = \sum_N N^2\ket{N}\bra{N}$, as well as the number-basis completeness relation $I = \sum_N \ket{N}\bra{N}$, giving:
\begin{align}
\label{eq:qubits-chargeQB-numberbasisH}
\mathcal{H} &= E_C \sum_N \br{N-N_g}^2 \ket{N}\bra{N} - \frac{E_J}{2} \sum_N \ket{N}\bra{N{+}1} + \ket{N{+}1}\bra{N}
\end{align}

\subsubsection{Case: $E_J=0$ --- no tunnel coupling}

Traditionally, charge qubits are operated in the regime where the coupling is much smaller than the charging energy ($E_J \ll E_C$).  It is therefore instructive to start by considering the case where there is no coupling at all, i.e., $E_J = 0$, where the Hamiltonian is just:
\begin{align}
\mathcal{H} &= E_C \sum_N \br{N-N_g}^2 \ket{N}\bra{N}
\end{align}
It is immediately obvious that $\mathcal{H}$ is already in diagonal form in this basis, which means that the number states, $\{\ket{N}\}$, are also simultaneously the energy eigenstates of the system%
\footnote{It is easy to verify this is the case by showing by direct calculation that $\mathcal{H}\ket{N} \propto \ket{N}$.}.
We can therefore write down, by inspection, that the eigenstate $\ket{E_N} = \ket{N}$ has eigenenergy $E_N = E_C \br{N-N_g}^2$.

If we now plot the eigenenergies as a function of gate offset ($N_g$), we see that we have a series of parabolas, where key energy scales and values for gate offset are defined by the points where different eigenenergy curves cross one another.  At lower energies, these key points are:
\begin{align}
N_g &= N \quad\gives E_N = 0 \\
N_g &= N\pm\frac{1}{2} \quad\gives E_N = \frac{E_C}{4} \\
N_g &= N\pm 1 \quad\gives E_N = E_C \\
N_g &= N\pm \frac{3}{2} \quad\gives E_N = \frac{9E_C}{4}
\end{align}
Since there is no tunnel coupling in this scenario, transitions between different eigenstates can only be caused by absorbing or emitting a microwave photon.

These results show that there are two important points in the energy diagramme in terms of gate offset charge $N_g$ which define two fundamentally different operating regimes for the CPB.

\emph{\textbf{Good ground state}}: At $N_g = N$, there is one low-energy level ($E_N = 0$) and two next--highest energy levels at $E_{N{\pm}1} = E_C$.  At this setting, we therefore have a good ground state which is relatively well separated from all other levels.  \emph{This allows good state preparation.}

\emph{\textbf{Good qubit system}}: By contrast, at $N_g = N+\half$, there are two degenerate lowest-energy levels ($E_N = E_{N{+}1} = E_C/4$), while the next-highest energy levels are now much higher at $E_{N{-}1} = E_{N{+}2} = 9E_C/4$.  We now have two close energy levels which are well separated from all other levels, making this a \emph{potentially good qubit system}.

\subsubsection{Case: $E_J \ll E_C$ --- small coupling regime}

Now that we have considered the zero coupling case, we next move on to the traditional charge-qubit regime of small tunnel coupling.  Intuitively, we can guess that, for sufficiently small $E_J$, the tunnel coupling must act as a small perturbation on the zero coupling case.  (Note that the only meaningful way to define ``small'' is in comparison with other energy scales, namely $E_C$.)  In other words, in the small coupling regime, the energy eigenstates should look almost exactly the same as the $E_J=0$ case, i.e., $\ket{E} \approx \ket{N}$ should still be true.  Furthermore, because $E_J$, the energy change arising from a tunnelling event, is small, the coupling term will only have a significant effect at places in the energy diagramme where different $E_J=0$ eigenstates have almost the same energy.  So all interesting deviations will occur near these ``crossing points''.

We will now focus on the qubit scenario, defining $N_g = N + \half + \Delta_g$.  At this point, since the energy landscape is periodic in $N_g$, we can choose to set $N=0$, for simplicity.  Provided $\Delta_g$ is sufficiently small, there are two low-energy levels ($\ket{0}$ and $\ket{1}$) and all other levels have much higher energy (at least 9 times $E_{0,1}$).  We therefore assume that the higher-energy levels play a negligible role in the system dynamics and simply ignore them in our calculations.  This is known as the ``two-level approximation'' or the ``qubit approximation''.

Making the substitution $N_g = \half + \Delta_g$ and discarding any Hamiltonian terms which involve states other than $\ket{0}$ and $\ket{1}$, we get the following simplified qubit Hamiltonian:
\begin{align}
\mathcal{H} &= E_C \br{\half + \Delta_g}^2 \ket{0}\bra{0} + E_C \br{\half - \Delta_g}^2 \ket{1}\bra{1} - \frac{E_J}{2} \br{ \ket{0}\bra{1} + \ket{1}\bra{0} } \\
&= E_C \br{\smfrac{1}{4} + \Delta_g^2} \br{ \ket{0}\bra{0} + \ket{1}\bra{1} } + E_C \Delta_g \br{ \ket{0}\bra{0} - \ket{1}\bra{1} } - \frac{E_J}{2} \br{ \ket{0}\bra{1} + \ket{1}\bra{0} } \\
& = E_C \br{\smfrac{1}{4} + \Delta_g^2} I + E_C \Delta_g \, \sigma_z - \frac{E_J}{2} \, \sigma_x
\end{align}
We can further simplify our calculations by noting that all states are eigenstates of the identity operator $I$, so this term just adds an equal energy offset of $E_C \br{\smfrac{1}{4} + \Delta_g^2}$ to all states.  We can therefore omit this term from our calculations and add it in at the end.  We are therefore left with the final qubit Hamiltonian:
\begin{align}
\label{eq:qubits-chargeQB-reducedH}
\mathcal{H} &= E_C \Delta_g \, \sigma_z - (E_J/2) \, \sigma_x \\
&= \sqbr{\begin{matrix} E_C \Delta_g & -E_J/2 \\ -E_J/2 & -E_C \Delta_g \end{matrix}}
\end{align}
which is characterised by two new energy scales $E_C \Delta_g$ and $E_J/2$.

\emph{\textbf{Case 1}}: $E_C \Delta_g \gg E_J / 2$ --- When $\Delta_g$ is large enough, the Hamiltonian is still approximately diagonal, so the energy eigenstates are still $\ket{0}$ and $\ket{1}$, exactly as in the $E_J = 0$ case%
\footnote{As we will see later, for the excited state $\ket{1}$, this is not strictly true when $\Delta_g$ approaches $1/2$, where its energy approaches the energy of the $E_J=0$ state $\ket{-1}$.  This situation is where the qubit approximation breaks down.}.

\emph{\textbf{Case 2}}: $\Delta_g = 0$ --- This is a special case, since the Hamiltonian is completely off-diagonal in form:
\begin{align}
\mathcal{H} = - (E_J/2) \, \sigma_x.
\end{align}
But the eigenvectors for the $\sigma_x$ Pauli matrix are already known to be $\ket{\lambda_\pm} = \ket{\pm}$, with corresponding eigenvalues $\lambda_\pm = \pm 1$.  Since the Hamiltonian is just proportional to $\sigma_x$, its eigenstates will also be $\ket{E_\pm} = \ket{\pm}$ and we can immediately note, by inspection, that the corresponding eigenenergies will be $E_\pm = (-E_J/2) \cdot (\pm 1) = \mp E_J/2$.  More completely, if we add back in the energy offset we omitted at the beginning, $E_\pm = E_C/4 \mp E_J/2$.

The first thing to note from these results is that there is now an energy gap between the two eigenstates of $\Delta E = E_J$, i.e., as a result of the coupling, the eigenstates are no longer degenerate.  In technical parlance, we say that the ``tunnel coupling has lifted the degeneracy of the eigenstates''.  Furthermore, at least for small couplings, the size of the eigenenergy splitting is proportional to the coupling strength.  The other important point to note is that the eigenstates are no longer states with a well-defined value of $N$, but are instead the symmetric and antisymmetric superposition states.  This kind of result is really ubiquitous in quantum mechanics---it comes up all the time, in fact anywhere there are two coupled basis states, such as a particle confined to a double-well potential.  We will see another example of this later in the section, when we look at flux qubits.

\emph{\textbf{Case 3}}: $E_C \Delta_g \lesssim E_J/2$ --- This is now the most general case within the scope of the qubit approximation.  To solve this case, we need to calculate the eigenenergies for the full Hamiltonian in Eq.~\ref{eq:qubits-chargeQB-reducedH}, following the usual approach:
\begin{align}
&\mathcal{H} \ket{E} = E \ket{E} \\
\gives &(H-EI) \ket{E} = 0
\end{align}
Nontrivial solutions of this equation require $\text{det}(H-EI) = 0$.  Therefore:
\begin{align}
0 &= \text{det}\br{ \sqbr{\begin{matrix} E_C \Delta_g - E & -E_J/2 \\ -E_J/2 & -E_C \Delta_g - E \end{matrix}} } \\
&= -(E_C \Delta_g - E) (E_C \Delta_g + E) -E_J^2/4 \\
&= E^2 - \br{E_C^2 \Delta_g^2 + E_J^2/4} \\
\gives E &= \pm \sqrt{E_C^2 \Delta_g^2 + E_J^2/4} \\
&= \pm \frac{E_J}{2} \sqrt{1+ \frac{4 E_C^2 \Delta_g^2}{E_J^2}}
\end{align}
From these values, we can again calculate the energy splitting
\begin{align}
\Delta E = E_J \sqrt{1+ \frac{4 E_C^2 \Delta_g^2}{E_J^2}}.
\end{align}

The first point to note is that $\Delta E \ge E_J$ for all values of $\Delta_g$, which shows that the energy gap at the crossing point ($\Delta_g=0$) is a minimum value.  Next, we note that there are two obvious energy regimes where we can simplify this equation:
\begin{align}
\frac{4 E_C^2 \Delta_g^2}{E_J^2} \ll 1 \quad\gives \Delta E &\approx E_J \br{1+ \frac{2 E_C^2 \Delta_g^2}{E_J^2}} \\
&= E_J + O(\Delta_g^2) \\
\frac{4 E_C^2 \Delta_g^2}{E_J^2} \gg 1 \quad\gives \Delta E &\approx E_J \frac{2 E_C \Delta_g}{E_J} \\
&= 2E_C \Delta_g
\end{align}
The first of these results tells us that the energy gap is independent of the gate offset charge, $N_g$, up to second (quadratic) order in $\Delta_g$.  As we will see later, this implies that the CPB qubit is somewhat immune to noise in $N_g$ at this extremal point ($\Delta_g=0$).  This offset reference point is usually called the ``sweet spot'' for exactly this reason.  Note that, although $4 E_C^2 \Delta_g^2/E_J^2 \ll 1$ in this regime, it is still also true that $E_J \ll E_C$.

To understand the significance of the second result, we need to make a comparison with the $E_J=0$ case:
\begin{align}
\Delta E = E_0 - E_1 &= E_C \br{0 - \sqbr{\half + \Delta_g}}^2 - E_C \br{1 - \sqbr{\half + \Delta_g}}^2 \\
&= E_C \br{\half + \Delta_g}^2 - E_C \br{\half - \Delta_g}^2 \\
&= E_C \br{2 \times 2\half\Delta_g} \\
&= 2E_C \Delta_g
\end{align}
In other words, when $4 E_C^2 \Delta_g^2/E_J^2 \gg 1$, the energy level separation is exactly the same as the $E_J=0$ case, so as we guessed earlier, when $E_J$ is small, the coupling is only effective near the crossing points.  Moreover, this tells us that the condition $4 E_C^2 \Delta_g^2/E_J^2 \sim 1$ in some sense defines the cross-over point between the coupled and uncoupled regime for the eigenstates.  Rearranging this condition therefore gives us the position of this crossover point in terms of $\Delta_g$:
\begin{align}
\label{eq:qubits-crossover}
\Delta_g \sim \frac{(E_J/2)}{E_C}
\end{align}
which is just the ratio of the energy scale coefficients in the initial full-system Hamiltonian of Eq.~\ref{eq:qubits-chargeQB-numberbasisH}.  So overall, the energy gap at the crossing point is defined by the Josephson coupling $E_J$ and the width of the coupling regime is defined by the ratio of $E_J$ to $E_C$.

\begin{figure}
\begin{center}
\includegraphics[width=0.5\textwidth]{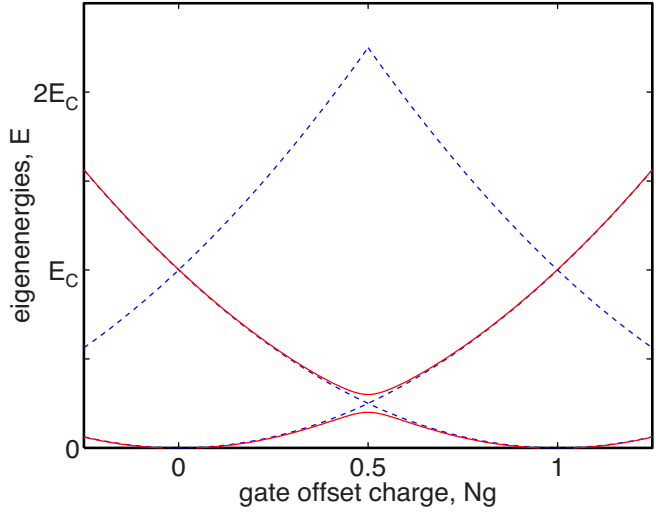}
\end{center}
\captionsetup{labelfont=bf,font={small},margin=6pt}
\caption{\emph{\textbf{Avoided crossing in a CPB charge qubit in the traditional $E_J \ll E_C$ regime.}}  This plot was generated with $E_J = 0.1 E_C$.}
\label{fig:avoidedcrossing}
\end{figure}

The exact eigenenergies from the full equation:
\begin{align}
E &= E_C \br{\smfrac{1}{4} + \Delta_g^2} \pm \sqrt{E_C^2 \Delta_g^2 + E_J^2/4}
\end{align}
are plotted in Fig.~\ref{fig:avoidedcrossing}.   The region around the $E_J=0$ crossing point is called an ``avoided crossing''.

\subsubsection{Higher-order coupling terms}

Written in matrix form, the full CPB Hamiltonian looks like:
\begin{align}
\mathcal{H} =
\sqbr{ \begin{matrix}
\ddots & \quad & \quad & \quad & \quad & \quad & \quad \\
\quad & 2^2 E_C (N_g{+}2)^2 & -E_J/2 & 0 & & \ddots & \\
& -E_J/2 & E_C (N_g{+}1)^2 & -E_J/2 & 0 & & \\
& 0 & -E_J/2 & E_C N_g^2 & -E_J/2 & 0 & \\
& & 0 & -E_J/2 & E_C (N_g{-}1)^2 & -E_J/2 & \\
& \ddots & & 0 & -E_J/2 & E_C (N_g{-}2)^2 & \\
& & & & & & & \ddots
\end{matrix} }
\end{align}
Solving the general dynamics of this full system is difficult in general.  For example, taking a brute force approach to diagonalise the Hamiltonian (and hence determine the eigenvectors and eigenvalues) would require solving a complex, higher-order polynomial.  We can, however, get a flavour of how this works by looking directly at the unitary evolution operator form of the Schr\"odinger equation.  Let us first write:
\begin{align}
\mathcal{H} = \mathcal{H}_C + \mathcal{H_J},
\end{align}
where $\mathcal{H}_C$ and $\mathcal{H}_J$ are the charging and Josephson terms, respectively, of Eq.~\ref{eq:qubits-chargeQB-numberbasisH}.  Using the by-now-familiar Taylor expansion of the unitary evolution operator, this gives:
\begin{align}
U(t) &= \sum \frac{1}{n!} \br{ \frac{-i\mathcal{H}t}{\hbar} }^n \\
&= \sum \frac{(-it/\hbar)^n}{n!} \br{ \mathcal{H}_C + \mathcal{H}_J }^n \\
&= \sum \frac{(-it/\hbar)^n}{n!} \br{ \mathcal{H}_C^n + \mathcal{H}_J^n + \text{ cross terms } }
\end{align}
The evolution operator, which maps input states to output states at an arbitrary time $t$, therefore contains some terms proportional to $\mathcal{H}_C^n$ and some terms proportional to $\mathcal{H}_J^n$, along with all possible cross terms which are just products of $\mathcal{H}_C^n$ with $\mathcal{H}_J^m$.  We can therefore get a good idea of what happens to an evolving system by understanding what $\mathcal{H}_C^n$ and $\mathcal{H}_J^n$ do independently.

The charging term is fairly straightforward to deal with.  By writing down the first couple of examples, it shouldn't be too difficult to convince yourself that%
\footnote{You can prove this more formally by induction.}:
\begin{align}
\mathcal{H}_C^n = E_C^n \sum_N (N-N_g)^{2n} \ket{N}\bra{N}
\end{align}
In other words, it always contains the same operator terms, just with different coefficients.  Another way of saying this is that powers of diagonal matrices will always be another diagonal matrix, with the diagonal terms raised to the appropriate power.

\begin{figure}
\begin{center}\begin{tabular}{ccc}
\includegraphics[height=50mm]{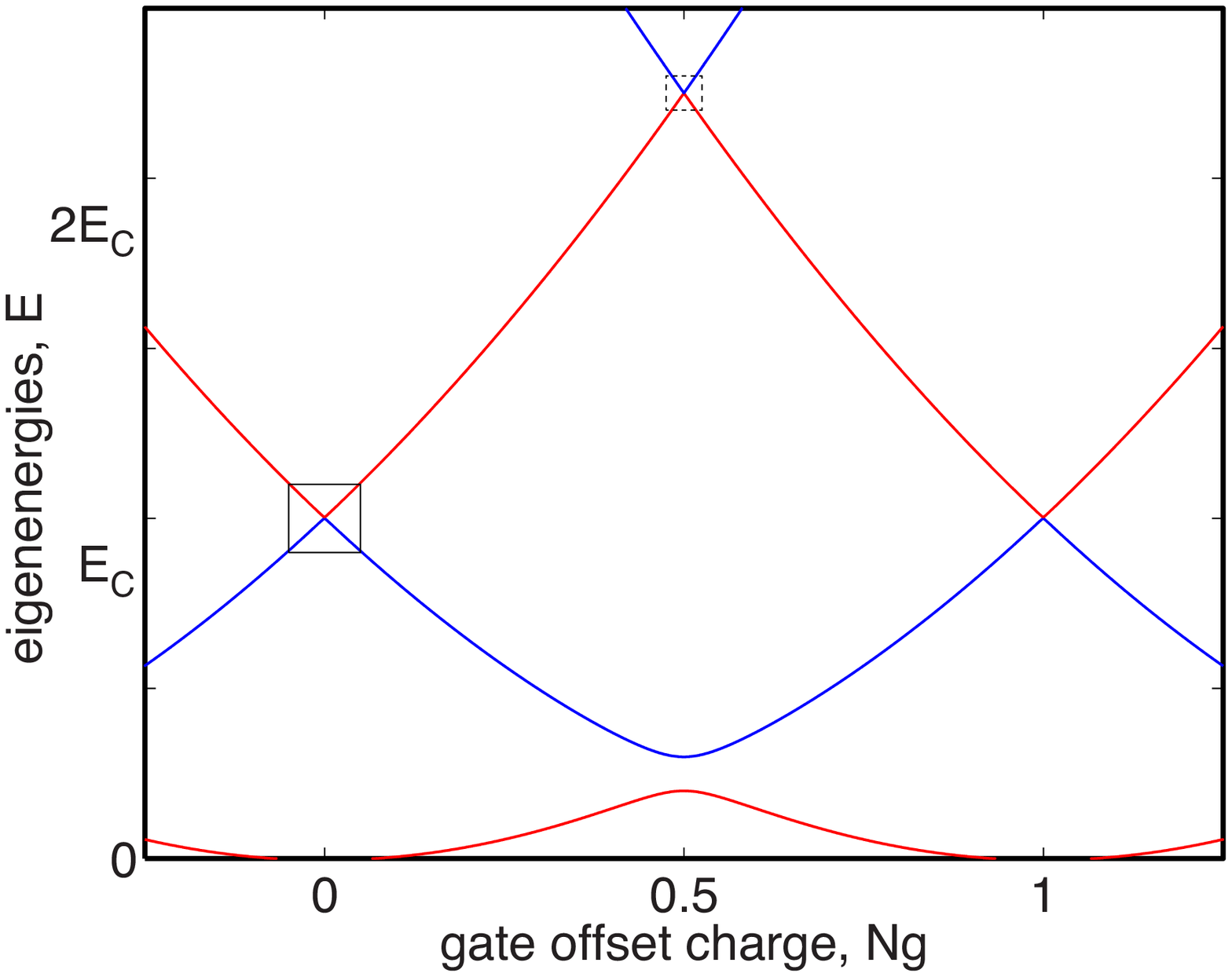} &
\includegraphics[height=50mm]{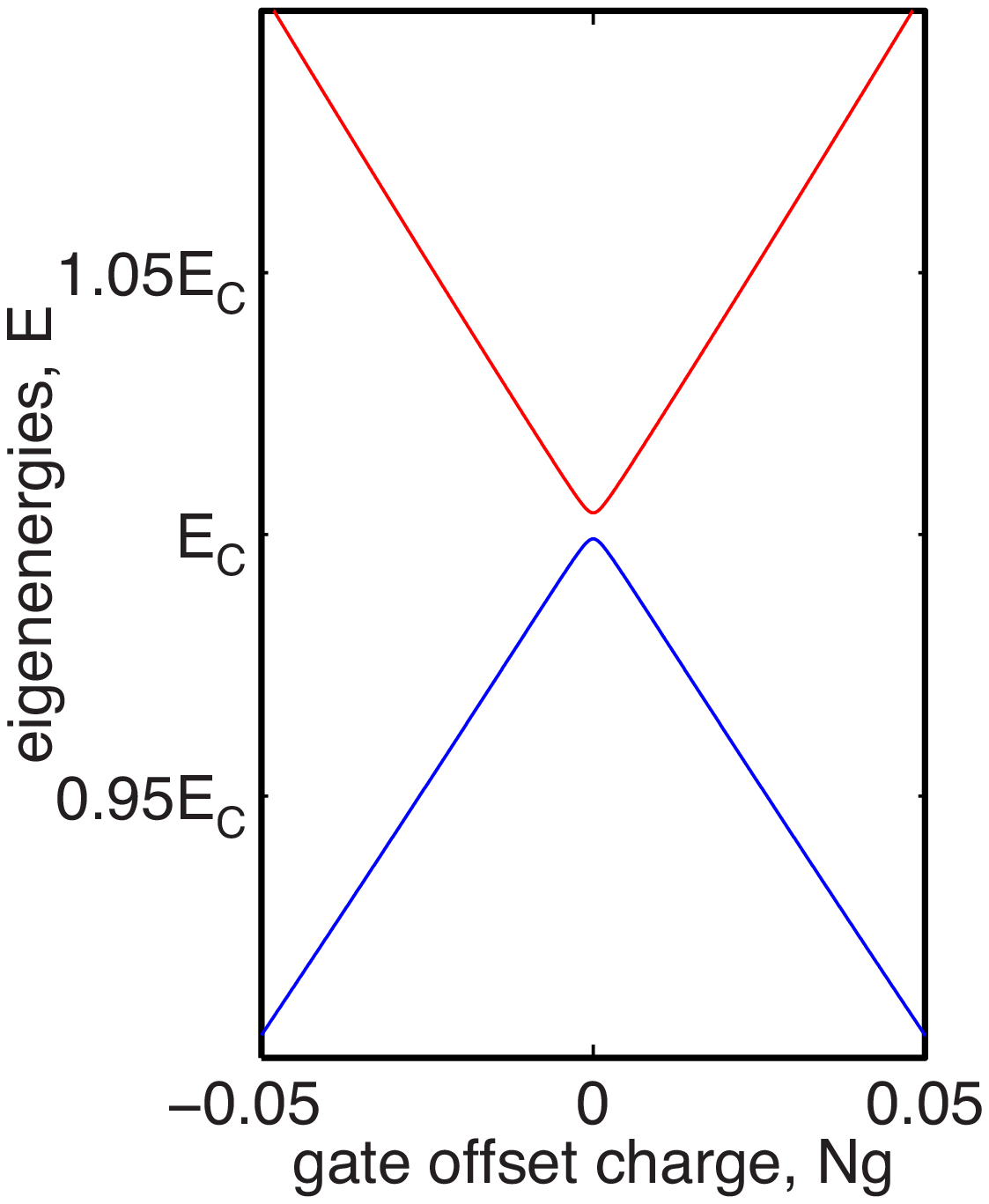} &
\includegraphics[height=50mm]{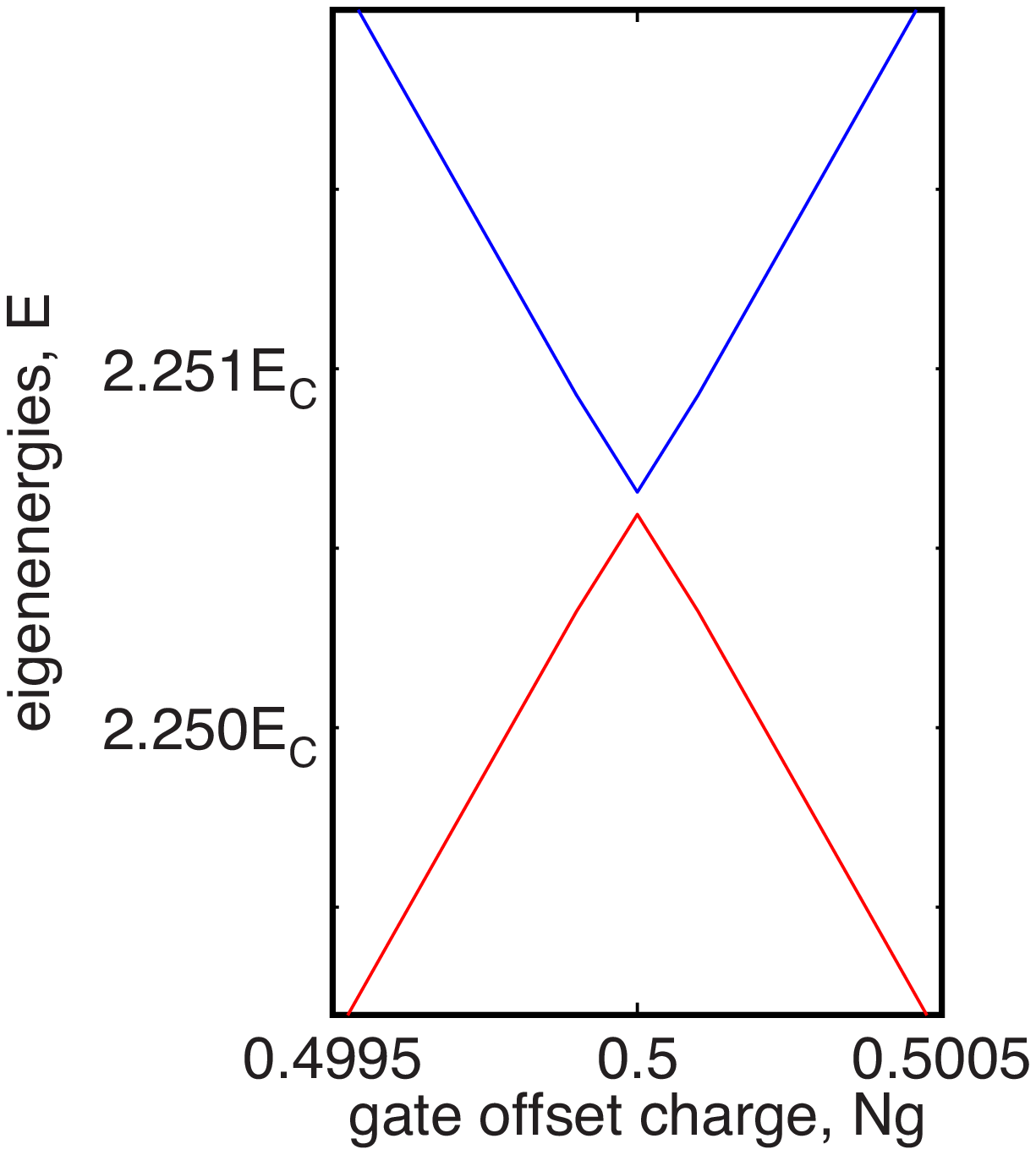} \\
(a) & (b) & (c)
\end{tabular}\end{center}
\captionsetup{labelfont=bf,font={small},margin=6pt}
\caption{\emph{\textbf{Energy diagramme of a CPB charge qubit in the traditional $E_J \ll E_C$ regime ($E_J = 0.1 E_C$).}} (a) Full energy diagramme. (b) Zoomed-in view of the $\ket{\pm 1}$ avoided crossing at $N_g = 0$ (zoomed section marked as a box in (a)). (c) Zoomed-in view of the $\ket{0}$--$\ket{2}$ avoided crossing at $N_g = 0.5$ (dashed box marked in (a)---box not representative of size, because actual size is too small to be seen in the full-sized diagramme).}
\label{fig:avoidedcrossing-higherorder}
\end{figure}

The Josephson term is a little more complicated, because it isn't diagonal and each term in the sum gives a binomial expansion.  Looking at the simplest nontrivial case ($n=2$), however, is very instructive:
\begin{align}
\mathcal{H}_J^2 = \br{ \frac{E_J}{2} }^2 \sum_N 2\ket{N}\bra{N} + \ket{N}\bra{N{+}2} + \ket{N{+}2}\bra{N}
\end{align}
The key thing to notice here is that we now have coupling terms between the $\ket{N}$ and $\ket{N{+}2}$ states in exactly the same form as we had previously in Eq.~\ref{eq:qubits-chargeQB-numberbasisH} between $\ket{N}$ and $\ket{N{+}1}$, which earlier led to the avoided crossing between the $E_J=0$ eigenstates of $\ket{0}$ and $\ket{1}$ (see Fig.~\ref{fig:avoidedcrossing}).  We therefore predict that this higher-order coupling term will lead to additional avoided crossings in the energy eigenstate diagramme, e.g., between the $E_J=0$ eigenstates of $\ket{0}$ and $\ket{2}$ at the crossing point we see near the right-hand edge of Fig.~\ref{fig:avoidedcrossing}.  However, the \emph{strength} of this higher-order coupling is proportional to $(E_J/2)^2$.  Consequently, in the traditional low-coupling regime of the charge qubit ($E_J \ll E_C$), this coupling will be much weaker than the first-order coupling, and the size of the avoided-crossing regime will therefore be proportionately smaller as well.  The energy eigenstates based on a numerical calculation are shown in Fig.~\ref{fig:avoidedcrossing-higherorder}a.  As expected, the overall form of the energy diagramme looks almost identical to the $E_J=0$ parabolas, but now there is an avoided crossing at every crossing point (zoomed-in sections of the two next crossings up are shown in Fig.~\ref{fig:avoidedcrossing-higherorder}b and Fig.~\ref{fig:avoidedcrossing-higherorder}c).

\subsubsection{Simple qubit operations with a CPB qubit}

\emph{\textbf{State preparation}}: The simplest operation you can do with a CPB qubit is state preparation%
\footnote{At least to first order.  Actually, while this simple routine works well, it's hard to get it to work \emph{very} well, and researchers have recently demonstrated much more sophisticated techniques to do this with better precision, e.g., involving advanced quantum control techniques.}.
Basically, because of the second law of thermodynamics, pretty much any quantum system, given enough time, wants to drop down into its lowest energy (ground) state, provided the temperature of the system is low enough%
\footnote{If the temperature is a significant fraction of the energy gap between the ground state and the excited states, then the system will instead decay to a thermal mixture of states with a probability determined by the Boltzmann distribution.  In this course, however, we are always assuming that the system is at approximately zero temperature by comparison with all other relevant energy scales.}.
This generally happens according to a random (spontaneous) relaxation process, although we won't worry here about the exact dynamics of this process.  To prepare a well-defined state, we therefore set the offset gate charge to $N_g=0$, where, as we discussed earlier, there is one very low energy level, with the next highest levels at substantially higher energies.  We then just need to wait long enough before each time we run the experiment, so that the system will have relaxed into the ground state $\ket{0}$ with a very high probability.

\emph{\textbf{Fast single-qubit gates}}: The next conceptually simplest operation to perform can be summarised as follows:
\begin{itemize}

\item Start by preparing the state in the ground state $\ket{N=0} (\equiv \ket{\psi(0)})$ by setting $N_g=0$ and waiting, as described above.

\item Now, at $t=t_0$, suddenly change the offset gate voltage to $N_g = 1/2$.  Provided this is done fast enough, it instantaneously changes the Hamiltonian (and therefore the ``direction'' the system is evolving) without change the state.  So while the state is still $\psi(t_0) = \ket{0}$, the Hamiltonian becomes $\mathcal{H}(t_0) = -(E_J/2) \sigma_x$.  Conceptually, this is the same as suddenly changing the force $F$ on a classical particle.  The acceleration on the particle changes immediately (which affects its future position), but its instantaneous, current position has not yet had a chance to be affected by the new conditions.

\item The eigenstates of the new Hamiltonian, however, are $\ket{\pm}$, not $\ket{0}$.  This describes the situation at the beginning of a traditional Rabi experiment (as you will have seen from the exercises).  Consequently, the instantaneous system state will now oscillate sinusoidally between $\ket{0}$ and $\ket{1}$, and the corresponding measurement probabilities will be:
\begin{align}
p_\zero (t) &= \frac{1}{2} \br{1 + \cos \br{\frac{E_J (t-t_0)}{\hbar}} } \\
p_\one (t) &= \frac{1}{2} \br{1 - \cos \br{\frac{E_J (t-t_0)}{\hbar}} }
\end{align}
Therefore, a range of different single-qubit gates (rotations around the $\sigma_x$ axis of the Bloch sphere by varying angles) can be implemented by specifying the appropriate delay time for this oscillation stage.  For example, setting $t = t_0 + \hbar \pi / 2E_J$ results in preparing the excited state $\ket{1}$.  Based on these equations, it can be seen that the speed of such a single-qubit gate is determined (limited) by the oscillation frequency, $E_j/h$.

\item After the chosen delay time, the system can either be measured, or the gate offset can be again rapidly to the value required for the next experiment.

\end{itemize}

\emph{\textbf{Adiabatic (slow) single-qubit gates}}: The other key technique for directly implementing single-qubit operations on a CPB qubit operates on an entirely different principle.  It relies on the so-called \emph{adiabatic theorem}:
\begin{description}[nosep,font=\it]
\item[Adiabatic Theorem:] If a system Hamiltonian is changed slowly enough, the system will remain in the Hamiltonian's instantaneous eigenstate \emph{at all times}.
\end{description}
The overall process can be summarised as follows:
\begin{itemize}

\item Start by preparing the state in the ground state $\ket{N=0} (\equiv \ket{\psi(0)})$ by setting $N_g=0$ and waiting, as described above.

\item Now, very slowly change the offset gate voltage from $N_g=0$ to $N_g=1/2$.  Provided the gate voltage is changed slowly enough, as a result of the adiabatic theorem, the final state at $N_g = 1/2$ will be $\ket{\psi}=\ket{+}$.

\end{itemize}
This technique therefore allows the reliable creation of superposition states.  The only cost is that the gate voltage needs to be changed slowly enough to satisfy the adiabatic theorem, which places a limit on the speed of gate operations.

\subsubsection{Effects of noise in CPB qubits}

As we have already seen in earlier sections and exercises, the free evolution dynamics of any qubit system depend critically on $\Delta E$, the gap between the qubit eigenstates.  Specifically, the phase oscillation frequency of superpositions of those eigenstates is determined directly by $\Delta E$.

In the CPB qubit, we have already seen that the energy gap changes as the offset gate voltage is varied, which in principle allows the experimenter to control the speed of the system dynamics.  \emph{However, it is not only the experimenter that can change the offset gate voltage}---there are also a number of other environmental effects that can affect the gate voltage and this can introduce noise into the system.

How significant is such an effect?  Well, as we discussed earlier, the critical conceptual feature of a CPB system is that the superconducting island should be small enough that adding one single Cooper pair onto the island makes a significant difference in the charging energy of the system.  In terms of the energy diagramme in Fig.~\ref{fig:avoidedcrossing-higherorder}a, the CPB moves through a complete period of the energy level structure as a result of just two extra electrons!  This means that the CPB is extremely sensitive to very small changes in the local charge environment, e.g., a single electron hopping randomly for some reason between different locations in the substrate underneath the qubit.  This is commonly known as ``charge noise''.

There are a range of mechanisms one can think of which could affect the offset gate voltage, such as:
\begin{itemize}[nosep]
\item Systematic instrument errors: that is, you may try to apply a certain gate offset and, perhaps simply because of internal calibration errors or drift, $N_g$ may not be what you think it is.
\item Stochastic instrument noise: you may be applying the right mean voltage, but any voltage source will have some associated noise which will fluctuate around randomly, e.g., Johnson noise from an internal resistance.
\item Noise from other local charges near the qubit: impurities or defects in the local environment may lead to changes in the local charge environment---such imperfections may easily arise during complex fabrication processes.
\end{itemize}
At the end of the day, identifying exactly what is the source of limiting noise sources in circuit QED experiments is still an open problem and one of the key challenges facing the field today.  This is perhaps particularly difficult, becase the CPB is so sensitive to even single-electron effects.

An important strategy that researchers have used over the last decade is therefore not just to try and identify and eliminate sources of noise from the experiments, but rather to design experiments so that the circuit elements are more immune and insensitive to known categories of noise, irrespective of what the underlying mechanism is.

As we have discussed, changes in the qubit energy gap directly affect the qubit's phase evolution when it is evolving freely.  If those changes arise as a result of a random noise process, then this can gradually erode the qubit's phase coherence.  Consequently, the more that $\Delta E$ is affected by fluctuations in the offset gate voltage, $N_g$, the more phase noise the qubit will experience and the more quickly that noise will degrade the qubit's coherence.  The strategy of operating the charge qubit at so-called ``sweet spots'' aims to take advantage of exactly this concept, by operating the CPB at points where the energy gap is at an extremal point, as a function of gate charge.

We already showed in the previous sections that near $N_g=1/2$, the qubit energy gap is at a minimum ($\Delta E = E_J$), meaning that the derivative of $\Delta E$ with respect to $N_g$ is zero.  This means that the energy gap is \emph{independent} of fluctuations in $N_g$ at linear order, and is only sensitive to fluctuations in $N_g$ at quadratic order.  So for small fluctuations $\Delta_g$ in $N_g$, the fluctuations in $\Delta E$, which are proportional to $\Delta_g^2$ are much smaller (they are quadratically suppressed).

As we've seen from Fig.~\ref{fig:avoidedcrossing-higherorder}a, the other point where $\Delta E$ has an extremum is near $N_g = 0$ (or $1$), where the energy gap is a maximum and approximately $E_C$ (for small coupling).  Following a similar, but simpler process to how we solved the full qubit Hamiltonian, we can set $N_g = 0 + \delta_g$.  Even for $E_J=0$, this already gives $E_0 = E_C \delta_g^2$.  And as we've seen, it turns out that the upper levels also reach extrema at this value of $N_g$, but the qubit excited state $\ket{1}$ is no longer an eigenstate, because of the higher-order Josephson coupling, so this adds extra complexity beyond the target simple qubit system.  This is another reason why this operating point is usually only used to provide good conditions for state preparation, so superpositions and therefore phase coherence are not so important here.

\subsubsection{Operating a CPB charge qubit in the transmon regime}

Pretty much everything we have investigated up until now has been in the traditional small-coupling regime, where $E_J \ll E_C$.  One of the most important advances of the last 5 years, however, was the development of the so-called \emph{transmon qubit}, which is a CPB charge qubit, but operated in the strong-coupling regime where $E_J \gg E_C$.

We have already seen (e.g., from Eq.~\ref{eq:qubits-crossover}) that the flattened avoided crossing region around the charge qubit ``sweet spot'' gets wider and flatter as $E_J$ increases, and we know that a flatter energy spectrum means that the energy gap, $\Delta E$, varies less dramatically with unwanted fluctuations in the gate charge $N_g$.   So what happens if we continue this trend?

Effectively, stronger Josephson coupling implies that more and more Cooper-pair number states will contribute to the eigenstates of the system at any given setting, and if the eigenstates involve lots of different number states anyway, this will make them less sensitive to changes in charge number, and hence also fluctuations in the local charge environment.  Unfortunately, however, this also makes the system harder to solve, because the qubit approximation becomes invalid, even very near $N_g=1/2$ sweet spot.  Fortunately, it turns out that more sophisticated techniques are still able to produce an analytical solution in terms of complex, but known polynomials, mainly because of the periodic symmetry of the system.  However, we won't go into these gory details in this course, because we can still paint a rough intuitive picture of what happens based on the simple scenarios we have already considered.

\begin{figure}
\begin{center}
\includegraphics[width=0.6\textwidth]{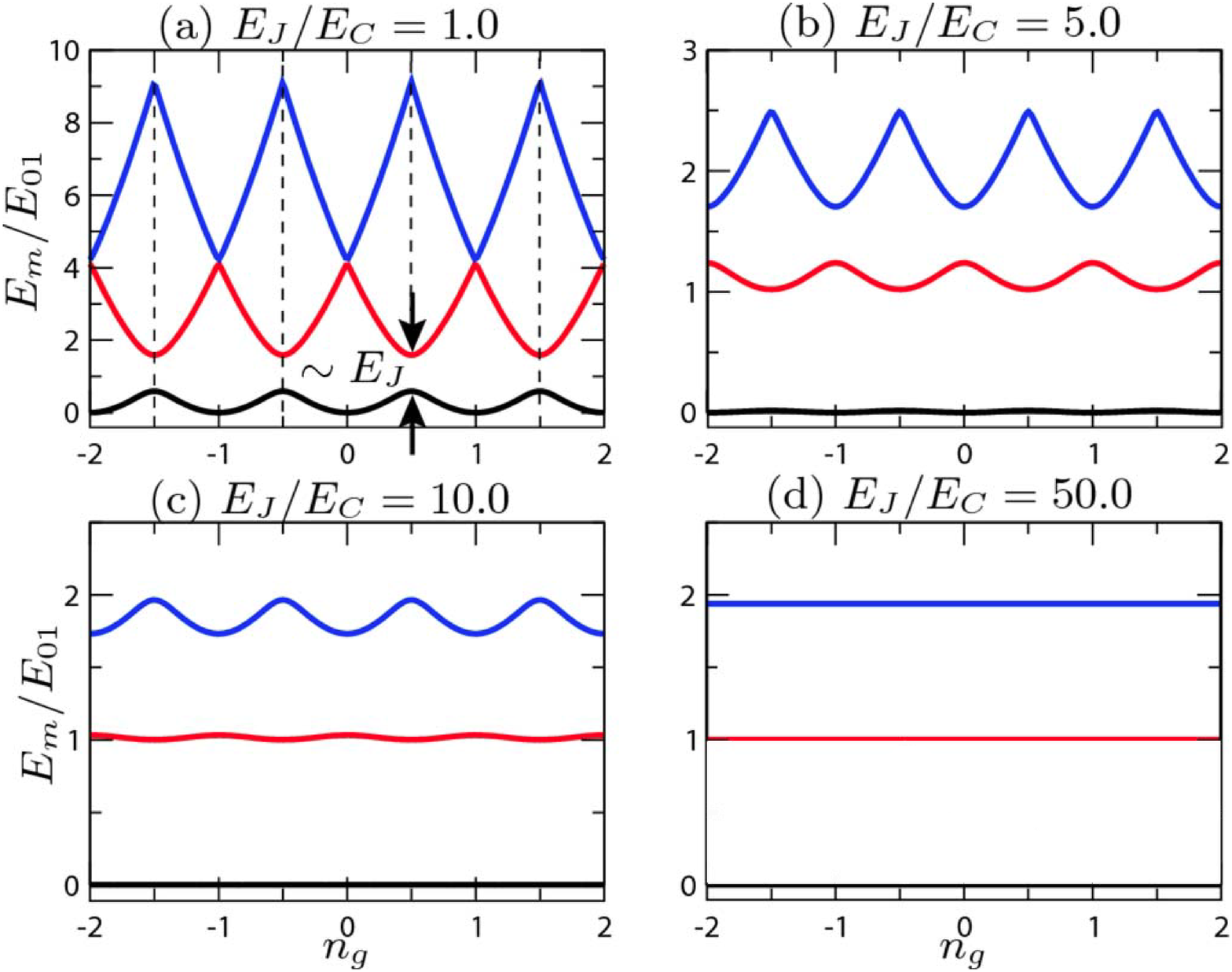}
\end{center}
\captionsetup{labelfont=bf,font={small},margin=6pt}
\caption{\emph{\textbf{Operating a CPB charge qubit in the transmon regime ($E_J \gg E_C$).}}  (Image taken, by permission, from Koch \emph{et al.}~\cite{KochJ2007cqd}.)}
\label{fig:transmonregime}
\end{figure}

Essentially, as the coupling gets bigger, the minimum energy gap increases and the avoided crossing region spreads out.  This has the effect of pushing the strongly periodic energy levels apart at the points where they approach each other.  Since this occurs from both above and below simultaneously, this causes them to gradually become flatter and eventually, the lower energy levels become effectively flat.  As a result, the energy gap $\Delta E$ essentially becomes independent of $N_g$ and the qubit is operating in a regime where there is a sweet spot everywhere!  This trend is illustrated in Fig.~\ref{fig:transmonregime}.

To create a transmon qubit, rather than trying to increase the Josephson coupling, the standard approach is to decrease the charging energy by increasing the junction capacitance.  However, playing with the junction itself would again affect the coupling, so instead a transmon is typically fabricated with an extra large capacitor in parallel with the tunnel junction.  As we already saw with the gate capacitance, this just creates an effective overall capacitance for the island which is the sum of all such capacitances (i.e, the junction capacitance, the gate capacitance and the extra transmon capacitance, the so-called ``shunt capacitance'').  Varying this extra capacitance provides full design control over the charging energy parameter, independently of the junction tunnel coupling.

The one drawback with this approach is that we actually initially added the gate capacitor in order to provide ourselves with a knob with which we could address and control the qubit system.  If the system is now effectively independent of the gate voltage, we no longer have access to this degree of freedom.  As a result, for example, we are no longer able to implement the single-qubit gates described above.

This is where the cavity part of circuit QED comes in.  Although the transmon is no longer sensitive to the gate voltage, it turns out that it still possesses a very strong electric dipole and this can interact very strongly with electromagnetic fields of the appropriate frequency when the qubit is embedded in a coplanar waveguide resonator.  Unfortunately, we don't have time to go into the mechanism for this coupling in detail in this course, but one of the exercises describes how a qubit system can be coupled to a field mode and how energy can be resonantly transferred between the two.  Furthermore, this cavity can even be used as an interconnect between two such qubits (called a ``cavity bus'') to enable multiqubit gates and interactions.

\subsection{Phase qubits: a.k.a. current-biased Josephson-junction qubits}

The alternative name for a phase qubit---current-biased Josephson-junction qubit---pretty much gives the game away about what they are.  Essentially, a phase qubit is a single Josephson junction which is ``shunted'' by a constant-current source, i.e., a current source is used to connect the two electrodes of a Josephson junction and drives a current through the junction.  This external current provides a built-in means for controlling the behaviour of the qubit.

As we have already seen, the overall Hamiltonian for this system is given by Eq.~\ref{eq:sc-JJ-currentbiased}, or it can also be rewritten in terms of the superconducting phase difference across the junction:
\begin{align}
\mathcal{H} = E_C \hat{N}^2 - I\varphi_0 \hat{\varphi} - I_0 \varphi_0 \cos \hat{\varphi}
\end{align}
which has a potential energy given by:
\begin{align}
\mathcal{U}(\hat{\varphi}) = - I\varphi_0 \hat{\varphi} - I_0 \varphi_0 \cos \hat{\varphi}
\end{align}

Because the circuit doesn't contain an isolated island, the number of Cooper pairs on one side of the junction (what was the island in the CPB qubit) is no longer completely determined by the process of tunnelling on and off the island.  As a result, the charge on the junction is no longer limited to an integer number of Cooper pairs, but is now a continuous number operator.  It also means the phase qubit isn't sensitive to charge noise---a simple intuitive argument is that charge fluctuations can be compensated through the shunting circuit without disturbing the junction phase.  Phase qubits are normally operated in the ultrastrong-tunnelling regime, with $E_J \sim 10^4$--$10^6 E_C$ (see, e.g.,~\cite{YouJQ2005scq}), and because the tunnelling provides such a strong link between the superconducting condensates on either side of the junction, it is no longer the number states, but the phase states which define the behaviour of the system, which is where the name comes from.

At a simple level, the phase qubit's sinusoidal potential is clearly not quadratic, which makes the system clearly nonlinear, as is required to turn the system into a qubit.  It is worth probing this idea a bit more closely, however, elaborating slighly on the results derived in the exercises.  In Eq.~\ref{eq:sc-JJ-currentrelation}, we showed that the position of the lowest-order potential minimum of the phase qubit is given by the Josephson relation.  We can therefore look at two interesting operating regimes, where the bias current is either very small ($I \ll I_0$) or only just smaller than the critical current ($I \lesssim I_0$).

In the small bias current regime, the Josephson relation tells us that the potential minimum occurs at $\varphi \sim 0$.  Now, provided the depth of the potential well is very large (i.e., large Josephson energy / criticial current), then the phase is likely to be confined very close to the bottom of this potential well.  Making a Taylor expansion around $\varphi = 0$, we can then show that the potential energy is approximately:
\begin{align}
\mathcal{U}(\varphi) \approx I_0 \varphi_0 \br{ -1 + \varphi^2/2 }
\end{align}
In other words, although the cosine potential is clearly nonlinear, it is approximately quadratic, provided the bias current is small and the potential well is deep enough.  And as we know, a system which has a quadratic potential behaves like a harmonic oscillator, with equally spaced energy levels, and therefore cannot be operated as a qubit.

We therefore need to consider an alternative operating regime, namely the large bias current regime, where the bias current is just below the critical current.  The Josephson relation tells us that, in this regime, the position of the potential well will approach the critical value of $\varphi=\pi/2$.  We therefore make a new Taylor expansion of the potential around this point, giving the following potential energy:
\begin{align}
\mathcal{U}(\varphi) \approx -\frac{\pi I_0 \varphi_0}{2} - (I_0-I) \varphi_0 \varphi - \frac{I_0 \varphi_0}{6} \br{\varphi - \frac{\pi}{2}}^3
\end{align}
which now clearly corresponds to a nonlinear system.  Interestingly, not only is the potential now a cubic function of $\varphi$, but the quadratic component is actually exactly zero in this regime, making the system purely nonlinear.  This now provides the right conditions to enable operation of a qubit, with distinguishable energy-level transitions.

\subsection{Flux qubits: a.k.a. persistent-current qubits}

Flux qubits provide the conceptual complement to the charge qubit.  Although they come in several different flavours, the simplest version of the charge qubit is simply a single Josephson junction embedded in a single superconducting loop or ring, much like a SQUID.  The superconducting loop also provides an inductance to the system.  The equivalent electronic circuit for this type of flux qubit is a single Josephson junction (Josephson energy, $E_J$, and capacitance, $C$) in parallel with a linear inductor (inductance $L$).

Because there is again no isolated island in the flux qubit system, it is again effectively insensitive to charge fluctuations.  As the name suggests, however, the flux qubit can be controlled by an applied magnetic flux, but this does now make it sensitive to flux noise from the local environment.

The full Hamiltonian for the flux-qubit system is given by:
\begin{align}
\mathcal{H} = \frac{\hat{Q}^2}{2C} + \frac{\hat{\Phi}^2}{2L} - E_J \cos \sqbr{ \frac{2\pi}{\Phi_0} \br{\hat{\Phi} - \Phi_{\rm ext}} }
\end{align}
with the corresponding potential energy:
\begin{align}
\mathcal{U}(\hat{\Phi}) = \frac{\hat{\Phi}^2}{2L} - E_J \cos \sqbr{ \frac{2\pi}{\Phi_0} \br{\hat{\Phi} - \Phi_{\rm ext}} }
\end{align}
Like the gate voltage in the charge qubit, we see that an externally applied flux appears in the Hamiltonian, which provides a controllable offset this time in the cosine potential term.  As discussed above, this provides both a control mechanism and a source of noise.  From these equations, we see the potential has a quadratic component (in $\Phi$) which is overlaid with a cosine oscillation term, the phase (orientation) of which is controlled by the external flux.  Provided the relative energy scales are chosen appropriately, this therefore also provides a strongly nonlinear system, which can be operated as a qubit.  The flux qubit is usually operated in the strong-tunnelling regime ($E_J \sim 10 E_C$~\cite{YouJQ2005scq}).

The operation of the flux qubit can essentially be understood in terms of the same fluxoid quantisation phenomenon in a closed superconducting loop that we derived in Eq.~\ref{eq:sc-fluxoid-quantisation}.  This said that, for any closed superconducting loop, the total \emph{fluxoid}, $\Phi\dash$, enclosed by the loop must be quantised in integer multiples of the superconducting flux quantum, $\Phi_0$:
\begin{align}
\Phi\dash = n\Phi_0
\end{align}
In the simple case of an uninterrupted superconducting ring we considered in Eq.~\ref{eq:sc-fluxoid-quantisation}, the enclosed fluxoid has contributions from the circulating current and from the externally applied field.  In the case of the single-junction flux qubit (as in the case of the tunable Josephson junction calculation from the exercises), there can also be a contribution from the junction itself.  Let us assume, however, that the critical current of the junction is much larger than the typical circulating currents that are generated during a normal experiment, and that the phase (flux) across the junction will therefore be negligibly close to zero.  We are therefore left with:
\begin{align}
n\Phi_0 = \Phi\dash \approx \text{flux from circulating current} + \text{externally applied flux}
\end{align}

There are many parallels between the operation of the flux qubit and the CPB charge qubit.  For example, there are two key operating points.  Let us first assume that the externally applied flux is itself already an integer multiple of $\Phi_0$, i.e., $\Phi_{\rm ext} = n \Phi_0$.  In this scenario, there will be one potential well at $\Phi = 0$, which will be substantially lower than the neighbouring wells.  (Effectively, this corresponds to the scenario where there is zero circulating current flowing around the flux-qubit loop.)  Consequently, there will be one low-energy level, while the next two energy levels (corresponding to occupation of the neighbouring potential wells) will be approximately degenerate with each other and substantially higher than the lowest ground state.  This scenario represents the equivalent of the $N_g=0$ case for the charge qubit, and like the charge qubit, because there is one low-energy level energetically separated from the next excited states, this enables \emph{good state preparation}.

Next, we will assume that the applied flux lies half-way between integer multiples of $\Phi_0$, i.e., $\Phi_{\rm ext} = (n+1/2) \Phi_0$.  In this case, the applied flux causes a current to circulate in order that the total flux can still be quantised, but this can happen in one of two ways.  Either the current can circulate in one direction to produce an extra flux component of $+\Phi_0/2$, or in the other direction (with equal magnitude) to produce a flux of $-\Phi_0/2$.  These two states correspond to small, quantised persistent currents flowing in opposite directions around the loop, which is why the flux qubit is sometimes called a persistent-current qubit.  However, if the applied flux is exactly half-way between these two cases, neither one of these solutions will be energetically favoured over the other.  This corresponds to two low, degenerate potential wells, with the next two potential wells substantially higher in energy.

This scenario is the equivalent of the $N_g=1/2$ case for the charge qubit, and provides the conditions required for a \emph{good qubit system}.  Moreover, as with the charge qubit, because the flux qubit is a quantum system, there will be some coupling between these states due to tunnelling between the potential wells.  Consequently, if you plotted the energy eigenstates as a function of $\Phi_{\rm ext}$, this coupling would again lead to an avoided crossing between the energy levels, similar to Fig.~\ref{fig:avoidedcrossing}.  Because the flux-qubit energy splitting goes through a minimum value at this point, this again provides an operational ``sweet spot'' for qubit operation, because near this value of $\Phi_{\rm ext}$, the flux qubit will be more insensitive to environmental flux noise.  Finally, eigenstates of the flux qubit at this sweet spot won't be the simple $\ket{\pm\Phi_0/2}$ basis states described above, but rather the symmetric and antisymmetric superpositions of these states, $(\ket{+\Phi_0/2} \pm \ket{-\Phi_0/2})/\sqrt{2}$.

\subsection{Exercises---qubits in circuit QED}

\subsubsection{Phase qubits or ``current-biased Josephson junction qubits''}

The full Hamiltonian for a phase qubit, which is simply a current-biased Josephson junction, when written in terms of number and phase operators, is given by:
\begin{align}
\hat{\mathcal{H}} = \frac{(2e)^2}{2C_J} \hat{N}^2 - \frac{I\Phi_0}{2\pi} \hat{\varphi} - \frac{I_0 \Phi_0}{2\pi} \cos \hat{\varphi}
\end{align}
where $I_0 = E_J / \phi_0$.  The potential energy is therefore:
\begin{align}
\mathcal{U}(\varphi) = - \frac{I\Phi_0}{2\pi} \varphi - \frac{I_0 \Phi_0}{2\pi} \cos \varphi
\end{align}
In Question 1, we showed that the position of the first potential minimum is given by the DC Josephson relation:
\begin{align}
\varphi = \arcsin \br{ \frac{I}{I_0} }
\end{align}
Remember that such a minimum only exists if $|I| \le I_0$.

\begin{enumerate}[label=(\alph*)]

\item Sketch a circuit diagramme for the phase qubit, labelling and defining all relevant parameters from the Hamiltonian in the context of the physical electronic circuit.  [Hint: Remember that a real-world Josephson junction can be described by a perfect tunnelling element in parallel with a capacitor.]

\item \emph{Small bias current regime} ($I \ll I_0$):  From the Josephson relation, in this regime the potential minimum is close to $\varphi=0$.  When $I \ll I_0$, show that the potential energy for the phase qubit can be written approximately as:
\begin{align}
\nn
\mathcal{U}(\varphi) \approx \frac{I_0 \Phi_0}{2\pi} \br{ -1 + \varphi^2/2 }
\end{align}
[Hint:  To do this, expand the cosine term in the potential energy out to third order around $\varphi=0$ using the general Taylor expansion formula (already written here to third order)
\begin{align}
\nn
f(x \text{ near } x_0) \approx f(x_0) + f\dash(x_0) (x-x_0) + f\ddash(x_0) (x-x_0)^2 / 2! + f^{\prime\prime\prime}(x_0) (x-x_0)^3 / 3!
\end{align}
where $f\dash(x_0)$ indicates, as usual, the first derivative of $f$ evaluated at $x=x_0$.]

\item \emph{Critical bias current regime} ($0 \ll I < I_0$):  As the bias current approaches the critical current, the Josephson relation shows that the potential minimum approaches $\varphi = \pi/2$.  In this regime, show that the potential energy for the phase qubit can be written approximately as:
\begin{align}
\nn
\mathcal{U}(\varphi) \approx -\frac{I_0 \Phi_0}{4} - \frac{(I_0-I) \Phi_0}{2\pi} \varphi - \frac{I_0 \Phi_0}{12 \pi} \br{\varphi - \frac{\pi}{2}}^3
\end{align}
[Hint:  Again use the Taylor expansion above to third order around the point $\varphi = \pi/2$.]

\item \emph{Qubit operating regime}:  Use the results from the previous two parts of the question to identify which of these regimes is good for qubit operation and which isn't.  Explain why.  [Hint:  Recall that constant offset terms in the potential energy don't affect the system dynamics.]

\end{enumerate}

\subsubsection{Charge qubits or ``Cooper-pair box qubits''}

Written in the Fock-state basis, the full Hamiltonian of a Cooper-pair box is given by:
\begin{align}
\hat{\mathcal{H}} = E_C \sum_{n{=}-\infty}^{+\infty} (n-n_g)^2 \ket{n}\bra{n} - \frac{E_J}{2} \sum_{n{=}-\infty}^{+\infty} \left[ \ket{n}\bra{n+1} + \ket{n+1}\bra{n} \right] \equiv \hat{\mathcal{H}}_C + \hat{\mathcal{H}}_J,
\end{align}
where $E_C = (2e)^2/2C_\Sigma$ and $C_\Sigma = C_J + C_g$.

\begin{enumerate}[label=(\alph*)]

\item Sketch a circuit diagramme for the charge qubit, labelling and defining all relevant parameters from the Hamiltonian in the context of the physical electronic circuit.  [Hint: As in the previous question, remember that a real-world Josephson junction can be described by a perfect tunnelling element in parallel with a capacitor.]

\item Sketch the energy eigenstates for the zero tunnelling case of $E_J=0$.  Calculate the eigenenergy values at the crossing points.

\item Identify which gate charge leads to a good qubit and which is good for state preparation and explain why. (Assume the traditional small-coupling regime, $E_J \ll E_C$.)

\item Expanding the unitary evolution operator form of the Schr\"odinger equation, $U(t) = \exp (-i\hat{\mathcal{H}}t/\hbar)$, to second order in Taylor form gives:
\begin{align}
U(t) \approx 1 - \br{\frac{it}{\hbar}} \hat{\mathcal{H}} + \frac{1}{2} \br{\frac{it}{\hbar}}^2 \hat{\mathcal{H}}^2.
\end{align}
Using the charging and coupling terms in the Hamiltonian, $\hat{\mathcal{H}}_C$ and $\hat{\mathcal{H}}_J$, defined above, show that:
\begin{align}
\nn
\hat{\mathcal{H}}_C^2 &= E_C^2 \sum_{n{=}-\infty}^{+\infty} (n-n_g)^4 \ket{n}\bra{n} \\
\nn
\hat{\mathcal{H}}_J^2 &= \br{\frac{E_J}{2}}^2 \sum_{n{=}-\infty}^{+\infty} \left[ 2\ket{n}\bra{n} + \ket{n}\bra{n+2} + \ket{n+2}\bra{n} \right]
\end{align}
Use these results to explain what happens to the energy eigenstates near higher-order crossing points.

\end{enumerate}

\subsubsection{Qubit-cavity coupling}

Consider a system which consists of a single superconducting qubit coupled to a single mode of a microwave cavity.  If the qubit is resonant with the cavity mode (i.e., the qubit energy gap is the same as the even spacing between energy levels in the microwave cavity), then the interaction between the two systems can generally be described by the simple coupling Hamiltonian (called the Jaynes-Cummings Hamiltonian):
\begin{align}
\mathcal{H} = \hbar g \br{ a \sigma_+ + a^\dag \sigma_- }
\end{align}
where $g$ is the coupling strength and, as usual, $a$ and $a^\dag$ are the annihilation and creation operators describing the harmonic oscillator mode of the microwave cavity, and $\sigma_\pm$ are the raising/lowering (excitation/de-excitation) operators for the qubit and are defined via the following identities:
\begin{align}
\sigma_+ \ket{\zero} &= \ket{\one} \\
\sigma_- \ket{\one} &= \ket{\zero} \\
\sigma_+ \ket{\one} &= \sigma_- \ket{\zero} = 0
\end{align}
As usual, the state of the cavity mode can be written in terms of the number-state basis, $\{ \ket{n} \}$, and the qubit state in terms of the computational basis, $\{ \ket{\zero}, \ket{\one} \}$. The overall state of the system can therefore be described as a tensor product of these two subsystems, e.g., $\ket{n,\zero} \equiv \ket{n} \otimes \ket{\zero}$.

\begin{enumerate}[label=(\alph*)]

\item By direct calculation, show that the following relations hold:
\begin{align}
\nn
\mathcal{H} \ket{n,\zero} &= \hbar g \sqrt{n} \ket{n{-}1, \one} \\
\nn
\mathcal{H} \ket{n,\one} &= \hbar g \sqrt{n+1} \ket{n{+}1, \zero}
\end{align}
[Hint: Remember that in the combined system, the cavity operators only act on the cavity states and the qubit operators only act on the qubit states, e.g., $a \sigma_+ \ket{n,\zero} \equiv \br{a \ket{n}} \otimes \br{\sigma_+ \ket{\zero}}$.]

\item Consider now the case where the cavity starts in the vacuum state ($\ket{0}$) and the qubit is initially in the excited state ($\ket{\one}$).  Use the previous results to show that:
\begin{align}
\nn
\mathcal{H} \ket{0,\one} &= \hbar g \ket{1, \zero} \\
\nn
\mathcal{H}^2 \ket{0,\one} &= (\hbar g)^2 \ket{0, \one}
\end{align}

\item Because the interaction Hamiltonian is time-independent, the Schr\"odinger equation can be solved in the usual way to give the unitary evolution operator:
\begin{align}
U(t) = \exp \br{-\frac{it}{\hbar} \mathcal{H}} = \sum_{j=0}^\infty \br{-\frac{it}{\hbar}}^j \frac{\mathcal{H}^j}{j!}.
\end{align}
Given the initial state $\ket{\psi(0)} = \ket{0,\one}$, use the Taylor expansion of the unitary evolution operator to show that the state of the overall system after time $t$ will be:
\begin{align}
\nn
\ket{\psi(t)} \equiv U(t) \ket{\psi(0)} = \cos (gt) \ket{0,\one} - i \sin (gt) \ket{1,\zero}
\end{align}
\emph{Note: This result shows that the qubit-cavity system behaves like a pseudo-single-qubit system when there is only one overall excitation in the system.}

\item The above equation shows that the excitation, initially in the qubit subsystem, periodically oscillates back and forth from the qubit to the cavity mode.  Calculate how much time is required for the excitation to transfer \emph{completely} from the qubit to the cavity mode.

\item Calculate the output state at time $t = \pi / 4g$.  Explain why this is a special type of state.

\end{enumerate}

\emph{Note: This simple Hamiltonian also describes how a qubit can be coupled to its environment.  As a result of this coupling, the qubit becomes entangled with its environment, leading to the qubit becoming mixed.}

\section{Noise and decoherence}

In this section, I want to discuss a bit about what makes a qubit good.  How do we know if a qubit is useful?  And how do we choose between different types of qubits?

To do this, we go back to some basic ideas of quantum information and the Nobel-prize motivation of achieving control of individual quantum systems.  Essentially, we can think of a quantum system as being ``good'' if it allows long-lived storage of quantum information, and in particular quantum coherence in the form of superpositions and entanglement.  The real world, however, is noisy and generally tries to destroy the coherence of quantum systems, which is called \emph{decoherence}.

Essentially, we want to know the answers to questions like how well can we control our qubits?  How well are they protected from noise?  How well are the isolated from the environment?

In particular, we're going to talk about measurements, mixture, entanglement, noise and the environment.  It turns out that these concepts are all related to each other and feed into each other when we try to understand about decoherence.

\subsection{Measurements---a quick refresher}

Let's assume that we have a machine (experiment) that spits out one qubit in the arbitrary pure state $\ket{\psi} = \alpha \ket{\zero} + \beta \ket{\one}$ every time we press a button.  We then take that qubit and measure it ``in the $\zero/\one$ basis'' and count how often we get either result.  Well, one of the basic postulates (rules) of quantum mechanics is that the probabilities we will measure a $\zero$ or $\one$ ($p_\zero$ and $p_\one$) are given by:
\begin{align}
p_\zero &\equiv |\alpha|^2 = |\braket{\zero}{\psi}|^2 \\
p_\one &\equiv |\beta|^2 = |\braket{\one}{\psi}|^2
\end{align}
In fact, this is true if we measure any state $\ket{\phi}$, namely that the probability of measuring the result $\phi$, given a state $\ket{\psi}$, is:
\begin{align}
\label{eq:pure-state-measurements}
p_\phi = |\braket{\phi}{\psi}|^2.
\end{align}
We can therefore often think of the ``bra'' vector $\bra{\phi}$ as a measurement of the outcome $\phi$.

As we discussed earlier, there are three important measurement bases for a qubit, which I often call the standard qubit bases: $\{ \ket{\zero}, \ket{\one} \}$, $\{ \ket{\pm} \}$ and $\{ \ket{{\pm}i} \}$.

\subsection{Mixture and ensembles}

Consider now the following simple experiment, involving two friends, Alice (A) and Bob (B), described by the following steps:
\begin{enumerate}[nosep,topsep=0mm,partopsep=0mm]
\item Alice sends Bob a qubit in one of the standard qubit basis states.
\item Bob then measures the qubit in one of the standard qubit bases.
\end{enumerate}
They do this for all different combinations of Alice's sent state and Bob's measurement bases and they find that the probability of Bob's measurement outcomes, determined by Eq.~\ref{eq:pure-state-measurements}, are:
{\renewcommand{\arraystretch}{1.3}
\begin{table}[h!]
\begin{center}
\begin{tabu}{c*{3}{|>{\centering}p{1cm}>{\centering}p{1cm}}|}
\multicolumn{1}{c}{} & \multicolumn{6}{c}{Bob} \\
Alice & $\bra{\zero}$ & \multicolumn{1}{c}{$\bra{\one}$} & $\bra{+}$ & \multicolumn{1}{c}{$\bra{-}$} & $\bra{{+}i}$ & $\bra{{-}i}$ \\
\hline
$\ket{\zero}$ & 100\% & 0\% & 50\% & 50\% & 50\% & 50\% \\
$\ket{\one}$ & 0\% & 100\% & 50\% & 50\% & 50\% & 50\% \\
\hhline{~|*{6}{-}}
$\ket{+}$ & 50\% & 50\% & 100\% & 0\% & 50\% & 50\% \\
$\ket{-}$ & 50\% & 50\% & 0\% & 100\% & 50\% & 50\% \\
\hhline{~|*{6}{-}}
$\ket{{+}i}$ & 50\% & 50\% & 50\% & 50\% & 100\% & 0\% \\
$\ket{{-}i}$ & 50\% & 50\% & 50\% & 50\% & 0\% & 100\% \\
\hhline{*{7}{-}}
\end{tabu}\end{center}
\caption{}
\end{table}}

In other words, whichever state Alice sends, there is always one measurement basis Bob can use where his measurement result is perfectly correlated to the state Alice sends.  If he chooses one of the other (``wrong'') bases, he will measure a random uncorrelated mix of different outcomes.

It is not too difficult to see that Alice could actually use these qubits to send Bob information.  If she chose a particular state, he could work out what choice she made, provided he measured in the correct basis (they could even discuss in advance which basis to use or communicate afterwards to work out whether he used the right basis).  They could therefore communicate one ``bit'' of information for each qubit she sent, with the information being contained in the correlations between Alice's choice of states and Bob's measurement outcomes.

Next, consider the following alternative experiment.  Alice now sends Bob a qubit in either $\ket{\zero}$ or $\ket{\one}$, but randomly decides which to send based on a coin toss and doesn't tell Bob the outcome.  We call this an ``ensemble mixture'' of different qubit states.  In other words, half of the time, Alice sends the state $\ket{\zero}$, in which case Bob's measurement probabilities are determined by the first row of the above table, and half of the time, she sends the state $\ket{\one}$, in which case his measurement probabilities are determined by the second row of the table.  However, because he doesn't tell Bob when she is sending which, Bob's overall measurement probabilities will be determined by the average of the probabilities in the two rows, i.e.:
{\renewcommand{\arraystretch}{1.3}
\begin{table}[h!]
\begin{center}
\begin{tabu}{c*{3}{|>{\centering}p{1cm}>{\centering}p{1cm}}|}
\multicolumn{1}{c}{} & \multicolumn{6}{c}{Bob} \\
Alice & $\bra{\zero}$ & \multicolumn{1}{c}{$\bra{\one}$} & $\bra{+}$ & \multicolumn{1}{c}{$\bra{-}$} & $\bra{{+}i}$ & $\bra{{-}i}$ \\
\hline
ensemble of $\ket{\zero}$ (50\%) and $\ket{\one}$ (50\%) & 50\% & 50\% & 50\% & 50\% & 50\% & 50\% \\
\hhline{*{7}{-}}
\end{tabu}\end{center}
\caption{}
\label{tab:oneQB-ensemble-mixture}
\end{table}}

To understand this in a bit more detail, contrast these results with the case where Alice sends qubits in the state $\ket{+}$.  This is also some kind of ``mixture'' of the states $\ket{\zero}$ and $\ket{\one}$, but it is a special type of coherent ``mixture'', called a superposition, as opposed to the random, ensemble mixture described here%
\footnote{In fact, because of the starkly different behaviour of these states, you should never actually describe a superposition state as being a mixture of different states, but only ever as a superposition of different states.  The term ``mixture'' is generally reserved exclusively for the sorts of ``ensembles'' we are discussing.}.
In the case of the superposition state, $\ket{+}$, we see that there is still one measurement basis ($\{\ket{\pm}\}$) in which Bob will only ever measure the outcome ``$+$'' and never ``$-$''.  But for the ensemble of states, Bob always measures a random mixture of both outcomes, ``$+$'' and ``$-$''.

These results allow us to make the following definition of a new type of state, called a mixed state:
\begin{description}[nosep,topsep=0mm,parsep=0mm,partopsep=0mm,leftmargin=1cm,style=sameline,font=\it]
\item[\textbf{Definition 1:}] \emph{Mixed states are random ensembles of pure states.}
\end{description}
The definition provides a ``classical'' idea of mixture---that is, at any given time, the system is in a specific state (possibly known to someone), but exactly which state may be selected randomly.  This interpretation of mixed states is often ``good enough'', and in fact, we often talk about mixed states as ``being classical''.

\emph{\textbf{Warning:}}  However, this definition contains a critical underlying assumption which can sometimes lead to problems---the assumption of \emph{realism}, that there is always a ``true'' pure state underneath the mixture.  Even if you don't happen to know what it is, it should be possible, at least in principle, for \emph{someone} to know the ``true state''.  This issue is in fact related to the one which lies at the heart of Schr\"odinger and Einstein's great debate over quantum mechanics, which lead to the so-called Einstein-Rosen-Podolsky (EPR) paradox.

We do have to be careful about using this definition.  In quantum physics, mixture is a more complex idea and we don't need to think about ensembles of states.  \emph{It is even possible for an individual quantum system to be mixed!}  To see some of the differences, we now need to look at two-qubit systems.

\subsection{Entanglement}

Let's assume we have a magic quantum box which spits out two-qubit quantum states, every time a third person, Xavier (X), presses a button.  We now consider the following experimental protocol:
\begin{enumerate}[nosep,topsep=0mm,partopsep=0mm]
\item Xavier prepares a two-qubit quantum state and sends ``one half of the state'' (one qubit) to Alice and the ``other half'' (the other qubit) to Bob.
\item Alice and Bob then each measure the qubit they have received in one of the standard qubit bases.
\end{enumerate}

When Alice sent a single qubit to Bob, we already saw that Bob could see correlations between the state Alice sent and the state he measured.  However, moving to systems with two (or more) qubits opens up the possibility for a new concept---that there might be correlations between the states of completely different systems (qubits).

To start with, suppose Xavier's device prepares either $\ket{\zero,\zero}$ or $\ket{\one,\one}$.  The simplest way to see what happens is to write the states in the different measurement bases, i.e., in the $\pm$ basis:
\begin{align}
\ket{\zero,\zero} &= \frac{1}{2} \br{ \ket{+} + \ket{-} } \br{ \ket{+} + \ket{-} } \\
&= \frac{1}{2} \br{ \ket{{+},{+}} + \ket{{+},{-}} + \ket{{-},{+}} + \ket{{-},{-}} } \\
\ket{\one,\one} &= \frac{1}{2} \br{ \ket{+} - \ket{-} } \br{ \ket{+} - \ket{-} } \\
&= \frac{1}{2} \br{ \ket{{+},{+}} - \ket{{+},{-}} - \ket{{-},{+}} + \ket{{-},{-}} }
\end{align}
and similarly, in the ${\pm}i$ basis:
\begin{align}
\ket{\zero,\zero} &= \frac{1}{2} \br{ \ket{{+}i} + \ket{{-}i} } \br{ \ket{{+}i} + \ket{{-}i} } \\
&= \frac{1}{2} \br{ \ket{{+}i,{+}i} + \ket{{+}i,{-}i} + \ket{{-}i,{+}i} + \ket{{-}i,{-}i} } \\
\ket{\one,\one} &= -\frac{1}{2} \br{ \ket{{+}i} - \ket{{-}i} } \br{ \ket{{+}i} - \ket{{-}i} } \\
&= -\frac{1}{2} \br{ \ket{{+}i,{+}i} - \ket{{+}i,{-}i} - \ket{{-}i,{+}i} + \ket{{-}i,{-}i} }
\end{align}
The key point to note is that, apart from sign (phase) differences, each term in these alternative bases has an equal probability of $(1/2)^2=1/4$.  In other words, as in the single-qubit case, if Alice and Bob measure in the ``right basis'' (the same one as the prepared state), they will measure the same outcome every time, but if they measure in the ``wrong basis'', they will measure a random combination of all possible outcomes.

Let us know again consider the slightly more complex scenario where Xavier randomly chooses which of the two above states to send based on a coin toss and doesn't tell Alice and Bob which he is sending---that is, he sends an ``ensemble mixture'' of the pure states $\ket{\zero,\zero}$ and $\ket{\one,\one}$.  As with the single-qubit case, because Xavier doesn't tell Alice and Bob which state he sends, their overall measurement probabilities will be determined by the average of the probabilities calculated from the equations above for $\ket{\zero,\zero}$ and $\ket{\one,\one}$%
\footnote{Note that, whereas in the qubit case above, each of the rows in each of the table ``cells'' had to sum to 100\% probability (because Bob always measured a definite outcome), the interpretation is slightly different here.  Now, each table ``cell'' describes a joint two-qubit measurement made by Alice and Bob.  This means that the probabilities in each cell (over both rows) must sum to 100\%.}:
{\renewcommand{\arraystretch}{1.3}
\begin{table}[h!]
\begin{center}
\begin{tabu}{c*{3}{|>{\centering}p{1cm}>{\centering}p{1cm}}|}
\multicolumn{1}{c}{} & \multicolumn{6}{c}{Bob} \\
Alice & $\bra{\zero}$ & \multicolumn{1}{c}{$\bra{\one}$} & $\bra{+}$ & \multicolumn{1}{c}{$\bra{-}$} & $\bra{{+}i}$ & $\bra{{-}i}$ \\
\hline
$\bra{\zero}$ & 50\% & 0\% &&&& \\
$\bra{\one}$ & 0\% & 50\% &&&& \\
\hhline{~|*{6}{-}}
$\bra{+}$ & && 25\% & 25\% && \\
$\bra{-}$ & && 25\% & 25\% && \\
\hhline{~|*{6}{-}}
$\bra{{+}i}$ & &&&& 25\% & 25\% \\
$\bra{{-}i}$ & &&&& 25\% & 25\% \\
\hhline{*{7}{-}}
\end{tabu}\end{center}
\caption{}
\end{table}}

The interesting point to note here is that, if Alice and Bob compare their measurement results with each other, they will see that their outcomes will be correlated if they are both measuring in the $\zero/\one$ basis: i.e., they will never see the opposite (anticorrelated) results of $\zero\one$ or $\one\zero$.  When they are looking in the other bases, however, they won't be able to see any correlations, because both states in the ensemble give the same flat distribution of measurement outcomes anyway.

We now consider a third scenario which isn't possible in the single-qubit case we discussed above.  Specifically, let's assume that Xavier's device prepares an entangled pure state.  It will prepare exactly the same state each time, but it will be an entangled superposition of $\ket{\zero,\zero}$ and $\ket{\one,\one}$, i.e., in the $\pm$ basis:
\begin{align}
\ket{\psi}  = \frac{1}{\sqrt{2}} \br{ \ket{\zero,\zero} + \ket{\one,\one} } \equiv \ket{\phi^+}
\end{align}
which is usually called the $\phi^+$ (phi-plus) Bell state.  As before, Xavier then sends one qubit to Alice and one to Bob.  Going back to Eq.~\ref{eq:pure-state-measurements}, we can see fairly easily what happens if Alice and Bob both measure in the $\zero/\one$ basis.  In this case, they will measure $\zero\zero$ and $\one\one$ each 50\% of the time, and will never measure $\zero\one$ or $\one\zero$.  In other words, they will observe the same sort of correlations that they would see above in the ensemble mixture we discussed above.  

The story looks quite different, however, when Alice and Bob measure in the other measurement bases.  We can again see what happens to their measurement results by writing the state in the different bases:
\begin{align}
\ket{\phi^+} &= \frac{1}{\sqrt{2}} \sqbr{ \frac{1}{2} \br{ \ket{+} + \ket{-} } \br{ \ket{+} + \ket{-} } + \frac{1}{2} \br{ \ket{+} - \ket{-} } \br{ \ket{+} - \ket{-} } } \\
&=  \frac{1}{\sqrt{2}} \sqbr{ \frac{1}{2} \br{ \ket{{+},{+}} + \ket{{+},{-}} + \ket{{-},{+}} + \ket{{-},{-}} } + \frac{1}{2} \br{ \ket{{+},{+}} - \ket{{+},{-}} - \ket{{-},{+}} + \ket{{-},{-}} } } \\
&= \frac{1}{\sqrt{2}} \br{ \ket{{+},{+}} + \ket{{-},{-}} }
\end{align}
and in the ${\pm}i$ basis:
\begin{align}
\ket{\phi^+} &= \frac{1}{\sqrt{2}} \sqbr{ \frac{1}{2} \br{ \ket{{+}i} + \ket{{-}i} } \br{ \ket{{+}i} + \ket{{-}i} } + \frac{-1}{2} \br{ \ket{{+}i} - \ket{{-}i} } \br{ \ket{{+}i} - \ket{{-}i} } } \\
&= \frac{1}{\sqrt{2}} \br{ \ket{{+}i,{-}i} + \ket{{-}i,{+}i} }
\end{align}
In other words, because the entangled state is a coherent superposition state, when it is rewritten in one of the other standard bases, some of the terms can cancel each other out.  From the resulting simple forms, we can again apply Eq.~\ref{eq:pure-state-measurements} and see that Alice and Bob will now also see correlated measurement outcomes when they are both measuring in either the $\pm$ or ${\pm}i$ bases.  In fact, it turns out that this is quite general---whichever basis Alice measures in, there will always be a corresponding basis which Bob can measure in which will give Alice and Bob correlated measurement outcomes.  The results for the standard qubit bases are summarised in the following table, where I have now included the results for the uncorrelated cases where Alice and Bob measure in different standard bases.
{\renewcommand{\arraystretch}{1.3}
\begin{table}[h!]
\begin{center}
\begin{tabu}{c*{3}{|>{\centering}p{1cm}>{\centering}p{1cm}}|}
\multicolumn{1}{c}{} & \multicolumn{6}{c}{Bob} \\
Alice & $\bra{\zero}$ & \multicolumn{1}{c}{$\bra{\one}$} & $\bra{+}$ & \multicolumn{1}{c}{$\bra{-}$} & $\bra{{+}i}$ & $\bra{{-}i}$ \\
\hline
$\bra{\zero}$ & 50\% & 0\% & 25\% & 25\% & 25\% & 25\% \\
$\bra{\one}$ & 0\% & 50\% & 25\% & 25\% & 25\% & 25\% \\
\hhline{~|*{6}{-}}
$\bra{+}$ & 25\% & 25\% & 50\% & 0\% & 25\% & 25\% \\
$\bra{-}$ & 25\% & 25\% & 0\% & 50\% & 25\% & 25\% \\
\hhline{~|*{6}{-}}
$\bra{{+}i}$ & 25\% & 25\% & 25\% & 25\% & 0\% & 50\% \\
$\bra{{-}i}$ & 25\% & 25\% & 25\% & 25\% & 50\% & 0\% \\
\hhline{*{7}{-}}
\end{tabu}\end{center}
\caption{}
\label{tab:entanglement-correlations}
\end{table}}

This actually provides quite a useful physical definition of entanglement:
\begin{description}[nosep,topsep=0mm,parsep=0mm,partopsep=0mm,leftmargin=1cm,style=sameline,font=\it]
\item[\textbf{Definition 2:}] \emph{Entangled states are states which give correlated measurement outcomes in many measurement bases.}
\end{description}
Thus, we see that quantum correlations (or entanglement) go beyond classical correlations.  This in fact lies at the heart of Bell violation experiments, which probe the EPR paradox.  ``Classical theories'' set a maximum value for the measurement correlations that can be observed in an experiment, but quantum physics breaks that maximum, allowing measurement correlations up to a higher value.

\subsection{Entanglement and mixture}

For our next thought experiment, we are still going to assume that Xavier prepares the $\phi^+$ entangled state and sends one half to Alice and one half to Bob.  Now, however, imagine that Alice and Bob are locked in different rooms and cannot communicate with each other in any way.  This means that Bob now doesn't know either which measurement basis Alice is using (which row of cells he is in within the above table) or which measurement outcome Alice observes (which row within each row of cells in the above table).  In other words, because Bob doesn't know Alice's measurement outcome, his measurement probabilities will just be the sum of probabilities of the two rows in each cell of the table (the top three rows of the following table).  Overall, however, because he doesn't even know which measurement basis Alice is measuring in, his total measurement probability will just be the average of the results he would in principle observe in each case (the last row in the following table).
{\renewcommand{\arraystretch}{1.3}
\begin{table}[h!]
\begin{center}
\begin{tabu}{c*{3}{|>{\centering}p{1cm}>{\centering}p{1cm}}|}
\multicolumn{1}{c}{} & \multicolumn{6}{c}{Bob} \\
Alice & $\bra{\zero}$ & \multicolumn{1}{c}{$\bra{\one}$} & $\bra{+}$ & \multicolumn{1}{c}{$\bra{-}$} & $\bra{{+}i}$ & $\bra{{-}i}$ \\
\hline
$\zero/\one$ & 50\% & 50\% & 50\% & 50\% & 50\% & 50\% \\
\hhline{~|*{6}{-}}
$\pm$ & 50\% & 50\% & 50\% & 50\% & 50\% & 50\% \\
\hhline{~|*{6}{-}}
${\pm}i$ & 50\% & 50\% & 50\% & 50\% & 50\% & 50\% \\
\hhline{*{7}{-}}
?? & 50\% & 50\% & 50\% & 50\% & 50\% & 50\% \\
\hhline{*{7}{-}}
\end{tabu}\end{center}
\caption{}
\end{table}}

As you can see from these results (which can be calculated from the results in Table~\ref{tab:entanglement-correlations}), if Bob can't communicate with Alice, he is not able to see any correlations in his measurement results, and just measures a random distribution of all possible measurement outcomes.  These results therefore look the same as what Bob would see if he was receiving an ensemble of pure single-qubit states (Tab.~\ref{tab:oneQB-ensemble-mixture}).  In other words, although an entangled state is a pure state, one half of an entangled state looks exactly like a mixed state.  In fact, it turns out that it doesn't matter whether Alice measures her qubit or not, as long as it is inaccessible to Bob.  Alice could simply throw her qubit away and Bob's results would look exactly the same.

This gives us the following, more rigorous and precise definition of mixture:
\begin{description}[nosep,topsep=0mm,parsep=0mm,partopsep=0mm,leftmargin=1cm,style=sameline,font=\it]
\item[\textbf{Definition 3:}] \emph{Any mixed state can be written (and interpreted) as a pure entangled state between the quantum system and another hidden system, which we call the ``environment''.}
\end{description}

The next interesting question to ask is how do states become mixed?  To answer this, let us now consider another experimental scenario, based on our original one-qubit experiment.  Suppose that Alice is trying, as before, to send a qubit to Bob, this time in the state $\ket{\zero}$.  \emph{But}, between Alice and Bob, another person, Eve (E), comes along and randomly flips the qubit 50\% of the time, introducing what we call a ``bit-flip error''.  As a result, Bob will receive a random mixture of qubits in the state $\ket{\zero}$ (50\%) and qubits in the state $\ket{\one}$ (50\%).  In other words, as far as Bob can tell, this is exactly the same situation as we considered earlier when Alice sent him an ensemble of randomly chosen pure states.  He will therefore not be able to measure any correlations between the state Alice sends and his own measurement outcome.  This thought experiment describes a very simple example of a \emph{noise} process.  We generally call all such processes \emph{decoherence}.

It is worth noting that if Alice were using these qubits to send information to Bob, then this decoherence process would corrupt or destroy any information she was trying to send.

Consider now another, related experimental scenario.  This time, suppose that Eve comes along between Alice and Bob and induces a special type of collision (interaction) between this qubit and a second qubit she has in her own possession, which then creates an entangled state between the two qubits.  At this point, Bob doesn't even know the second qubit exists.  Therefore, what he gets is again something that looks exactly like a mixed state, and there is no way Bob could ever distinguish between the two situations we have just discussed.  This allows us to make a useful and quite general definition of decoherence:
\begin{description}[nosep,topsep=0mm,parsep=0mm,partopsep=0mm,leftmargin=1cm,style=sameline,font=\it]
\item[\textbf{Definition 4:}] \emph{Decoherence is a process in which a quantum system interacts with its (often inaccessible) environment in some way which creates entanglement between it and its environment.  Alternatively, in slightly more abstract quantum information terms, decoherence is any process which leads to the loss of the quantum information stored in a quantum system (e.g., in the coefficients of an arbitrary pure state, $\ket{\psi} = \alpha \ket{\zero} + \beta \ket{\one}$).}
\end{description}

In the real world, decoherence is not just an ``all or nothing'' process.  We can also have partial decoherence (partial entanglement or partial mixture).  Decoherence can be fast or slow, continuous in time (the typical scenario for circuit QED) or happen at discrete intervals.

\subsection{Decoherence}

Hopefully, our discussion so far has shown that mixture, entanglement, decoherence, noise, measurement and the environment are concepts that are all tangled together.  There are two main types of decoherence that we need to consider: \emph{decay} and \emph{dephasing}.  I will now briefly describe these two types of decoherence and how to measure them.

\subsubsection{Decay}

Decay is a decoherence process where energy leaks into the environment and is often driven by a resonant interaction with surrounding environmental modes.  Decay occurs via spontaneous emission, exactly like radioactive decay, and is induced by vacuum fluctuations in the environment.

Essentially, a qubit in an excited state has a certain probability per unit time of emitting a photon and ``relaxing'' (dropping) into the ground state.  In parallel with radioactive decay, this results in an excited state population (probability) as a function of time, $p_e(t)$, which decays exponentially.  The lifetime for this process, the \emph{decay} or \emph{relaxation lifetime}, is referred to as $T_1$, in exact analogy with the language of NMR.

Measuring the decay lifetime is straightforward via the following protocol:
\begin{enumerate}[nosep,topsep=0mm,parsep=0mm,partopsep=0mm,style=sameline]

\item Prepare the qubit in the ground state, $\ket{\zero}$.
\item At time $t=0$, apply a bit flip ($\sigma_x$) operation to prepare the qubit in the excited state, $\ket{\psi(0)} = \ket{\one}$.
\item Wait a variable time, $t$, and then measure the qubit in the $\zero/\one$ basis ($\bra{\one}$).
\item Repeat this many times for each value of $t$ and estimate the excited-state population, $p_e(t)$, as the fraction of trials where the final measurement outcome is $\one$ (i.e., the qubit is found in the excited state at the end of the trial).
\item Repeat for many different values of $t$ and plot $p_e(t)$ to measure the exponential decay rate.

\end{enumerate}

\subsubsection{Dephasing}

Dephasing is a non-dissipative decoherence process where no energy is exchanged with the environment.  It is often driven by a non-resonant or dispersive interaction with the environmental modes and it is the non-resonant nature of this interaction which prevents energy exchange.  Effectively, dephasing can usually be described as a coupling between the qubit phase and the environment.  Instead of random emission events (as in decay), dephasing can be thought of as random phase kicks on the qubit system, induced by dispersive environmental coupling.

The \emph{dephasing lifetime}, called $T_2$, again by analogy with NMR spectroscopy, describes how long a qubit superposition can remain (i.e., can remain coherent).  The easiest way to understand dephasing is to understand how you measure the dephasing lifetime.

\subsection{Ramsey experiments: measuring the dephasing lifetime}

The basic Ramsey measurement protocol can be described as follows:
\begin{enumerate}[nosep,topsep=0mm,parsep=0mm,partopsep=0mm,style=sameline]

\item Prepare the qubit in the ground state, $\ket{\zero}$.
\item At time $t=0$, prepare the qubit in the equal superposition state, $\ket{+}$ (by applying the appropriate $\sigma_y$-based operation).
\item Allow the state to evolve freely for a variable time, $t$, and then measure the state in the $+/-$ basis ($\bra{+}$).
\item Repeat many times for each value of $t$ to calculate the the probability of finding the qubit in the $\ket{+}$ state.
\item Repeat for many different values of $t$ and plot $p_+(t)$.

\end{enumerate}

So how do the superpositions evolve?  Let us assume for the moment that the initial state is in fact the equal superposition state:
\begin{align}
\ket{\psi(0)} = \frac{1}{\sqrt{2}} \br{ \ket{\zero} + e^{i\phi} \ket{\one} }
\end{align}
where I have included an arbitrary phase offset for later.  Let us also assume that the qubit's free evolution Hamiltonian is given by $\mathcal{H} = -\half \hbar \Delta \sigma_z$, which corresponds to a system with an energy gap, $\hbar \Delta$.

Using the same techniques we have already applied in earlier sections and in tutorials, we can calculate the unitary evolution operator:
\begin{align}
U(t) = \sqbr{\begin{matrix} \exp(\half i \Delta t) & 0 \\ 0 & \exp(-\half i \Delta t) \end{matrix}}
\end{align}
which in turn allows us to calculate the evolved state:
\begin{align}
\ket{\psi(t)} = e^{i\phi/2} \sqbr{\begin{matrix} \exp(\half i \Delta t - \half i \phi) \\ 0 & \exp(-\half i \Delta t + \half i \phi) \end{matrix}}
\end{align}
which shows that the superposition state undergoes continuous phase evolution during the unitary free-evolution stage according to $\phi(t) = \Delta t - \phi(0)$.  This finally gives the measurement probability:
\begin{align}
p_+(t) = |\braket{+}{\psi(t)}|^2 = \half + \half \cos (\Delta t - \phi)
\end{align}
where the initial offset phase $\phi$ now simply gives a phase shift in the fringe pattern.

In order to develop a further intuition about what is happening, it's useful to think about what this evolution looks like on the Bloch sphere (Fig.~\ref{fig:blochsphere}).  Because the state always remains an equal superposition state, it always stays on the equator of the Bloch sphere.  The continuous phase evolution therefore looks like continuous rotation of the state around the $\sigma_z$ access at a constant velocity determined by the qubit gap or resonant frequency, $\Delta$.

As discussed above, dephasing noise can be thought of as small random phase kicks to the superposition state.  In relation to the above calculations, this results is gradual phase diffusion around the expected (mean) evolving phase value.  In other words, at $t=0$ (just after preparation of the initial superposition state), the phase offset will be very well defined (ideally $\phi(0) = 0$, if the state preparation is correct).  As $t$ evolves, however, the phase offset gradually gets smeared out.  As you average over many different trials, the measured fringe pattern will therefore start to look like a sum of many fringe patterns with slightly different phase offsets.  What effect does this have on the observed oscillations?

A basic intuition of this can be obtained from a simple calculation of what happens if the observed fringe pattern is the sum of just two fringe patterns with different phase offsets $\phi=0$ and $\phi=\delta$:
\begin{align}
p_+(t) &= \frac{1}{2} \br{ \frac{1}{2} + \frac{1}{2} \cos (\Delta t) + \frac{1}{2} + \frac{1}{2} \cos (\Delta t - \delta) } \\
&= \frac{1}{2} + \frac{1}{4} \sqbr{ \cos (\Delta t - \delta/2 + \delta/2) + \cos (\Delta t - \delta/2 - \delta/2) } \\
\nn
&= \frac{1}{2} + \frac{1}{4} [ \cos(\Delta t {-} \delta/2) \cos(\delta/2) - \sin(\Delta t {-} \delta/2) \sin(\delta/2) \\
&\qquad + \cos(\Delta t {-} \delta/2) \cos(\delta/2) - \sin(\Delta t {-} \delta/2) (-\sin(\delta/2)) ] \\
&= \frac{1}{2} + \frac{1}{2} \cos(\Delta t - \delta/2) \cos(\delta/2)
\end{align}
In other words, the resulting is just a single sinusoidal fringe pattern with phase offset $\phi=\delta/2$, but not with a reduced amplitude given by $\cos(\delta/2)$.  If $\delta$ is small, then this term is just close to one and the measured oscillation amplitude is close to its maximum value.  As $\delta$ increases, however, the amplitude gets smaller.  Similarly, as you add more and more terms in this sum, the amplitude will again get smaller.  In fact, in the extreme case, where the phase is completely random from shot to shot, it should be fairly easy to convince yourself that there will be no observed oscillations at all.

\begin{figure}
\begin{center}
\includegraphics[width=120mm]{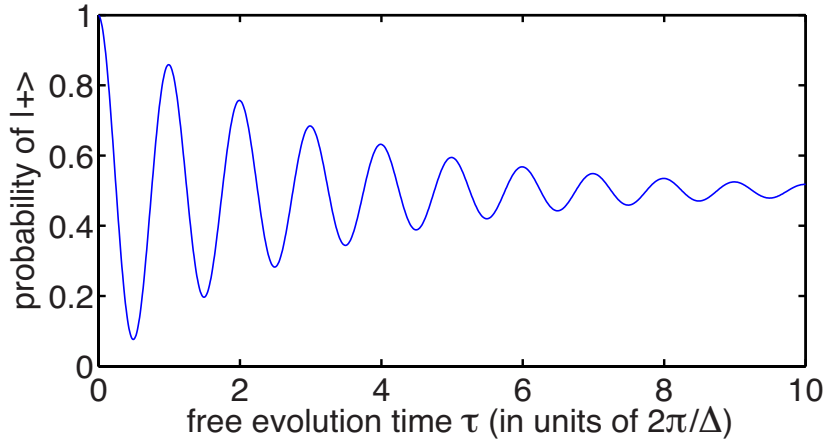}
\captionsetup{labelfont=bf,font={small},margin=6pt}
\caption{\emph{\textbf{Measurement fringes from a typical Ramsey experiment.}}}
\label{fig:ramseyfringes}
\end{center}
\end{figure}

In the general case, the amplitude of the fringe pattern (the oscillating population) will itself decay exponentially as a function of time, with a lifetime given by $T_2$.  Figure~\ref{fig:ramseyfringes} shows a typical example of exponentially decaying Ramsey fringes.

\subsection{Good qubits}

Now that we understand the different types of decoherence, we are now in a position to answer the question of whether we have a good qubit or not.  To do this, we go back to our original motivation, namely that we want to build a quantum system which can store coherent quantum information for long enough that we can control and manipulate that information.  In other words, we want to be able to create, manipulate and measure superposition states, before the coherence vanishes.

It turns out, therefore, that this requires both $T_1$ (decay lifetime) and $T_2$ (dephasing lifetime) to be long.  In the case of decay, if you create a superposition state, $\ket{+}$, the coherence can clearly only survive while there is population in both $\ket{\zero}$ and $\ket{\one}$ states.  Spontaneous decay which causes the excited state to relax into the ground state ($\ket{\one} \rightarrow \ket{\zero}$) will therefore also destroy the superposition, because there can't be any coherent superposition if the qubit has completely relaxed into the ground state.  On the other hand, dephasing by definition destroys the coherence of the superposition, even if the fractional populations of the ground and excited states remains unaffected.

But how long is long enough?  For this we need to think back to how qubit gates (operations) can be implemented.  In the fast-gate scenarios, the frequency of the gate operations is determined by the energy splitting between the qubit levels at the chosen operating point---exactly the same as the oscillation frequency of the Ramsey fringe experiment.  A $\pi$ phase shift therefore requires at least $t=\pi/\Delta$ to be implemented.

As a result, these decay lifetimes set an upper limit on the number of system operations that can be implemented before any qubit superpositions decohere (vanish).  A good quantum system needs many phase oscillations to be possible within the decay lifetimes, both $T_1$ and $T_2$.  If $T_1$ and $T_2$ are much larger than the Ramsey oscillation period (as determined by $\Delta$), this means many operations are possible before any quantum information encoded in the system is lost.

\subsection{Combined decoherence processes}

Understanding the two different types of decoherence, we can also ask whether the lifetimes of the two noise processes are linked in any way.

On the one hand, it is fairly straightforward to explain that $T_2$ doesn't influence $T_1$.  To see this, simply consider what happens if you excite a qubit into the state $\ket{\one}$ and watch how fast it decays.  Since phase noise only affects superposition states, adding phase noise doesn't make any different to the excited state decay rate.  Spontaneous decay only creates a mixture of excited and ground states, not a superposition, so there is no coherence for the dephasing to destroy.

This is not true for the reverse direction, however.  As we discussed above, the coherence in a superposition state, $\ket{+}$, can only survive while there is population in both $\ket{\zero}$ and $\ket{\one}$ states.  If you add in extra spontaneous decay, then once the excited state has completely relaxed into the ground state ($\ket{\zero}$), there can also no longer be any coherence left.  In fact, it turns out that $T_2 \le 2T_1$.  This can be seen roughly by considering the free evolution of a general superposition state:
\begin{align}
\ket{\psi(0)} = \cos\theta \ket{\zero} + e^{i\phi} \sin\theta \ket{\one}
\end{align}
Using the same techniques as above, it can be shown that the Ramsey fringe measurement probability is
\begin{align}
p_+(t) = \half + \half \sin 2\theta \cos (\Delta t - \phi)
\end{align}
whereas the excited state probability is
\begin{align}
p_\one(t) = \sin^2 \theta
\end{align}
For small $\theta$, the oscillation amplitude is therefore proportional to $\theta$, while the excited state population is proportional to $\theta^2$.  This means that oscillations ($T_2$) should vanish half as fast as populations ($T_1$) for a system which is only affected by decay (i.e., where there is no dephasing and the Ramsey fringe decay results only from population decay), which is the limiting case of the general rule that $T_2 \le T_1$.  (This is because if $T_2=2T_1$, then $\exp (-t/T_1) = \exp (-2t/T_2) = \sqbr{\exp (-t/T_2)}^2$.)

\subsection{Exercises---entanglement and decoherence}

\subsubsection{Entanglement and the Bell state basis}

The standard basis to use when talking about a two-qubit system is the computational basis, defined by the states: \{$\ket{\zero,\zero}$, $\ket{\zero,\one}$, $\ket{\one,\zero}$, $\ket{\one,\one}$\} (where, e.g., $\ket{\zero,\zero} \equiv \ket{\zero}_a\otimes\ket{\zero}_b$).  Another useful basis is defined by the four special entangled states known as the Bell states:
\begin{align}
\ket{\phi^\pm} &= \frac{1}{\sqrt{2}} \br{ \ket{\zero,\zero} \pm \ket{\one,\one} } \\
\ket{\psi^\pm} &= \frac{1}{\sqrt{2}} \br{ \ket{\zero,\one} \pm \ket{\one,\zero} }
\end{align}

\begin{enumerate}[label=(\alph*)]

\item Show that the Bell states form a basis by verifying that they are orthogonal to each other.

\item For each Bell state, rewrite them using the alternative standard bases for both subsystems, i.e., $\{ \ket{\pm, \pm} \}$ and $\{ \ket{{\pm}i, {\pm}i} \}$.

\item Consider now the experiment where one half of the Bell state (i.e., one subsystem) is sent each to Alice ($a$) and Bob ($b$), who randomly measure the system they receive in one of the standard qubit bases and record the results.  This is repeated many times and Alice and Bob use the results to estimate measurement probabilities.  Calculate the expected probabilities for two scenarios: (i) Alice and Bob tell each other what measurement bases they used after they record their measurements, and (ii) Alice and Bob don't tell each other anything.  As an example, do this for the Bell state $\ket{\Phi^-}$ and record the results in a table like the one below:
{\renewcommand{\arraystretch}{1.3}
\begin{table}[h!]
\begin{center}
\begin{tabu}{c*{3}{|>{\centering}p{1cm}>{\centering}p{1cm}}|>{\centering}p{1cm}|}
\multicolumn{1}{c}{} & \multicolumn{7}{c}{Bob} \\
Alice & $\bra{\zero}$ & \multicolumn{1}{c}{$\bra{\one}$} & $\bra{+}$ & \multicolumn{1}{c}{$\bra{-}$} & $\bra{{+}i}$ & $\bra{{-}i}$ & ? \\
\hline
$\bra{\zero}$ & & & & & & & \\
$\bra{\one}$ & & & & & & & \\
\hhline{~|*{7}{-}}
$\bra{+}$ & & & & & & & \\
$\bra{-}$ & & & & & & & \\
\hhline{~|*{7}{-}}
$\bra{{+}i}$ & & & & & & & \\
$\bra{{-}i}$ & & & & & & & \\
\hhline{*{8}{-}}
? & & & & & & & \multicolumn{1}{c}{} \\ \hhline{-------}
\end{tabu}\end{center}
\end{table}}

\item In scenario (ii) above, what other type of state do these results look like?

\item Now suppose that Alice's subsystem is instead entangled with an unknown, inaccessible environment.  How do these results relate to decoherence in quantum systems?

\end{enumerate}

\subsubsection{Frequency noise and dephasing}

From the notes above and previous exercises, we know that, for a qubit with energy gap $\hbar \Delta_0$ in an arbitrary initial state $\ket{\psi(0)} = \alpha \ket{\zero} + \beta \ket{\one}$, the freely evolving state as a function of time (evolving according to the Hamiltonian $\mathcal{H} = -\hbar \Delta_0 \sigma_z / 2$) will be of the form:
\begin{align}
\ket{\psi(t)} = \alpha e^{+i\Delta_0 t/2} \ket{\zero} + \beta e^{-i\Delta_0 t/2} \ket{\one}
\end{align}
In other words, free evolution under a Hamiltonian of the form given above doesn't change the balance of the superposition, but just gives a time-dependent relative phase shift between the two energy eigenstates in the superposition.

\begin{enumerate}[label=(\alph*)]

\item Consider now a qubit where the energy gap is subject to some random noise fluctuations, $\Delta(t) = \Delta_0 + \delta\Delta(t)$, distributed around the previous value of $\Delta_0$.  The free Hamiltonian for this system is now $\mathcal{H}(t) = -\hbar \sqbr{\Delta_0 + \delta\Delta(t)} \sigma_z / 2$.  Because this is now a time-dependent Hamiltonian, to solve the dynamics of this system, we need to go back to the differential form of the Schr\"odinger equation:
\begin{align}
\frac{d}{dt}\ket{\psi(t)} = -\frac{i}{\hbar} \mathcal{H} \ket{\psi(t)}
\end{align}

\begin{enumerate}[label=\roman*.]
\item Writing the instantaneous state at time $t$ as $\ket{\psi(t)} = \alpha(t) \ket{\zero} + \beta(t) \ket{\one}$ and substituting this into the Schr\"odinger equation, derive the following differential equations for the coefficients:
\begin{align}
\nn
\frac{d\alpha}{dt} &= \frac{1}{2} i\sqbr{\Delta_0 + \delta\Delta(t)} \alpha(t) \\
\nn
\frac{d\beta}{dt} &= -\frac{1}{2} i\sqbr{\Delta_0 + \delta\Delta(t)} \beta(t)
\end{align}
\item Based on the form of the solution for the time-independent case, we can guess that the solution for $\alpha(t)$ and $\beta(t)$ will be of the form:
\begin{align}
\nn
\alpha(t) &= \alpha(0) \exp \sqbr{A(t)} \\
\nn
\beta(t) &= \beta(0) \exp \sqbr{B(t)}
\end{align}
Substituting this into the differential equations above, derive the following differential equations for the unknown functions in the exponent:
\begin{align}
\nn
\frac{dA}{dt} &= \frac{1}{2} i\sqbr{\Delta_0 + \delta\Delta(t)} \\
\nn
\frac{dB}{dt} &= -\frac{1}{2} i\sqbr{\Delta_0 + \delta\Delta(t)}
\end{align}
\item Integrating these equations, show that the final solution for $\alpha(t)$ and $\beta(t)$ is:
\begin{align}
\nn
\alpha(t) &= \alpha(0) \exp \sqbr{\frac{1}{2} i\br{\Delta_0 t + \int_0^t \delta\Delta(\tau) d\tau}} \\
\nn
\beta(t) &= \beta(0) \exp \sqbr{-\frac{1}{2} i\br{\Delta_0 t + \int_0^t \delta\Delta(\tau) d\tau}}
\end{align}
\end{enumerate}

\item Suppose this noisy system is now used in a Ramsey experiment, where the initial state is $\ket{\psi(0)} = \ket{+}$ and the $\ket{+}$ state is measured again after a variable time $t$.

\begin{enumerate}[label=\roman*.]
\item Using the above solution for $\ket{\psi(t)}$, show the final measurement probability is:
\begin{align}
\nn
p_+(t) \equiv |\braket{+}{\psi(t)}|^2 = \frac{1}{2} \sqbr{ 1 + \cos \br{ \Delta_0 t + \int_0^t \delta\Delta(\tau) d\tau} }
\end{align}
\item This result shows that noise in a qubit's energy gap leads to phase offset noise and thus reduced amplitudes in Ramsey oscillations.  Briefly discuss how this relates to avoided crossings and operating ``sweet spots'' in charge or flux qubits.
\end{enumerate}

\end{enumerate}

\bibliographystyle{refsallauthors}
\bibliography{abbreviations,langford,lectures}

\end{document}

%% file: lecturesdefinitions.tex
\newlength{\manualrowheight}
\newcommand{\manualrowspace}[1][1]%
  {\makebox[0mm]{\normalsize\setlength{\manualrowheight}{#1\baselineskip}
  \vphantom{\rule{0mm}{\manualrowheight}}}}

\newcommand{\zero}{\ensuremath{\mathbf{0}}}
\newcommand{\one}{\ensuremath{\mathbf{1}}}

\newcommand{\undertilde}[1]{\rlap{\smash{\lower 2ex \hbox{$\tilde{\hphantom{#1}}$}}}#1}
\newcommand{\lundertilde}[1]{\rlap{\smash{\lower 2.2ex \hbox{$\tilde{\hphantom{#1}}$}}}#1}
\newcommand{\bvec}[1]{\ensuremath{{\bf #1}}}		
\newcommand{\uvec}[1]{\ensuremath{\undertilde{#1}}}		

\newcommand{\bra}[1]{\langle#1|}		
\newcommand{\ket}[1]{|#1\rangle}		
\newcommand{\braket}[2]{\langle#1|#2\rangle}		
\newcommand{\matelem}[3]{\langle#1|#2|#3\rangle}		
\newcommand{\expect}[1]{\ensuremath{\langle#1\rangle}}  
\newcommand{\br}[1]{\left(#1\right)}		
\newcommand{\sqbr}[1]{\left[#1\right]}		
\newcommand{\cbr}[1]{\left\{#1\right\}}		

\newcommand{\dash}{^{\prime}}		
\newcommand{\ddash}{^{\prime\prime}}		
\newcommand{\dasq}[2]{\frac{\partial^2#1}{\partial#2^2}}		
\newcommand{\smfrac}[2]{\begin{array}{@{}c@{}}\frac{#1}{#2}\end{array}}		
\newcommand{\half}{\smfrac{1}{2}}		
\newcommand{\nn}{\nonumber}
\newcommand{\vcurl}{\nabla\times}		
\newcommand{\dint}{\int\!\!\!\!\int}			
\newcommand{\real}[1]{\text{Re}\cbr{#1}}			
\newcommand{\imag}[1]{\text{Im}\cbr{#1}}			
\newcommand{\gives}{\Rightarrow\quad}			

